\documentclass[sigconf]{acmart}

\usepackage{acmart-taps}
\AtBeginDocument{%
  \providecommand\BibTeX{{%
    \normalfont B\kern-0.5em{\scshape i\kern-0.25em b}\kern-0.8em\TeX}}}

\usepackage{xspace}
\usepackage{enumitem}
\usepackage{microtype}

\aptLtoXcmd{\newcommand{\topicdef}[1]{{\leavevmode\apttarget{topic:#1}{\textcolor{brown}{\textbf{#1}}}}\space}
\newcommand{\requdef}[1]{{\leavevmode\apttarget{requ:#1}{\textcolor{blue}{\textbf{#1}}}}\space}

\newcommand{\topic}[1]{{\aptlink{topic:#1}{\textcolor{brown}{\textbf{#1}}}}}
\newcommand{\requ}[1]{{\aptlink{requ:#1}{\textcolor{blue}{\textbf{#1}}}}}
\newcommand{\goal}[1]{{\aptlink{goal}{\textcolor{violet}{\textbf{#1}}}}}
\newcommand\itema{\item[]{\leavevmode\apttarget{goal}{\textbf{\textcolor{violet}{G1}}}}\space}
\newcommand\itemb{\item[]\textbf{\textcolor{violet}{G2}}\space}
\newcommand\itemc{\item[]\textbf{\textcolor{violet}{G3}}\space}
}{\newcommand{\topicdef}[1]{\textcolor{brown}{\textbf{#1}}}
\newcommand{\requdef}[1]{\textcolor{blue}{\textbf{#1}}}

\newcommand{\topic}[1]{\hyperref[topic:#1]{\textcolor{brown}{\textbf{#1}}}}
\newcommand{\requ}[1]{\hyperref[requ:#1]{\textcolor{blue}{\textbf{#1}}}}
\newcommand{\goal}[1]{\hyperref[goal]{\textcolor{violet}{\textbf{#1}}}}
\newcommand\itema{\item[\textbf{\textcolor{violet}{G1}}]}
\newcommand\itemb{\item[\textbf{\textcolor{violet}{G2}}]}
\newcommand\itemc{\item[\textbf{\textcolor{violet}{G3}}]}}

\newcommand{\tool}{\textsc{AutoVis}\xspace}
\copyrightyear{2023} 
\acmYear{2023} 
\setcopyright{rightsretained} 
\acmConference[CHI '23]{Proceedings of the 2023 CHI Conference on Human Factors in Computing Systems}{April 23--28, 2023}{Hamburg, Germany}
\acmBooktitle{Proceedings of the 2023 CHI Conference on Human Factors in Computing Systems (CHI '23), April 23--28, 2023, Hamburg, Germany}
\acmDOI{10.1145/3544548.3580760}
\acmISBN{978-1-4503-9421-5/23/04}

\begin{document}

%%
%% The "title" command has an optional parameter,
%% allowing the author to define a "short title" in page headers.
%\title[\tool]{\tool: Combining In-Situ and Ex-Situ Visual Analytics for Analyzing Automotive User Interface Interaction Studies}
%\title[\tool]{\tool: Enabling Collaborative Mixed-Immersive Analysis of Automotive User Interface Interaction Studies}
\title[\tool]{\tool: Enabling Mixed-Immersive Analysis of Automotive User Interface Interaction Studies}

\author{Pascal Jansen}
\email{pascal.jansen@uni-ulm.de}
\orcid{0000-0002-9335-5462}
\affiliation{%
  \institution{Institute of Media Informatics, Ulm University}
  \city{Ulm}
  \country{Germany}
}

\author{Julian Britten}
\email{julian.britten@uni-ulm.de}
\orcid{0000-0002-2646-2727}
\affiliation{%
  \institution{Institute of Media Informatics, Ulm University}
  \city{Ulm}
  \country{Germany}
}

\author{Alexander Häusele}
\email{alexander.haeusele@uni-ulm.de}
\orcid{0000-0001-7005-1485}
\affiliation{%
  \institution{Institute of Media Informatics, Ulm University}
  \city{Ulm}
  \country{Germany}
}

\author{Thilo Segschneider}
\email{thilo.segschneider@uni-ulm.de}
\orcid{0000-0002-9511-9252}
\affiliation{%
  \institution{Institute of Media Informatics, Ulm University}
  \city{Ulm}
  \country{Germany}
}

\author{Mark Colley}
\email{mark.colley@uni-ulm.de}
\orcid{0000-0001-5207-5029}
\affiliation{%
  \institution{Institute of Media Informatics, Ulm University}
  \city{Ulm}
  \country{Germany}
}

\author{Enrico Rukzio}
\email{enrico.rukzio@uni-ulm.de}
\orcid{0000-0002-4213-2226}
\affiliation{%
  \institution{Institute of Media Informatics, Ulm University}
  \city{Ulm}
  \country{Germany}
}

%%
%% By default, the full list of authors will be used on the page
%% headers. Often, this list is too long and will overlap
%% other information printed in the page headers. This command allows
%% the author to define a more concise list
%% of authors' names for this purpose.
\renewcommand{\shortauthors}{Jansen et al.}

%%%%%%%%%%%%%%%%%%%%%%%%%%%%%%%%%%%%%%%%%%%%%%%%%%%%%%%%%%%%%%%%%%%%%%%%%%%%%%%%%%%
%%%%%%%%%%%%%%%%%%%%%%%%%%%%%%%%%%%%%%%%%%%%%%%%%%%%%%%%%%%%%%%%%%%%%%%%%%%%%%%%%%%
%% The abstract is a short summary of the work to be presented in the
%% article.
\begin{abstract}
% 150 words or less
Automotive user interface (AUI) evaluation becomes increasingly complex due to novel interaction modalities, driving automation, heterogeneous data, and dynamic environmental contexts. 
Immersive analytics may enable efficient explorations of the resulting multilayered interplay between humans, vehicles, and the environment.
However, no such tool exists for the automotive domain.
With \tool, we address this gap by combining a non-immersive desktop with a virtual reality view enabling mixed-immersive analysis of AUIs.
We identify design requirements based on an analysis of AUI research and domain expert interviews (N=5).
\tool supports analyzing passenger behavior, physiology, spatial interaction, and events in a replicated study environment using avatars, trajectories, and heatmaps.
We apply context portals and driving-path events as automotive-specific visualizations.
To validate \tool against real-world analysis tasks, we implemented a prototype, conducted heuristic walkthroughs using authentic data from a case study and public datasets, and leveraged a real vehicle in the analysis process.
%We demonstrate \tool's potential of improving AUI analysis through interactive exploration and re-experience.

% What is the large scope and problem space? Why should we care? Motivation
% What is the specific problem addressed? Problem
% Why is the problem Important? Why was this work carried out?
% What have we done? Solution
% What did we find out? What are the concrete results?
% What are the implications on a larger scale? How does it change the bigger picture?
\end{abstract}


%%
%% Keywords. The author(s) should pick words that accurately describe
%% the work being presented. Separate the keywords with commas.

%% A "teaser" image appears between the author and the affiliation
%% information and the body of the document, and typically spans the
%% page.
\begin{teaserfigure}
  \includegraphics[width=\textwidth]{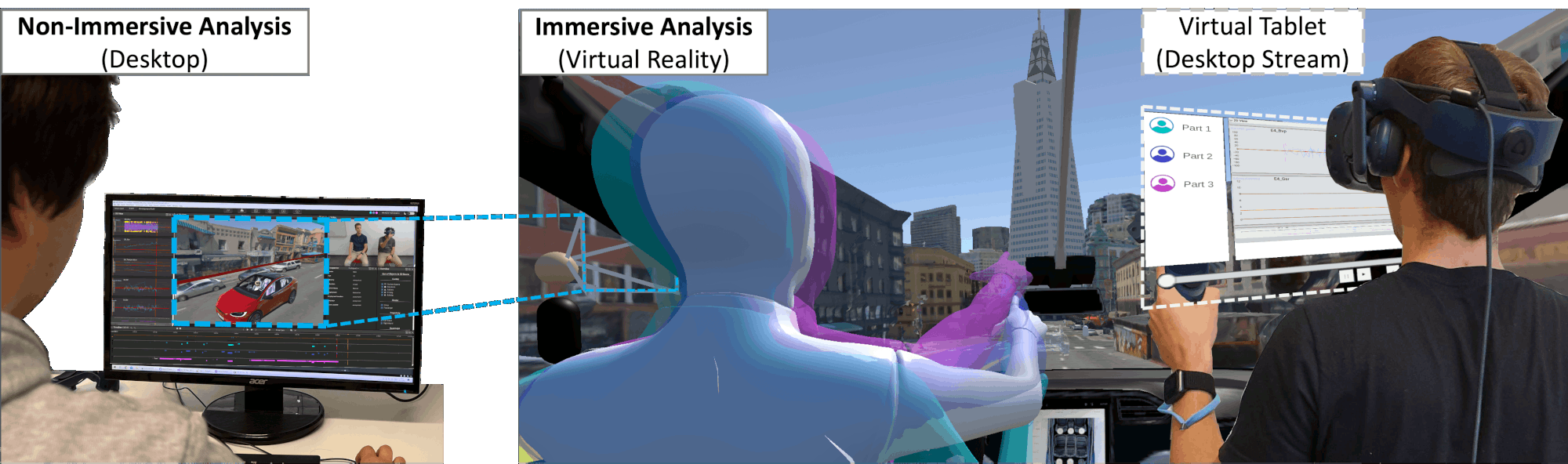}
  \caption{\tool combines an immersive VR view with a synchronized non-immersive desktop view for analyzing automotive user interface studies. In VR, analysts can re-experience and analyze studies in an immersive environment. On the desktop, they can analyze aggregated data. A virtual tablet attached to the controller visualizes the desktop view in VR.}
  \Description{The image is split into two parts: On the left, there is a real human sitting in front of a computer using the desktop version of \tool. This is the non-immersive view. On the right, a virtual vehicle interior in a 3D city environment can be seen. In the foreground on the left side, there are multiple semi-transparent and opaque humanoid avatars on the driver's seat pointing at a building in front of them. In the right foreground is a real human wearing a virtual reality headset and holding a VR controller. On the VR controller, a cutout of the \tool desktop view is visible. This is the immersive view of \tool. }
  \label{fig:teaser}
\end{teaserfigure}

%%
%% This command processes the author and affiliation and title
%% information and builds the first part of the formatted document.
\maketitle

%%%%%%%%%%%%%%%%%%%%%%%%%%%%%%%%%%%%%%%%%%%%%%%%%%%%%%%%%%%%%%%%%%%%%%%%%
%%%%%%%%%%%%%%%%%%%%%%%%%%%%%%%%%%%%%%%%%%%%%%%%%%%%%%%%%%%%%%%%%%%%%%%%%
\section{Introduction}
\label{introduction}
% Motivation
With the increasing technological advances of driving automation (enabling non-driving related activities \cite{detjen_wizard_2020}) and vehicle systems (e.g., touch screens \cite{detjen_driving_2021}, embedded sensors \cite{kim_seated_2020}, or speech assistance \cite{lin_adasa_2018}), human-vehicle interaction analysis becomes more complex.
Today's AUI development often considers numerous factors, such as the vehicle environment \cite{bethge_vemotion_2021,liu_empathic_2021}, novel input and output modalities \cite{jansen_design_2022}, multimodal interaction \cite{aftab_multimodal_2019}, the temporal sequence of interactions, user behavior \cite{liu_empathic_2021}, and user physiology \cite{trenta_advanced_2019}.
Besides, other road users~\cite{10.1145/3411764.3445480} can interact with the vehicle via external human-machine interfaces (eHMIs) \cite{de_external_2019, colley_investigating_2021, colley_user_2022, colley2022effects, colley_towards_2020}.
Consequently, practitioners gather large bodies of heterogeneous, spatio-temporal data of movements, user interactions, audio/video recordings, and other events that need to be visualized and analyzed to gain the desired insights into how people use novel AUIs.
%Tools that directly relate in-vehicle passenger behavior to the environment may benefit the efficient and more effective analysis of such data.
Efficient and effective data analysis may benefit from tools that directly relate in-vehicle passenger behavior to the environment.
%Besides, immersive interactions with virtual replication of studies can facilitate the development of future AUIs and inspire design explorations.

% Problem
However, today’s analysis tools for AUIs (e.g., \cite{tavakoli_harmony_2021, ebel_visualizing_2021}) are non-immersive, 2D, bounded to desktops, and distancing the analysts from the often complex environments outside and inside the vehicle that might have a critical impact on passenger behavior.
As a result, analysis can be time-consuming (e.g., see \cite{jakobsen_upclose_2014,dey_pedestrian_2017,hecht_what_2020,madigan_understanding_2019}), and data patterns may be invisible during analysis from a fixed point-of-view (POV) or are unclear without the original environmental context.
Likewise, due to the difficult instrumentation of participants (e.g., on-body sensors) and costly external camera systems (e.g., LiDAR \cite{lidar}), datasets may contain only videos of the driving environment or the interior (e.g., \cite{jegham_mdad_2019}).
Therefore, (manual) analysis of public datasets might be incomplete or require extensive post-processing. 
%For this, dataset preprocessing could use machine/deep learning to automatically infer relevant events, emotions, and objects.
Besides, current tools do not support each driving automation level defined by the Society of Automotive Engineers (SAE) taxonomy J3016 \cite{sae_levels}.
However, automated driving significantly affects passengers \cite{dettman_comfort_2021, 10.1145/3409120.3410648, 10.1145/3534609, 10.1145/3411764.3445351} and other road users \cite{xu_when_2021}.
Furthermore, current tools often do not adequately support the various forms of multimodal in-vehicle interaction (e.g., \cite{aftab_multimodal_2019, kim_cascaded_2020,roider_just_2018}), which combine input modalities such as gaze, gesture, speech, and touch.

In contrast, immersive analytics enable the spatial analysis of interaction and movement data in augmented (AR) and virtual reality (VR) within a replicated study environment \cite{buschel_miria_2021,kloiber_immersive_2020,lilija_who_2020}.
However, immersive tools, similar to \cite{reipschlager_avatar_2022,hubenschmid_relive_2022}, are currently nonexistent in the automotive research domain.
This prevents an outside-in view that may be needed to get an overview of the data \cite{buschel_miria_2021}.
%On the other hand, ex-situ tools are often well-suited for overview tasks \cite{kraus_assessing_2020}, in-situ tools allow analysts to re-experience the context of the original setting and facilitate the analysis of 3D data \cite{lilija_who_2020,hubenschmid_relive_2022}.
However, non-immersive tools are well-suited for overview tasks, such as quickly finding relevant events \cite{kraus_assessing_2020}.
Therefore, in line with \citet{hubenschmid_relive_2022}, we argue that effective and efficient analysis of AUI studies requires immersive and non-immersive tools.
In combination, these approaches compensate for their drawbacks and have the potential to overcome barriers between practitioners, their data, and the tools they use for analyzing and understanding human-vehicle interaction.

%Research Questions:
%More specifically, which aspects of the analysis process benefit from immersive representation, and which aspects are better served by using a non-immersive visual analytics interface?
%How can we transfer context information and support users when switching from, e.g., an immersive VR device to a non-immersive desktop environment and back?

% Solution
To unleash this potential, we propose \tool, a tool for the mixed-immersive analysis of AUI studies.
\tool combines an immersive VR with a non-immersive desktop view (see \autoref{fig:teaser}) to enable in-depth visualization of passenger states and interactions in- and outside the vehicle, aiming to create a strong link between data and environment.
The VR view replicates a scenario in an interactive 3D environment explorable via VR head-mounted displays (HMDs).
This view combines trajectory-based visualizations with interactive humanoid 3D avatars, providing a detailed representation of passengers’ movements and interactions.
Additionally, heatmaps aggregate passengers' gaze, pointing, or speech references in the vehicle or environment.
%Additional aggregated visualizations (e.g., heatmaps) embedded into the 3D environment show where passengers interacted with the vehicle/environment by gaze, pointing, or referencing via speech.
To further visualize the link between in-vehicle and environmental contexts, we employ context portals.
In VR, analysts can walk in the vehicle's surroundings, sit on the passengers' seats, relive behavioral, physiological, or interaction-related events, and re-experience the study data within its original context.
In addition, analysts can leverage a real vehicle using passthrough VR.
A virtual tablet (see \autoref{fig:teaser}) enables an overview of physiological and event data to aid explorations.

Complementary to VR, a desktop view serves as a visual analysis workbook to summarize, link, explore, and compare details of different study sessions' spatio-temporal, event, and nominal data.
The desktop view also offers a 2D window into the current 3D VR scene.
View synchronization enables control of tool-wide playback.
Moreover, \tool enables transitions between VR and desktop view, therefore, representing a hybrid UI (see \cite{hubenschmid_relive_2022, reipschlager_avatar_2022}).
In addition, the interplay between the VR and desktop view enables collaborative use of \tool in different levels of immersion, time, and space (see \ref{autovis_interplay_vr-desktop}).
%\tool can scale to several study scenarios in the AUI domain (see \ref{autovis_literature_analysis}), such as takeovers, conversational systems, eHMIs, and the ideation and validation of novel interactive systems.
By applying deep learning (DL) approaches for the automatic event, emotion, and object inference, \tool can convert datasets with low context details (e.g., only containing driving/passenger videos) into datasets of high context fidelity.

We evaluated a prototype of \tool to identify the advantages and challenges of combining immersive and non-immersive views for analyzing AUI studies.
For this, we investigated three use cases to systematically validate our concepts against current AUI research topics.
First, we conducted a use case study on multimodal interaction in automated vehicles (AVs).
%This use case covers various modalities (e.g., touch, gesture, distant pointing, and gaze) and different contexts (e.g., ride-sharing, sightseeing, or unexpected events).
Second, we demonstrate the conversion and visualization of a real-world dataset (Drive\&Act \cite{driveact}) via \tool.
Finally, we leveraged a real vehicle in the analysis process.
In the evaluation, we apply Olsen's heuristics \cite{olsen_evaluating_2007} to investigate how \tool supports analysis tasks.
Our evaluation focuses on the applicability of our analysis concept, the interplay between immersive and non-immersive views, and how immersion fosters an effective analysis.

\textit{Contribution Statement:}
(1) The concept of mixed-immersive analysis of AUI studies, utilizing 3D avatars, motion trajectories, and aggregated visualizations embedded in virtually replicated environments.
(2) Automotive domain-specific 3D visualization concepts using context portals and driving-path events, and leveraging a real vehicle for analysis via passthrough VR.
(2) A prototype implementation of our concepts. We share a demo\footnote{The \tool demo website: \url{https://autovis.onrender.com}} and the open-source repository\footnote{The \tool repository: \url{https://gitlab.com/Pascal-Jansen/autovis}} with the research community to enable the design of future interactive systems.
(3) Insights and research implications derived from a heuristic evaluation of three use cases highlighting how \tool can be used to analyze AUI interactions.

%- “Enabling” contributions are resources that facilitate the development of future interactive systems and inspire future interface design explorations.
%- “Enabling” contributions include datasets, tools, libraries, infrastructure, and languages.
%- These contributions will be judged by how well they support the construction, engineering, or validation of interactive systems and how well they can be shared among the research community to design future interactive systems.

%This should adapt \inlinegraphics{example-image-a}.
%Lorem ipsum \highlight{\inlinegraphics{example-image-a} dolor} sit amet, \highlightr{consectetuer} adipiscing elit, \highlightg{consectetuer} adipiscing elit, \highlightb{consectetuer} adipiscing elit

%\printinunitsof{cm}\prntlen{\textwidth}

%%%%%%%%%%%%%%%%%%%%%%%%%%%%%%%%%%%%%%%%%%%%%%%%%%%%%%%%%%%%%%%%%%%%%%%%%
%%%%%%%%%%%%%%%%%%%%%%%%%%%%%%%%%%%%%%%%%%%%%%%%%%%%%%%%%%%%%%%%%%%%%%%%%
\section{Related Work}
\label{related_work}
\tool builds upon work from several research domains.
Therefore, we look into:
(1) immersive and non-immersive analytics and spatio-temporal data visualizations, and
(2) related work on current desktop-based analysis and visualization of in-vehicle interactions.
%(3) approaches for sensing environmental, behavioral, and physiological signals in automotive research studies.

\subsection{Immersive \& Non-Immersive Analytics}
%%%% Non-Immersive Analytics %%%%
A common method for user study analysis is to use \textbf{non-immersive} tools, such as Tableau \cite{tableau} and Spotfire \cite{spotfire}, or toolkits like D3.js \cite{d3} and R \cite{r}.
%Computing notebooks and visual analysis workbooks are often used complementary (e.g., \cite{badam_vistrates_2018,perez_ipython_2007}), combining code, explanations, and visualizations.
Previous work also employed non-immersive visualizations of spatio-temporal data.
For example, augmented top-down views \cite{brudy_eagle_2018,chittaro_vuflow_2006,decamp_immersive_2010,tang_vistaco_2010,zadow_giant_2017} and 3D views \cite{bruning_pamocat_2012,decamp_immersive_2010,prilla_analysis_2018} with various visualizations, such as trajectories \cite{brudy_eagle_2018,bruning_pamocat_2012,chittaro_vuflow_2006,decamp_immersive_2010,tang_vistaco_2010}, heatmaps \cite{brudy_eagle_2018,chittaro_vuflow_2006,zadow_giant_2017}, and field of view frustums \cite{brudy_eagle_2018,chittaro_vuflow_2006,prilla_analysis_2018}.
Often, timelines (e.g., \cite{brudy_eagle_2018,bruning_pamocat_2012,chittaro_vuflow_2006,decamp_immersive_2010,marquardt_excite_2015,zadow_giant_2017}) annotated by events \cite{brudy_eagle_2018,bruning_pamocat_2012,marquardt_excite_2015,zadow_giant_2017} control these visualizations.
Non-immersive visualizations are mostly complemented with videos to synchronize movement visualizations with the recordings (e.g., \cite{brudy_eagle_2018,bruning_pamocat_2012,decamp_immersive_2010,marquardt_excite_2015,zadow_giant_2017}).
Still, non-immersive tools may require time-consuming adaptions to custom scenarios (e.g., AUIs) and do not allow re-experiencing the data in the original environmental context.
In contrast, \tool is well-suited towards multivariate data, such as video, audio, sensor, and nominal data gathered in such quantity and variety, specifically in the AUI domain.

%%%% Immersive Analytics %%%%
Recent work has presented various \textbf{immersive} toolkits facilitating the analysis of spatio-temporal data, for example, obtained from mixed-reality studies \cite{buschel_miria_2021,nebeling_mrat_2020,kloiber_immersive_2020,lilija_who_2020}.
They mostly focused on placing classic visualizations, such as scatter plots or bar charts, in immersive environments \cite{butcher_vria_2021,filho_evaluating_2020,hubenschmid2021stream,rosenbaum2011involve,saenz2017reexamining,wagner2018immersive,zhang_using_2015}.
However, these approaches represent only two spatial dimensions.
There are also 3D trajectories and 3D point plots to visualize the position and speed of participants’ heads and hands \cite{kloiber_immersive_2020,lilija_who_2020,nebeling_mrat_2020}, gaze cues \cite{prilla_analysis_2018}, tracked objects \cite{nebeling_mrat_2020,lilija_who_2020,buschel_investigating_2017}, and events \cite{nebeling_mrat_2020}.
In contrast to these approaches, where the visualizations have little relation to their original environmental context, \tool places visualizations directly in a virtually replicated study environment.
%Similarly, Kloiber et al. \cite{kloiber_immersive_2020}, Lilija et al. \cite{lilija_who_2020}, and GhostAR \cite{cao_ghostar_2019} visualized human motion in VR using 3D trajectories and proxy objects of the hands and heads of multiple users.
%The 3D trajectories and proxies are visualized in the same virtually replicated environments in which they were originally captured.
Similarly, MIRIA \cite{buschel_miria_2021} and ReLive \cite{hubenschmid_relive_2022} enable in-situ movement and interaction data analysis by rendering 3D trajectories, proxy objects, and in the case of MIRIA, additional visualizations such as heatmaps on walls or floors.
AvatAR \cite{reipschlager_avatar_2022} increased the movement and interaction detail by replicating user postures in AR.
To further enrich their 3D visualizations, \citet{reipschlager_avatar_2022} placed 2D visualizations (e.g., heatmaps) on a physical tablet.
However, all these works on immersive analytics target mixed-reality studies that consider (multiple) HMD users interacting in static areas (e.g., rooms) with constant environmental contexts.
They do not support the automotive context (without requiring extensive adaptions) with its unique combination of moving interaction area (the vehicle) and dynamic environment with numerous actors (e.g., pedestrians, bicycles, or cars).
In contrast, \tool provides immersive visualizations tailored to the requirements of AUI research.

%All these works target similar problems as AvatAR, but they do not offer a detailed analysis of a person’s posture, being limited to either a single position per user [11] or the positions of the head and both hands [18, 34, 40] for each time frame.
%In contrast, AvatAR aims to improve upon these approaches by providing a much more detailed replication of a user's posture to facilitate the in-depth analysis of their behavior.
%In contrast, Kraus et al. [35] work provides detailed 3D models of people and a 3D environment replicated from camera images, which can be explored in VR.
%However, being a VR environment, it lacks the in-situ aspect AvatAR aims for and instead relies on teleportation to explore the environment.

Overall, AUI researchers can use non-immersive or immersive tools to analyze user studies and experiments.
Non-immersive tools offer flexibility and reproducibility (e.g., via computational notebooks).
Immersive tools can support decision-making \cite{reichherzer_bringing_2021} and increase understanding of spatial data \cite{kraus_impact_2020,reipschlager_avatar_2022,hubenschmid_relive_2022}, physiological data \cite{lee_data_2021}, and environmental context \cite{buschel_miria_2021,reipschlager_avatar_2022,hubenschmid_relive_2022}.
However, discomfort issues (e.g., HMD weight \cite{yan_effects_2019} or simulator sickness \cite{kennedy_simulator_1993}) can make these immersive approaches unattractive.
Therefore, we propose an analysis of AUI studies using immersive and non-immersive approaches, enabling to choose and transition between different immersion levels based on the analysis task.
A similar tool exists for analyzing mixed-reality studies, see \cite{hubenschmid_relive_2022}.
However, there is currently no approach enabling the mixed-immersive visual analysis of AUI interaction studies.

%Yet, most existing tools only support one or the other; users have to either choose one or spend significant effort in migrating their current analysis workflow to a different reality. 

\subsection{Visualization of In-Vehicle Interactions}
Practitioners in the AUI domain commonly use non-immersive desktop tools, such as Tableau \cite{tableau}, Dovetail \cite{dovetail}, or R \cite{r}, to analyze study data, for example, video \cite{dey_pedestrian_2017}, audio \cite{wang_conversational_2022}, physiological \cite{dillen_keep_2020}, behavioral \cite{hernandez_autoemotive_2014}, or tabular data \cite{kim_design_2018}.
However, for many research topics (e.g., multimodal interaction, conversational systems, or eHMIs), such analysis is time-consuming and relevant insights on user behavior may be superficial or overlooked.

Besides, current tools are often specific to a research topic and, therefore, not (directly) applicable to other topics.
For example, Blickshift Analytics \cite{blickshift} primarily enables analysis of eye-based interaction via scan paths and gaze heatmaps without immersive replication of the study environment.
In contrast, \citet{ebel_visualizing_2021} presented an approach for analyzing passenger behavior.
Their tool visualizes the events, time on tasks, and interaction sequences for touchscreens.
Still, there is no relation with the vehicle environment or the passenger state (e.g., physiology) and little relation to the in-vehicle space, which is only textually represented (e.g., button names).
However, such visual relations would facilitate in-depth analysis of user behavior, yielding more relevant insights faster \cite{nebeling_mrat_2020,hubenschmid_relive_2022,reipschlager_avatar_2022}.
\citet{tavakoli_harmony_2021} presented a first promising approach.
They fused context with driver-specific measures such as heart rate to understand the interplay of the vehicle environment and the driver's state.
However, their tool only focuses on manual driving (SAE 0), and their non-interactive visualizations provide no spatial information on passenger movements, impeding detailed analysis.

Altogether, there is no tool for interactive immersive and only insufficient tools for non-immersive analysis of human-vehicle interactions.
Besides, solutions providing 3D environments that relate in-vehicle interactions with the vehicle environment are missing.
With \tool, we aim to fill this gap for any driving automation level (SAE 0-5).

%%%%%%%%%%%%%%%%%%%%%%%%%%%%%%%%%%%%%%%%%%%%%%%%%%%%%%%%%%%%%%%%%%%%%%%%%
%%%%%%%%%%%%%%%%%%%%%%%%%%%%%%%%%%%%%%%%%%%%%%%%%%%%%%%%%%%%%%%%%%%%%%%%%
\section{\tool: Process \& Requirements}
\label{autovis_process}
The \tool development process is based on the Design Study Methodology proposed by \citet{sedlmair_design_2012} and the development of novel visualization concepts (see \cite{langner_marvis_2021, buschel_miria_2021}).
The process consists of (1) Ideas \& Scope, (2) Requirements, (3) Concepts, (4) Prototype \& Demonstration, and (5) Evaluation.
This section explains the design rationale behind the \tool concepts (Section \ref{autovis_concepts}).
We provide results from a literature analysis of AUI research (see \ref{autovis_literature_analysis}), report results from five expert interviews (see \ref{autovis_expert_interviews}), and describe the derived toolkit requirements (see \ref{autovis_requirements}).

%%%%%%%%%%%%%%%%%%%%%%%%%%%%%%%%%%%%%%%%%%%%%%%%%%%%%%%%%%%%%%%%%%%%%%%%%
\subsection{Analysis of AUI Research}
\label{autovis_literature_analysis}
As part of the \tool development process, we analyzed existing AUI research.
For this, literature reviews served as a starting point, see \cite{jansen_design_2022,ayoub_10years_2019,detjen_how_2021}.
We then retrieved relevant publications via backward chaining with a depth of two.
For further inclusion, publications should:
(1) capture heterogeneous (spatio-) temporal data, or
(2) consider in-vehicle and environment contexts, or
(3) investigate interactions with novel modalities.
Based on this analysis, we identified AUI research topics that would benefit from tool support:
%In the following, each research topic summarizes how an immersive analytics tool could improve existing workflows:

\paragraph{\topicdef{T1} Ideation of Novel In-Vehicle Interaction}
\label{topic:T1}
In AUI research, practitioners create novel UI interaction concepts and suggest future research directions from creative processes, for example, using design spaces \cite{jansen_design_2022} or brainstorming \cite{dey_interface_2018}.
They also design novel input and output modalities leveraging, for example, olfactory \cite{dimitrenko_caroma_2020} or thermal \cite{hernandez_autoemotive_2014} sensations.
We found that early work on proof-of-concept prototypes often conducted user studies.
Here, an immersive analysis tool could provide insights into the spatio-temporal usage of novel interactions (e.g., olfactory) to inform further iterations about their interplay with other modalities and the environment.

\paragraph{\topicdef{T2} Evaluation of In-Vehicle UIs}
\label{topic:T2}
In general, AUI evaluations are part of user-centered design processes containing user interviews \cite{funk_nonverbal_2020}, field studies \cite{wintersberger_man_2018}, and lab studies \cite{normark_design_2015}.
The goal is to measure, for example, the usability \cite{qi_dynamic_2022}, trust \cite{colley_increasing_2021}, acceptance \cite{wintersberger_fostering_2018}, or perceived safety \cite{osswald_predicting_2012} of UIs.
Evaluations use driving simulators of varying fidelity (e.g., \cite{colley_swivrcarseat_2022, hock2022vampire}) or real vehicles \cite{goedicke_vroom_2018}.
We found that multiple evaluation methods are combined due to complex dynamics between passengers, vehicles, and the environment.
For example, quantitative logs (e.g., passenger video, audio, and vehicle telemetry) combined with qualitative measurements (e.g., interviews).
A tool could ease such combined analysis and thus enable more efficient and effective AUI evaluations by visualizing relations between the results of different evaluation methods.

\paragraph{\topicdef{T3} Know Thy Passenger}
\label{topic:T3}
Sensors, cameras, and wearables collect large amounts of real-time information about passengers, such as heart rate \cite{pakdamanian_deeptake_2021}, skin conductance \cite{dillen_keep_2020}, or pupil size \cite{benedetto_driver_2011}.
This enables context-adaptive applications \cite{scholkopf_conception_2021} and facilitates driver assistance \cite{kundinger_performance_2021} or infotainment system interactions \cite{rittger_adaptive_2022}.
Various variables indicate the passenger's state, such as cognitive load, emotion, fatigue, stress, drowsiness, attention, distraction, and situation awareness.
Therefore, manually inferring actions (e.g., a button click) from a passenger state is challenging and often automated using DL (e.g., \cite{ahmad_predictive_2018}).
However, DL and user modeling can be complex and, therefore, impractical in early development.
Automatic conversion of data (e.g., tabular or video) into AUI domain-tailored visualizations could enable manual pattern detection to determine data worth further investigation.
In addition, the impact of in-vehicle, environmental, and social factors on passengers' physiological and behavioral signals and vice versa is difficult to determine.
An immersive tool could support such analysis by providing spatio-temporal context visualizations paired with passenger sensor data.

\paragraph{\topicdef{T4} Driver Distraction}
\label{topic:T4}
In today's vehicles, drivers can operate infotainment functions while driving, but this should not affect safety.
As a result, distraction is still one of the primary problems of infotainment systems \cite{schmidt_driving_2010}.
Research focuses on reducing display glances, for example, employing speech dialogue systems \cite{chang_usability_2009} or auditory output \cite{kun_glancing_2009}.
However, we found a challenge in determining in-vehicle and environmental distraction factors.
Besides, distracted driver interactions with AUIs can only be assessed via manual coding of video recordings and interaction logs.
A tool could help identify distraction factors without explicitly logging them based on prior assumptions by visually aggregating related events.

\paragraph{\topicdef{T5} Multimodal In-Vehicle Interaction}
\label{topic:T5}
Multimodal UIs provide multiple modes of interaction between passengers and vehicles. 
This can be sequential, requiring mode switches or allowing multiple modes simultaneously (see \cite{jansen_design_2022}).
%Using more than a single human perception channel for information transport can support information processing \cite{}.
Practitioners presented various approaches for multimodal in-vehicle interaction, such as combining gaze with speech \cite{roider_just_2018}, gestures \cite{gomaa_studying_2020}, or touch \cite{kim_cascaded_2020}.
We found that researchers are mainly interested in usable combinations of modalities at different interaction locations.
However, modalities' usage order, time interval, selection, and environmental contexts are challenging to determine without tool support, as spatio-temporal information is difficult to assess from tabular, video, and audio files.

\paragraph{\topicdef{T6} Conversational and Speech-Based UIs}
\label{topic:T6}
%Speech can convey rich content and emotion.
Speech enables hands- and glances-free interactions during driving \cite{weng_conversational_2016}.
Conversational systems support drivers in assisted (SAE 1-2), automated driving (SAE 3-5), and non-driving-related activities.
For example, practitioners proposed interactions with artificial speech \cite{kim_predicting_2019,guo_driving_2021} and voice commands \cite{winzer_intersection_2018}.
Speech interactions visualized as graphs or dialog structures often fail to provide insight into speech-referred in-vehicle and environment contexts.
A tool could overcome this by visualizing speech interactions as events in relation to other events and the driving environment.

\paragraph{\topicdef{T7} Control Transitions Between Vehicle and Driver}
\label{topic:T7}
Current automation systems have operational design domains requiring control transitions between driver and vehicle \cite{morales_takeover_2020}.
Practitioners presented various methods for safe and comfortable transitions.
For example, they assessed driver's readiness \cite{chen_takeover_2018} and used various modalities for transition requests, such as visual \cite{radlmayr_how_2014}, auditory \cite{politis_language_2015}, and tactile \cite{BAZILINSKYY201882} modalities.
We found that researchers often measure the transition quality using driving performance logs and driver reaction time.
However, the driving environment which impacts the transition quality can only be determined if the scenario is set up in advance.
A tool replicating the driving environment could enable insights into traffic density, road conditions, and shared driver-vehicle situation awareness for various transition scenarios.

\paragraph{\topicdef{T8} External Human Machine Interfaces}
\label{topic:T8}
When semi- and fully AVs are introduced, vehicles may feature eHMIs to communicate (safety-related) information to other drivers \cite{colley2022investigating} and vulnerable road users, such as cyclists, pedestrians, and impaired people \cite{colley_towards_2020, haimerl2022evaluation}.
For example, practitioners proposed concepts for safe road crossing \cite{eisma_external_2020} and automation mode indications \cite{lee_understanding_2019}.
However, we found that a challenge in assessing eHMIs' visibility, placement, and content design is the limited access to other road users' POVs in real-world settings.
A tool replicating the driving environment in 3D could provide these POVs.

%%%%%%%%%%%%%%%%%%%%%%%%%%%%%%%%%%%%%%%%%%%%%%%%%%%%%%%%%%%%%%%%%%%%%%%%%
\subsection{AUI Domain Expert Interviews}
\label{autovis_expert_interviews}
In the next development step, we invited five AUI researchers (PhD students) from our institute for individual interviews.
The goal was to reflect upon the research topics (see \ref{autovis_literature_analysis}) and elicit common challenges in analyzing AUI studies to identify promising design choices for a mixed-immersion tool.
All experts have developed, conducted, and analyzed AUI studies.
The five semi-structured interviews lasted approx. one hour.
Similar to \citet{langner_marvis_2021}, we provided a list of typical visualization tasks (the interaction categories by \citet{yi_towards_2007}) to guide the discussion towards immersive analytics, as the interviewees were no experts in designing such systems.
In addition, we presented the research topics \topic{T1} - \topic{T8} obtained from the literature analysis (see \ref{autovis_literature_analysis}) as an impulse for discussion and to set the scope.
However, we omitted the description of how each research topic could benefit from tool support to prevent biasing the interviewees.

The interviews comprised elements of brain-writing.
First, interviewees (E1 - E5) collected their thoughts on each research topic as a written text.
After approx. 32 minutes (four minutes for each topic), both parties discussed the written thoughts for the remaining time.
The interviewer took notes for later analysis.
Three authors discussed, labeled, and coded these notes into themes (see Appendix \ref{interview_appendix}).
The themes are: visualize object positions and movements (E1, E2, E3, E4), enable collaborative analysis (E2, E3, E4, E5), visualize data interdependencies (E2, E4), allow data annotations (E4), enable data filtering (E1, E5), include a real vehicle in the analysis (E1, E3, E4), enable mixed-immersion analysis and transitions between desktop and VR (E2, E3, E4, E5), link in-vehicle and environmental contexts (E3), and preprocess data automatically (E4, E5).
The themes guided the specifications for the toolkit requirements, the concept design, and the prototype.

In addition, we derived four challenges specific to the AUI domain:
\textbf{Large Distances Between Objects of Interest:} A driving environment can span several kilometers with a large distance between objects of interest (e.g., stores or landmarks), referenced in natural interactions (e.g., gesture, speech, or gaze).
\textbf{Volatile In-Vehicle and Environmental Contexts:} Passengers and the vehicle can interact bidirectionally. If, in addition, these interactions relate to the environment, also hosting object interactions (e.g., between cars and pedestrians), volatile interactions between in-vehicle and environmental contexts emerge.
\textbf{Ubiquitous Physiological Measures:} As interior embedded sensors always surround passengers, physiological measures are ubiquitous. However, the resulting data is not spatial. Nevertheless, effective analysis of AUI studies requires a link to spatio-temporal data, such as passenger and vehicle movements.
\textbf{Different Actors:} In- and outside the vehicle, there is an interplay of diverse actors (e.g., passengers, drivers, other vehicles, bicyclists, or pedestrians), each with different goals, characteristics, and behaviors in a traffic situation.

%%%%%%%%%%%%%%%%%%%%%%%%%%%%%%%%%%%%%%%%%%%%%%%%%%%%%%%%%%%%%%%%%%%%%%%%%
\subsection{Toolkit Requirements}
\label{autovis_requirements}
%Each research topic (T1-T8) may benefit from a detailed analysis.
%Therefore, we suggest that 2D and 3D visualizations combined in an ex-situ and in-situ environment may assist with various AUI research tasks.
%We adapted toolkit requirements from prior immersive analytics \cite{buschel_miria_2021} to the AUI domain.
After analyzing AUI research (T1 - T8, see \ref{autovis_literature_analysis}) and interviewing domain experts (E1 - E5, see \ref{autovis_expert_interviews}), we derived recurring themes and challenges.
Based on these challenges, we list functional requirements (R1 - R9) that a system, such as \tool, should address.
We integrate the categories by \citet{yi_towards_2007} into the requirements to ground them in widely used interaction techniques in information visualization.
The categories are \textit{Select} (mark something as interesting), \textit{Explore} (show me something else), \textit{Reconfigure} (show me a different arrangement), \textit{Encode} (show me a different representation), \textit{Abstract/Elaborate} (show me more or less detail), \textit{Filter} (show me something conditionally), and \textit{Connect} (show me related items).

\paragraph{\requdef{R1} \textbf{Visualization of Position and Movement Data}}
\label{requ:R1}
AUI research uses positions and movements over time to investigate behavior (\topic{T1}, \topic{T2}, \topic{T4}, \topic{T5}, \topic{T7}) (E1 - E4).
%Such spatio-temporal movement data could be used in novel interactions, such as automatic switching from touch to speech when the passenger has no hands free or suggesting design improvements, for example, when a button is not reachable in specific situations.
In \textit{Abstract/Elaborate} interactions, analysts should be able to retrieve movement trajectories, positions of other road users (\topic{T8}) (E2, E4), and gaze/gesture directions (E2), needed to describe UI interactions.
Besides, spatio-temporal object positions should enable understanding the context in \textit{Connect} and help to reveal movement patterns in \textit{Select} interactions.

\paragraph{\requdef{R2} \textbf{Visualization of Event Data}}
\label{requ:R2}
In addition to spatio-temporal data, AUI studies also gather event data, including interactions (e.g., gesture or speech events) (\topic{T1}, \topic{T2}, \topic{T4} - \topic{T7}), application events (e.g., mode switches or task completions) (\topic{T1} - \topic{T7}), and passenger events (e.g., emotional or cognitive state) (\topic{T1} - \topic{T8}).
In \textit{Connect} interactions, analysts should be able to obtain event order and (co-)occurrence, as they may reveal patterns and dependencies.
Moreover, analysts should be able to explore these patterns in detail over time within \textit{Explore} and \textit{Filter} interactions.

\paragraph{\requdef{R3} \textbf{Visualization of Data Interdependence}}
\label{requ:R3}
Researchers often measure physiological states to understand passengers (\topic{T1} - \topic{T7}).
However, temporal data may be difficult to interpret without the context of events or user behavior (E2, E4, and see \cite{lohani2019review}).
Therefore, temporal should be visualized close to spatio-temporal and event-based data.
In \textit{Connect} interactions, analysts should directly compare and discover dependencies with other temporal data streams, events, and passenger behavior (E2, E4).

\paragraph{\requdef{R4} \textbf{Linking In-Vehicle and Environment Contexts}}
\label{requ:R4}
In-vehicle interactions and the environment can be highly related (E1 - E5).
For example, passengers may refer to environment objects via gaze or pointing \cite{gomaa_studying_2020} (E3).
Therefore, AUI study analysis (in \topic{T1} - \topic{T7}) might benefit from merged virtual replications of both contexts.
%This should help analysts to understand in which contexts passengers interact with in-vehicle UIs.
Besides, sensor data gathered in (field) studies should be virtually replicated and linked to the in-vehicle context (E3).
For example, for \textit{Connect} interactions aiming to analyze eHMI interactions (\topic{T8}).

\paragraph{\requdef{R5} \textbf{Filtering, Flexibility, and Gradual Control over Visualizations}}
\label{requ:R5}
When evaluating and comparing interaction data in \topic{T1} - \topic{T8}, filtering allows focusing on subsets (E1).
Analysts should be able to perform \textit{Filter} interactions for sessions, study conditions, participants, and visualizations' time and location to account for spatio-temporal data.
Depending on the study data, the appropriate visualization placement may require flexibility and reconfigurability (E5).
For example, changing visualizations' visibility and arrangement during \textit{Reconfigure} and controlling their detail via \textit{Abstract/Elaborate} interactions.
There should also be approaches to overcome analysts' disabilities, such as colorblindness. %, to include a diverse community.

\paragraph{\requdef{R6} \textbf{Data Annotations}}
\label{requ:R6}
A typical AUI study analysis task is video or audio coding (e.g., see \cite{dey_pedestrian_2017}) to annotate situations or behaviors in \topic{T1} - \topic{T8} (E4).
Such \textit{Select} interactions support highlighting undefined events and labeling datasets, for example, for DL (E4).
Moreover, such annotations should relate to space and time.

\paragraph{\requdef{R7} \textbf{Integration of Video and Audio Recordings}}
\label{requ:R7}
Most AUI studies in \topic{T1} - \topic{T8} record video and audio to evaluate participant behavior.
Audio becomes particularly relevant for speech interactions (\topic{T1}, \topic{T2}, \topic{T5}, \topic{T6}).
Therefore, such recordings should be integrated into a single UI to prevent frequent program window switches to display different data types (E5).

\paragraph{\requdef{R8} \textbf{Leveraging a Real Vehicle}}
\label{requ:R8}
A real vehicle increases the external validity in \topic{T1} - \topic{T8}.
In such field studies, interactions highly depend on the vehicle interior (e.g., center console layout or seat configurations).
Besides, study supervisors often cannot be part of the study, for example, in naturalistic driving studies, and want to relive the experiments (E4).
Therefore, an immersive tool should integrate a real vehicle into the analysis.
In \textit{Connect}, \textit{Encode}, \textit{Filter}, and \textit{Select} interactions, analysts should be able to tangibly analyze AUIs and leverage the real vehicle (E1) via passthrough VR.

\paragraph{\requdef{R9} \textbf{Enabling Collaborative Analysis}}
\label{requ:R9}
Practitioners usually work in larger groups (E2, E3, E4, E5) and collaboratively share, evaluate, and iterate over data.
In addition, they collaborate across different locations and time zones (E2). 
Therefore, a tool should provide a shared and persistent analysis environment (E3).
Collaborators should be able to analyze the data at hand (a)synchronously in different locations, with different available technologies (e.g., VR and desktop, see \ref{expert_results_enabling-mixed-immersion}).

%%%%%%%%%%%%%%%%%%%%%%%%%%%%%%%%%%%%%%%%%%%%%%%%%%%%%%%%%%%%%%%%%%%%%%%%%
%%%%%%%%%%%%%%%%%%%%%%%%%%%%%%%%%%%%%%%%%%%%%%%%%%%%%%%%%%%%%%%%%%%%%%%%%
\section{\tool: Concepts}
\label{autovis_concepts}
Based on \requ{R1} - \requ{R9}, we propose \tool: a tool enabling the mixed-immersive and interactive analysis of in-vehicle interactions, passenger behavior, and physiology combined with a replicated vehicle environment and study context.
To the best of our knowledge, \tool is the first immersive analysis tool for the AUI domain.

The core of \tool is the interplay of a desktop and VR view (see cross-device interactions \cite{sereno2022hybrid,wang2020towards}).
This enables seamless transitions in single- and multi-user analysis (see \ref{autovis_collab_taxonomy}).
The desktop view provides a non-immersive overview of AUI studies.
In the immersive VR view, 2D (e.g., heatmaps on in-vehicle surfaces) and 3D visualizations (e.g., passenger avatars) may improve the understandability of spatial data.
Besides, a virtual tablet with an overview of the desktop view supports relating the VR environment with non-spatial data, such as physiological data.
Moreover, \tool integrates video and audio playback into one layout.
Thus, \tool can provide study context details that may not otherwise be found in more abstract visualizations.

Although \tool provides two analysis views (desktop and VR), analysts are not intended to switch between tools, as switching between a hybrid system can be cumbersome \cite{hubenschmid_relive_2022}.
Instead, they can use the desktop view in the preparation or wrap-up of the analysis process and use the VR as the main view during analysis.
In case of unavailable VR devices, the desktop can also serve as a main view with limited immersion.

This section describes the data processing pipeline, the views, and features of \tool based on the requirements informed by our development process (see \ref{autovis_process}), and the interplay between desktop and VR view in single and multi-user scenarios.

%%%%%%%%%%%%%%%%%%%%%%%%%%%%%%%%%%%%%%%%%%%%%%%%%%%%%%%%%%%%%%%%%%%%%%%%%
\subsection{Data Processing Pipeline}
\label{autovis_processing_pipeline}
To address requirements \requ{R5}, \requ{R7}, and \requ{R9}, we designed \tool's processing pipeline to be flexible and versatile in supporting a wide range of data, configurations, and applications.
We describe \tool's processing regarding the allowed data specification and the automatic detection of objects, events, and driving environments.

%\begin{figure*}[t!]
%        \centering
%        \includegraphics[width=.24\textwidth]{example-image-a}
%        \includegraphics[width=.24\textwidth]{example-image-b}
%        \includegraphics[width=.24\textwidth]{example-image-a}
%        \includegraphics[width=.24\textwidth]{example-image-b}
%    \caption{A horizontal (small height) figure similar to the pipelines of MIRIA and ReLive}
%    \Description{A horizontal (small height) figure similar to the pipelines of MIRIA and ReLive}
%    \label{fig:processing_pipeline}
%\end{figure*}

\begin{figure*}[ht!]
        \centering
        \includegraphics[width=\textwidth]{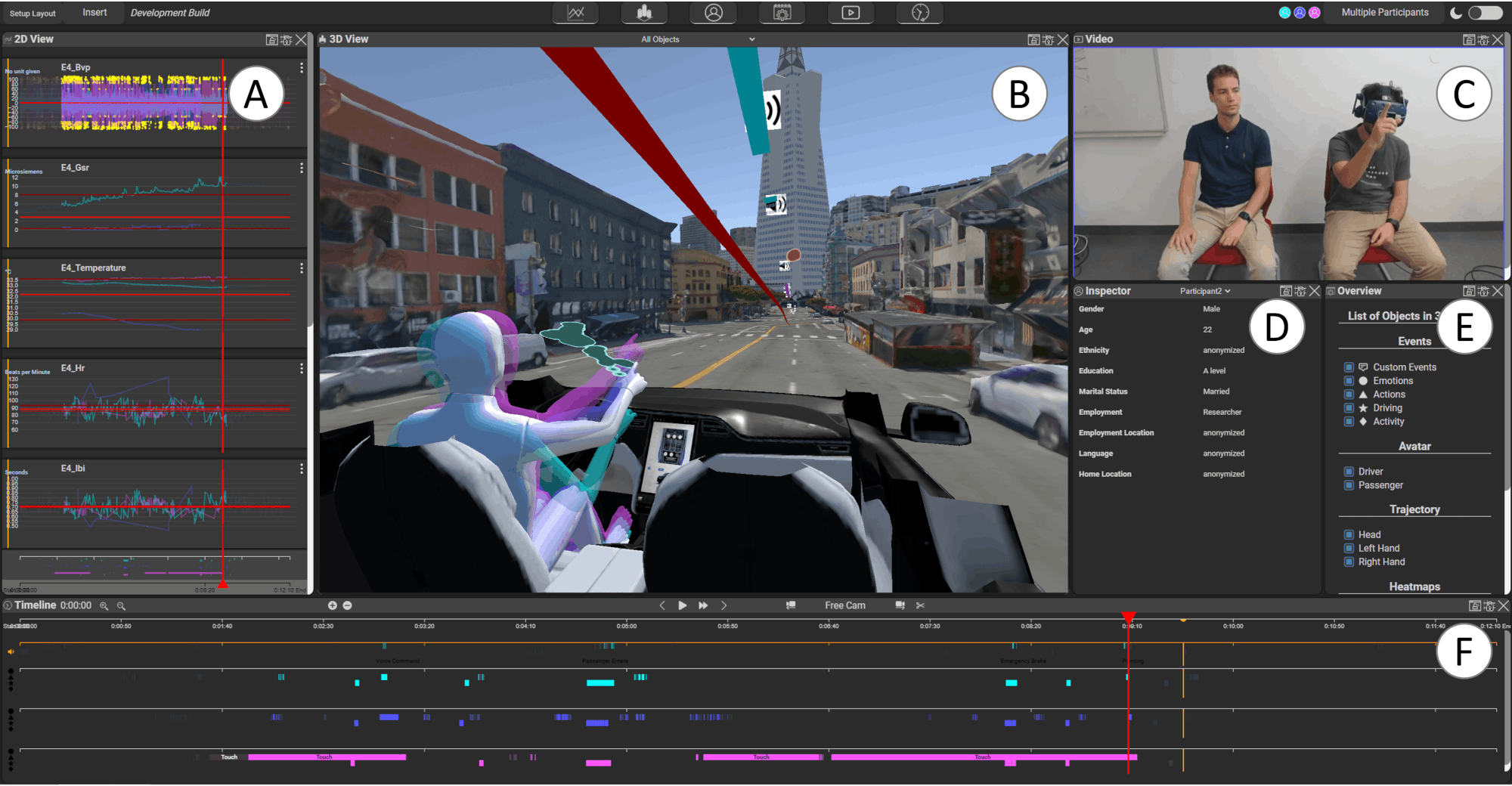}
    \caption{The \textbf{\tool desktop view} combines non-immersive analysis of aggregated data with a 3D view replicating the study environment. Additional data such as videos, events, and audio allows for a study overview. A timeline hosts (top to bottom) playback controls, an audio line with the participants' audio tracks, and multiple event lines for each participant.}
    \Description{A desktop view of a dashboard of AutoVis. The dashboard is split into different panels. On the left, multiple graphs are organized in rows within one column labeled "2D View". In the center, a panel named "3D View" shows a 3D visualization of the scene. A video panel is located in the top right. Two panels are directly below the video panel. The first is labeled "Inspector" and presents data as a table. The other panel, called Overview, contains multiple checkboxes that control other panels on the dashboard. At the bottom, a horizontal timeline can be found. On it, multiple rectangular boxes represent the start and endpoints of events.}
    \label{fig:desktop_view}
\end{figure*}

\subsubsection{Data Sources \& Specification}
\label{data-specification}
AUI interactions can be explicit (e.g., touch) or implicit (e.g., heart rate) \cite{jansen_design_2022}.
Therefore, practitioners leverage heterogeneous data sources, which provide temporal or spatio-temporal data.
For accurate study replication (\requ{R1} - \requ{R4}) and to address \requ{R5}, \tool allows using five data types: \textit{physiological} (e.g., electrodermal activity \cite{dillen_keep_2020}), \textit{behavioral} (represented as events, e.g., head movement \cite{broy_3d_2014} or drowsiness \cite{albadawi_review_2022}), \textit{vehicular} (e.g., acceleration or steering angle), \textit{environmental} (e.g., positions of other road users), and \textit{personal} (e.g., age or preferences).
\tool minimally requires behavioral data (i.e., passenger posture and movements) to visualize in-vehicle avatars and trajectories and environmental data (i.e., video data) to reconstruct the vehicle surroundings.
If spatio-temporal data, such as movements and GPS positions, is missing, \tool can still be used as a non-immersive desktop tool to visualize temporal data (e.g., sensor recordings) and events.

\subsubsection{Data Preprocessing}
\label{preprocessing}
AUI study datasets are often heterogeneous and processed to different degrees.
They may contain data in raw formats (e.g., sensor streams) or abstract classes and events.
Therefore, a companion tool embedded in the pipeline preprocesses data according to the \tool data specification (see \ref{data-specification}).
%Analysts can import their captured study data (e.g., timestamps, events, passenger movements, and interactions such as touch events) into \tool.

First, \tool converts file formats and automatically detects events in a dataset, preventing manual labeling or the cumbersome use of external programs.
We leverage open-source state-of-the-art DL approaches for automatic event detection.
With this, the companion converts \textit{low-level} physiological signals (e.g., electrodermal activity or pupil size) into \textit{higher-level} information, such as stress, cognitive load, distraction, or drowsiness.
Besides, our preprocessing concept employs image-based recognition of objects \cite{yolo4}, pose/motions \cite{openpose}, gestures \citet{shen_gesture_2022}, and emotions \cite{deepface}, to obtain spatial and contextual information about the vehicle environment and passengers.
Based on this, \tool automatically places 3D models of detected objects at their inferred positions in a 3D environment (cities, roads, and landscapes) queried via GPS (e.g., via OpenStreetMap).
Our preprocessing allows for omitting expensive and vast external vehicle sensors in study setups.
\tool supports various (partly) unprocessed datasets from the automotive domain, such as Drive\&Act \cite{driveact}, MDAD \cite{jegham_mdad_2019}, or HARMONY \cite{tavakoli_harmony_2021}.

Finally, the input files that already matched the \tool data specification are merged with the preprocessing results into a single config file containing all relevant study metadata.
Accordingly, the config file provides information about the study conditions and sessions, events, tracked scene entities, such as object positions for each time step, and physiological data.
Using timestamps, any recorded video and audio data can also be loaded and played synchronously with the replicated 3D environment and study context.
\tool uses the same config file for desktop and VR views, which enables seamless view transitions and multi-user scenarios. 

%%%%%%%%%%%%%%%%%%%%%%%%%%%%%%%%%%%%%%%%%%%%%%%%%%%%%%%%%%%%%%%%%%%%%%%%%
\subsection{Non-Immersive Desktop View}
\label{desktop_view}
%- desktop view (all views in short, adapted from previous works)

%(Figure option?: - new figure replacing single Desktop and VR view figures:
%    - left: Desktop view, 4:3 format
%    - right: VR view, 4:3 format)

The \tool desktop view provides a non-immersive overview of study data (see \autoref{fig:desktop_view}).
Inspired by related tools \cite{nebeling_mrat_2020,zadow_giant_2017,hubenschmid_relive_2022}, our concept utilizes a freely adjustable panel layout (\requ{R5}).
The desktop view is divided into five panels (see \autoref{fig:desktop_view}): \textit{2D panel} (A), \textit{3D scene panel} (B), \textit{video} (C), \textit{inspector} (D), \textit{overview} (E), and \textit{timeline} (F).

Similar to ReLive \cite{hubenschmid_relive_2022}, analysts can examine events (\requ{R2}), control the tool-wide audio and video playback (\requ{R7}), and annotate events (\requ{R6}) using the \textbf{timeline} (see \autoref{fig:desktop_view} F).
We added line diagrams in the \textbf{2D panel} (see \autoref{fig:desktop_view} A) for quickly identifying relevant sequences (\requ{R3}) in non-spatial temporal data, such as physiological data.
To address \requ{R1}, the \textbf{3D scene panel} replicates the original study environment, such as buildings or other road users, and visualizes the study vehicles' movements using a virtual ego-vehicle (see \autoref{fig:desktop_view} B) and the interplay of passengers' in-vehicle interactions and the environment (\requ{R4}).
For desktop panels' details, see Appendix \ref{desktop_view_details}.
%As we focus our concept on the immersive (passthrough) VR view, we describe the desktop panels' details in the Appendix \ref{desktop_view_details}.

\begin{figure*}[ht!]
        \centering
        \includegraphics[width=\textwidth]{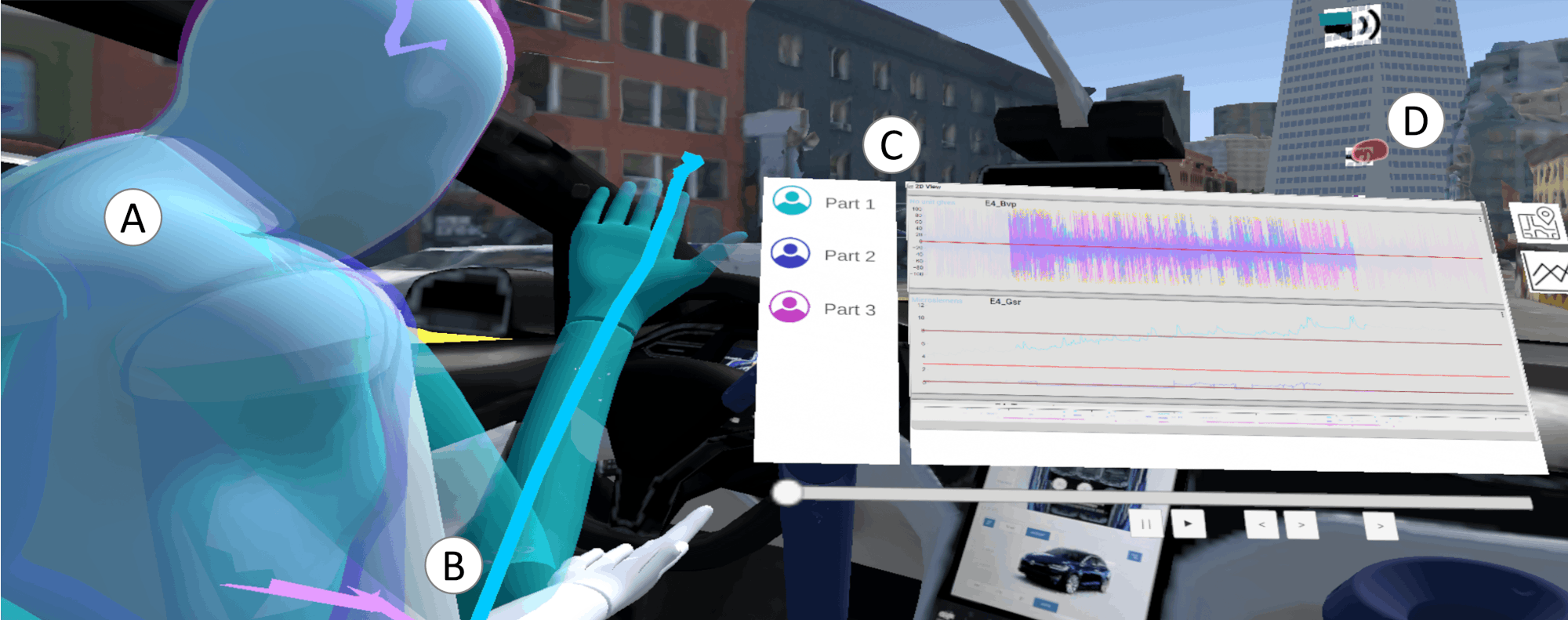}
    \caption{The \textbf{\tool VR view} provides an interactive study replication. A virtual tablet showing physiological data is attached to the left VR controller with a timeline for scene playback. The VR view hosts a virtual replication of the ego vehicle, passenger movements as avatars and trajectories, and gaze, touch, and pointing as heatmaps in the interior and environment.}
    \label{fig:vr_view}
    \Description{The picture shows the inside of a vehicle in a virtual environment. The picture is taken from the co-driver seat. On the driver's seat, multiple avatars can be seen as well as colored trajectories that represent their movement over time. Heatmaps can be found on the surfaces and windows of the vehicle. A virtual tablet used by the human analyst can be seen on the right. It allows the analyst to control the timeline, switch between participants and look at graphs.}
\end{figure*}

%%%%%%%%%%%%%%%%%%%%%%%%%%%%%%%%%%%%%%%%%%%%%%%%%%%%%%%%%%%%%%%%%%%%%%%%%
\subsection{Immersive VR View}
\label{vr_view}
%Figure: - additional figure showing the four new interactions (context portal: pointing, gaze, speech (close), speech (far))
%            - object highlighting and gaze/pointing ray are also visible in these images

%In contrast to the desktop view, the VR view provides immersive analytics (see \autoref{fig:vr_view}).
Inspired by related work \cite{hubenschmid_relive_2022,buschel_miria_2021,reipschlager_avatar_2022}, the immersive VR view enables the interactive re-experience of AUI studies (see \autoref{fig:vr_view}).
For this, \tool replicates the ego-vehicle, other road users, passenger behavior, and environmental context (see \ref{preprocessing}).

\begin{figure*}[t!]
        \centering
        \includegraphics[width=\textwidth]{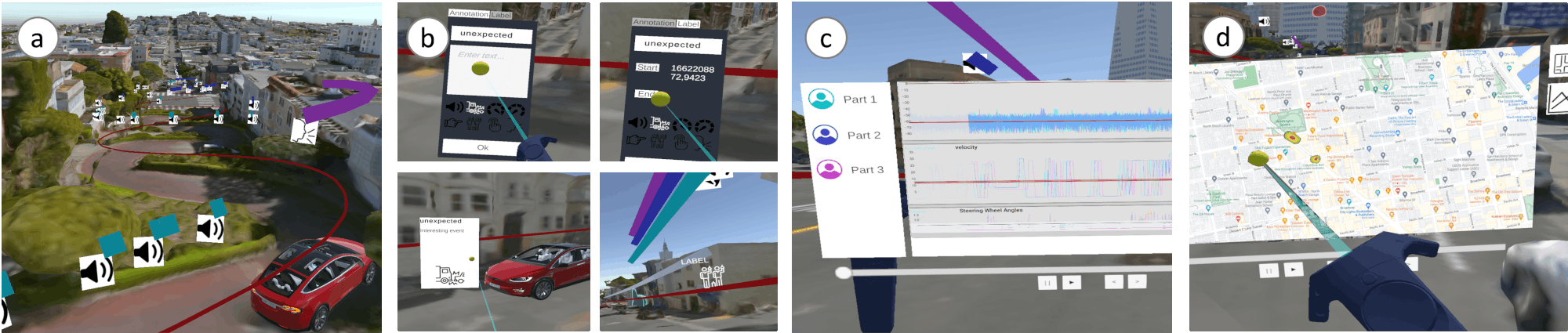}
    \caption{\textbf{(a)} Visualizations of driving-path events, colored for each participant and the timeline in red. \textbf{(b)} Placement and visualization of comments and labels attached to the driving-path event line. \textbf{(c)} Virtual tablet visualizing physiological and event data. \textbf{(d)} Virtual mini-map displaying a top-down overview of the analysis environment.}
    \Description{The figure is split into four sub-figures. The first picture shows a birds-eye view of the tool. A car is driving down Lombard street. Events that were recorded during a study are marked in the 3D environment. The second sub-figure shows the user placing annotations in VR as well as labels attached to the 3D event timeline. The third figure shows the virtual tablet which can be used to look at physiological and event data. The fourth picture shows a map of the location the user is currently at on the virtual tablet.}
    \label{fig:3D_visualizations_domain-specific}
\end{figure*}

Analysts can interact with the environment using their VR controllers for direct touch (tracked by Unity GameObject collisions) or interact with distant objects via raycast.
Analogous to object selections in the desktop 3D scene panel, analysts can interact with avatars, trajectories, heatmaps, events, and annotations.
The object selection via direct touch places a context menu next to the selected component in VR.
For example, next to an avatar's head.
The context menu provides the same features as the desktop view's inspector and overview panel (see \ref{desktop_view}).
The (distant) object selection via raycast opens the context menu in the virtual tablet attached to the left controller (see \autoref{fig:vr_view} C).
This ensures their readability regardless of low VR resolutions and prevents unnecessary approaching of distant objects.
In addition, the tablet displays scene controls, the timeline, study-related metrics, the 2D panel, the event line, and a mini-map (\requ{R3}) (see \autoref{fig:3D_visualizations_domain-specific} c and d).

The VR view hosts 3D visualizations that are adapted from related immersive analytics tools \cite{hubenschmid_relive_2022,buschel_miria_2021,reipschlager_avatar_2022} and novel approaches to overcome AUI domain-specific challenges (see \autoref{fig:vr_view}): \textit{avatars} (A), \textit{trajectories} (B), \textit{in-vehicle}, and \textit{environment} (D) heatmaps.

\paragraph{Spatio-Temporal Events \& Annotations}
AUI study analysis considers not only the event duration but also their location.
Inspired by \citet{buschel_miria_2021}, we propose to visualize such spatio-temporal events in the immersive VR view (see \autoref{fig:3D_visualizations_domain-specific} a), indicating the location and orientation of (inter)action, emotion, driving, and activity events, addressing \requ{R2} - \requ{R4}.
This enables discovering spatial relationships between interactions with in-vehicle UIs, driving environment, context, and passenger states (e.g., emotion or stress).
However, automotive events can visually overlap on a vertical axis if study vehicles drove the same route.
To overcome this, inspired by \citet{fouche2022timeline}, we propose a vertical axis \textit{explode} view for events of individual participants, triggered via direct touch or raycast.
However, events can be distributed across large distances (e.g., several kilometers, see the challenge in \ref{autovis_expert_interviews}) and hidden between replicated 3D buildings and trees.
Therefore, analysts can visualize on hover (e.g., via raycast) events of the same type (e.g., emotion) on a visual layer of higher priority than the remaining environment to peek through 3D objects.

\tool supports creating, editing, and persistent sharing of annotations in VR, addressing \requ{R6} and \requ{R9}.
Analysts can place annotations in space, similar to 3D markers in MIRIA \cite{buschel_miria_2021} and MRAT \cite{nebeling_mrat_2020}, by moving to a specific position and open the edit menu via the controller.
Such annotations are spatio-temporal \textit{labels} or \textit{comments} that are linked in space and to the timeline of a dataset (see \autoref{fig:3D_visualizations_domain-specific} b).
\tool visualizes this link by automatically placing \textit{labels} on the spatial event line.
Analysts can use the \textit{labels} to annotate their dataset, for example, for supervised DL.
In contrast, \textit{comments} can be set and edited anywhere in the 3D environment, for example, to leave hints, descriptions, and opinions about the analysis for oneself (when switching views) or collaborators. % in a multi-user scenario.

\begin{figure*}[t!]
        \centering
        \includegraphics[width=\textwidth]{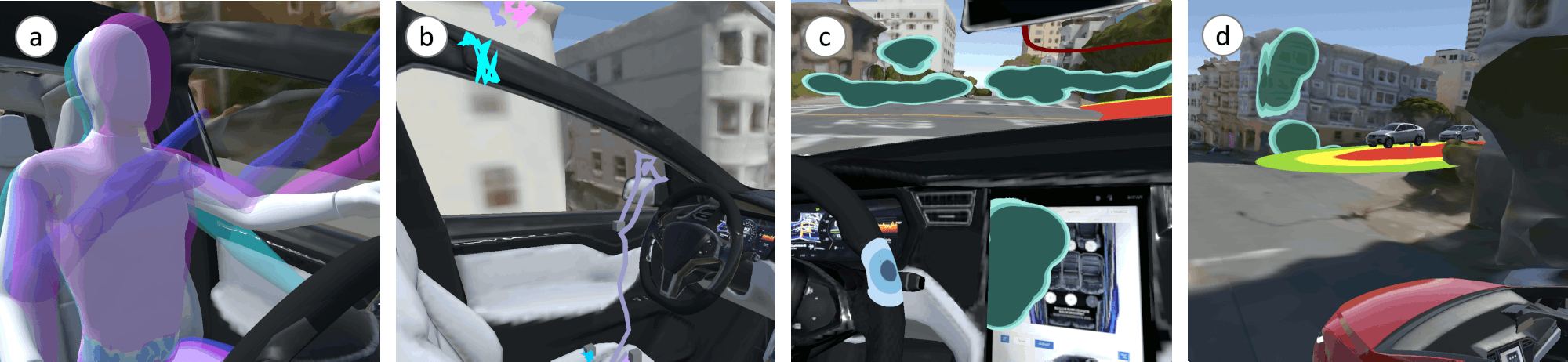}
    \caption{\textbf{(a)} An avatar (white) visualizes the aggregated movements of distinct participant avatars (turquoise, purple, and blue). \textbf{(b)} Head and hand movement trajectories. \textbf{(c)} In-vehicle heatmaps visualizing gazes (green) and touches (blue). \textbf{(d)} Environment heatmaps visualizing participants' gazes (green), pointing targets (magenta), and other road users' positions (red/yellow).}
    \label{fig:3D_visualizations}
    \Description{This figure is split into four sub-figures. The first shows three semi-transparent and one opaque humanoid avatars on the driver's seat in a virtual vehicle. The second shows colored lines representing trajectories of the avatars' head, left-hand, and right-hand movements. The third shows color-coded gaze and touch heatmaps on in-vehicle surfaces and windows throughout the interior. The last shows color-coded heatmaps of gazes and pointing at buildings and the stress in the vicinity. A heatmap on the street shows the positions of other vehicles.}
\end{figure*}

\paragraph{Avatars}
In \tool, avatars replicate passengers (see \autoref{fig:vr_view} A) from pre-recorded 3D skeleton data of participant movements (\requ{R1}).
Inspired by AvatAR \cite{reipschlager_avatar_2022}, \tool updates the avatars in each playback time step.
Free VR movement around avatars enables exploration of posture, relation to the vehicle environment, and movement patterns.
Analysts can enter an avatar's POV to gain first-person insights into how passengers interact with their surroundings.
Such an embodied analysis is impossible using non-immersive analysis tools.
%Analysts can also select avatars by direct touch or controller raycast to open a context menu (e.g., for setting visibility or color) in space or in the virtual tablet.
\tool displays a distinct avatar for each participant. %activated in the overview panel.
In contrast to AvatAR \cite{reipschlager_avatar_2022}, where avatars replicated room-scale movements, the \tool avatars have the same positions (e.g., sitting on driver and passenger seats).
Therefore, we propose an \textit{aggregated avatar}, which aggregates the positions and movements of the individual avatars to increase visual clarity (see \autoref{fig:3D_visualizations} a). 
For the \textit{aggregated avatar}'s skeleton, \tool calculates the average position and rotation of the individual avatars' joints per frame.
Using the \textit{aggregated avatar}, analysts can explore similar passenger behaviors on a meta-level (\requ{R3}).
To further reduce visual clutter, the avatars are semi-opaque, and their colors correspond to the participants' tool-wide colors.

\paragraph{Trajectories}
Similar to \cite{buschel_miria_2021,hubenschmid_relive_2022,reipschlager_avatar_2022}, we employ 3D trajectories.
The trajectories (see \autoref{fig:vr_view} B) correspond to an avatar, providing a different representation of movements.
They replicate hands and head movements for a selected time frame (\requ{R1}).
The trajectories are colored lines matching their avatar's color (see \autoref{fig:3D_visualizations} b).
They provide an overview of the passenger movements for a specific time frame without the need for playback in real-time. %having to play the scene back and forth or watch the playback in real-time.
%Analogous to avatar selection, selecting a trajectory (via direct touch or controller raycast) opens a context menu.

\paragraph{In-Vehicle Heatmaps}
In-vehicle heatmaps provide an overview of interactions with interior surfaces (\requ{R1}), such as windshield display, center console, or dashboard.
Instead of classic 2D heatmaps (e.g., as in \cite{reipschlager_avatar_2022}), \tool employs heatmap textures that accurately map the 3D interior mesh (see \autoref{fig:3D_visualizations} c).
%This preserves spatial information about gazes or touches.
Gaze heatmaps visualize where passengers looked, for example, to investigate glances at the dashboard or center console.
This can help determine passenger states, such as distraction and cognitive load.
Likewise, analysts can use touch heatmaps.
Combined touch and gaze heatmaps may indicate modality interdependencies (\requ{R3}) (see \autoref{fig:3D_visualizations} c).
%This allows gaining insights into passenger behavior unrelated to interactions with in-vehicle UIs that are otherwise not visible when only UI interactions are logged.

\paragraph{Environment Heatmaps}
%In the vehicle environment, \tool displays visualizations separated from the interior (see \autoref{fig:3D_visualizations} d).
%This allows focused analysis of in-vehicle or environment interactions or both.
According to our data specification (see \ref{data-specification}), passengers can interact with the vehicle environment via gaze, pointing, and speech.
Therefore, \tool employs gaze and pointing heatmap textures accurately mapping the replicated 3D meshes of buildings and other road users (see \autoref{fig:3D_visualizations} d).
These heatmaps can help to determine, for example, driver distractions or detect movement and gesture patterns (\requ{R1}).
It also highlights correlations between environmental context and gaze/pointing interactions (\requ{R4}).
Moreover, a traffic heatmap displays the positions of other road users (\requ{R1}) (see \autoref{fig:3D_visualizations} d).
This enables inferring the current driving context and traffic flow.
In addition, analysts can modify the distinct color schemes of each heatmap (\requ{R5}).

\begin{figure*}[ht!]
        \centering
        \includegraphics[width=\textwidth]{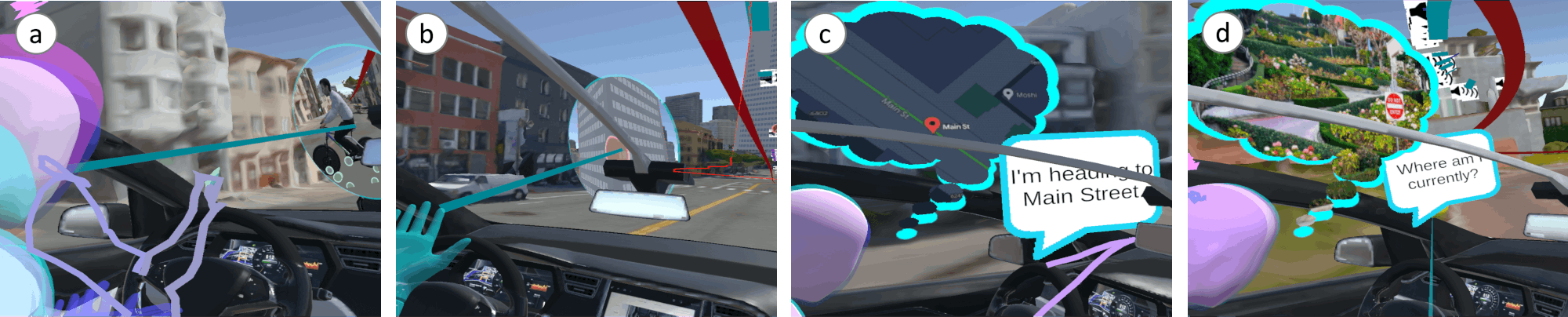}
    \caption{Automotive domain-specific visualizations of \textit{direct} and \textit{indirect} links between in-vehicle interactions and environment: The \textit{direct} portal visualizes \textbf{(a)} a gaze target and \textbf{(b)} a pointing target, including heatmaps and highlighting. The speech-referenced context is visualized via \textbf(c) \textit{indirect} portal, if present in the environment, and otherwise via \textbf{(d)} render image in a \textit{direct} portal.}
    \Description{The figure is split into four sub-figures. The first two images depict the gaze vector of a participant. Because the gaze target was far away, a portal working as a magnifier is used to make the target appear close and larger. The third and fourth figures show events where the participant gave speech input. Speech bubbles depict what was said, thought bubbles depict the context referenced in the speech input. }
    \label{fig:3D_visualizations_context_portals}
\end{figure*}

\paragraph{Context Portals}
%- new automotive domain-specific immersive visualization of driving context during (multimodal) interaction with in-vehicle UIs
Large distances between objects of interest and volatile in-vehicle and environmental contexts are common challenges in AUI study analysis (see \ref{autovis_expert_interviews}).
We propose context portals to overcome these challenges and to address \requ{R3} and \requ{R4}. 
Context portals provide a glimpse of the referenced context (i.e., a location or object) in interactions by showing a spatial portal next to an avatar's finger or head (e.g., as a thought bubble).
The portal shows the object or location up close using a render image from an additional virtual camera (see \autoref{fig:3D_visualizations_context_portals}).
Analysts can activate a context portal by selecting an \textit{(inter)action} event (see timeline in Appendix \ref{desktop_view_details}) on the event line in 3D or on the virtual tablet.
However, only one portal can be visible at a time.
We distinguish two context portal modes: (1) \textit{direct} and (2) \textit{indirect}.

\tool displays a \textbf{(1) \textit{direct}} context portal when participants referenced objects or locations in the vicinity using gaze, pointing, or speech.
The portal then shows a zoomed view of the referenced entity from the avatar's POV, for example, enabling to determine an object's visibility time during an interaction.
To explicitly visualize gaze and pointing targets, there is an additional ray and hit point visualization that reaches through the \textit{direct} context portal (see \autoref{fig:3D_visualizations_context_portals} a and b).
Regardless of interacting modality, the referenced object's outline is highlighted to make it stand out against the environment (see \autoref{fig:3D_visualizations_context_portals} b).

The interaction modality determines the 3D position of the \textit{direct} context portal.
In a pointing interaction, the \textit{direct} context portal is displayed in an extended line of two meters from the fingertip of the respective avatar (see \autoref{fig:3D_visualizations_context_portals} b).
Analogously, the avatar's eyes are the reference point for positioning the portal for gaze interactions.
However, for speech interaction, the portal is a thought bubble next to the respective avatar's head.
In addition, a speech bubble is displayed underneath, which contains the utterance for the inspected time frame (see \autoref{fig:3D_visualizations_context_portals} d).

An \textbf{(2) \textit{indirect}} context portal visualizes referenced objects or locations that are not present in the environment.
\tool queries the missing information, for example, from Google Maps, and shows a screenshot of the result in the \textit{indirect} context portal (see \autoref{fig:3D_visualizations_context_portals} c).
Since passengers can only reference objects that are not in the vicinity with speech, a thought bubble displays the query result.
The \textit{direct} and \textit{indirect} context portals circumvent searching for referenced objects or locations (e.g., a landmark) in the environment.
Otherwise, this search can be time-consuming and challenging if the driving environment is unknown or may result in a barely visible distant object or location.

%%%%%%%%%%%%%%%%%%%%%%%%%%%%%%%%%%%%%%%%%%%%%%%%%%%%%%%%%%%%%%%%%%%%%%%%%
\subsection{Immersive Passthrough VR View}
\label{autovis_pass-through}
\begin{figure*}[t!]
        \centering
        \includegraphics[width=\textwidth]{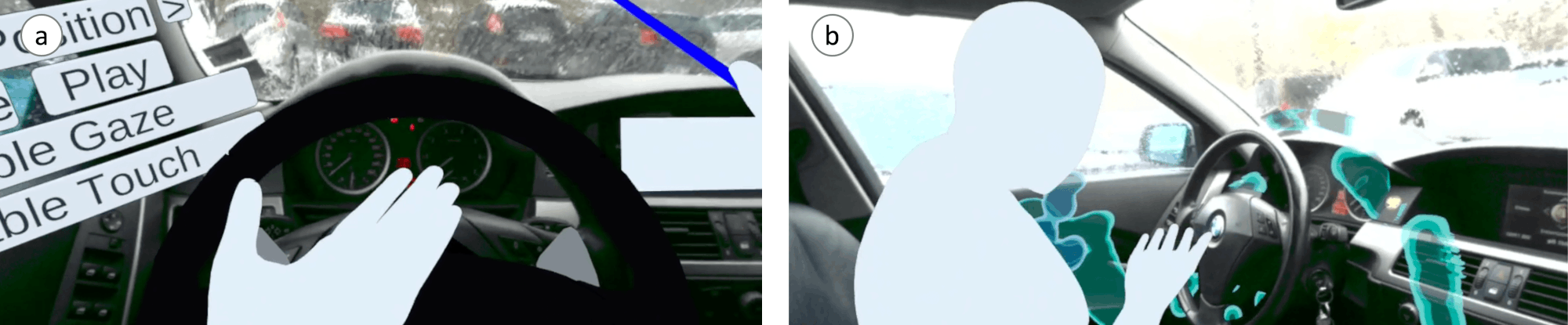}
    \caption{Leveraging a real vehicle in the analysis via passthrough VR. \textbf{(a)} Selective passthrough of the real interior for tangible analysis at the steering wheel and center console. \textbf{(b)} Overlay of the virtual in-vehicle visualizations on the real vehicle.}
    \Description{The figure depicts a real vehicle in the analysis process via passthrough VR. On the left selective passthrough can be seen, for example at the steering wheel or center console. On the right, the overlay of the virtual 3d in-vehicle visualizations can be seen.}
    \label{fig:pass-through_vr_view}
\end{figure*}

%Figure: - two to four images showing the novel pass-through VR view using a real vehicle
The passthrough VR view enables leveraging a real vehicle's interior, layout, and haptics for an immersive AUI domain-specific analysis concept, addressing \requ{R8}.
Using a VR HMD with external cameras, analysts can capture the real vehicle, for example, while sitting in the passenger seat.
\tool displays this camera stream as a separate visual layer of lower priority than the in-vehicle visualizations to augment the real vehicle with the 3D visualizations.
Besides, the virtual ego-vehicle's 3D model is substituted with the real vehicle (see \autoref{fig:pass-through_vr_view} a).
\tool places the 3D in-vehicle visualizations (avatars, trajectories, and heatmaps) at their intended positions within the real vehicle inferred from the dataset.

However, datasets in the automotive domain may not contain vehicle 3D meshes.
\tool can still infer the mesh by retrieving a pre-defined 3D model once for a given real-vehicle model name and number in a dataset (e.g., a 2004 BMW 525i Touring).
AUI practitioners could also integrate the real-vehicle mesh into their dataset using an HMD's spatial understanding.
The continuously updated infrared depth scans provide the required interior mesh to augment the 3D visualizations.
Using either approach, \tool can place avatars at their respective seats and map heatmaps to the real-vehicle interior (see \autoref{fig:pass-through_vr_view} b).

Outside the real vehicle, the 3D analysis environment is still visible as in the (normal) VR view.
Although possible, we decided against augmenting the real environment with 3D visualizations to prevent the real environment from interfering with the analysis environment (see experts' concerns on a passthrough VR mode in the Appendix \ref{expert_comment_real_vehicle}).
Therefore, this view is only usable during parking.
To activate the passthrough VR view, analysts can select an area within the real vehicle (e.g., dashboard or driver's seat) via direct touch or raycast.
Also, this enables real-vehicle passthrough only in certain areas of the virtual analysis environment (\requ{R5}).

%%%%%%%%%%%%%%%%%%%%%%%%%%%%%%%%%%%%%%%%%%%%%%%%%%%%%%%%%%%%%%%%%%%%%%%%%
\subsection{Interplay between Desktop and VR View}
\label{autovis_interplay_vr-desktop}

% Aspects such as the seamless synchronization between the two analysis tools (or investigating which one is better for which tasks) could have been investigated more closely.
% In addition, I would like to understand better, whether the immersive or desktop version work better for given tasks, and for which analysis goals in detail.
%→ Add a description about the envisioned use cases for each view to highlight their strengths and weaknesses

%In the following, 
We describe the interplay between desktop and VR in single and multi-user scenarios.
Besides, we explain specific visualizations and interaction concepts to foster collaborative analysis in \tool.

\subsubsection{Multi-User Scenario}
\label{autovis_collab_taxonomy}
Analysts can collaborate in multi-user scenarios (\requ{R9}) covering any combinations of desktop, VR, and passthrough VR views.
Similar to the space-time taxonomy of collaborative visualization by \citet{isenberg2011collaborative} and the definition of collaborative immersive analytics by \citet{billinghurst2018collaborative}, \tool supports collaborations across different levels of immersion (same- vs. mixed-immersion), times (synchronous vs. asynchronous), and spaces (co-located vs. distributed).

%Three dimensions span our taxonomy (see \autoref{fig:collaboration_taxonomy}): \textbf{D1: Immersion}, \textbf{D2: Time}, and \textbf{D3: Space}.
%Each dimension consists of two levels.
%\textbf{D1} is split into \textit{Same Immersion} (all collaborators use the same views, either desktop or VR) and \textit{Mixed-Immersion} (at least one desktop user and at least one VR user).
%However, for our taxonomy, we do not distinguish between VR users in a lab setting and pass-through VR users in a real vehicle.
%\textbf{D2} is split into \textit{Synchronous} (analysts work together at the same time) and \textit{Asynchronous} (collaboration over time one after another).
%Finally, \textbf{D3} is split into \textit{Co-Located} (analysts work together in the same lab space) and \textit{Distributed} (remote collaboration).

\tool enables hosting a persistent analysis environment as the basis for collaboration, addressing \requ{R9}.
The host shares the analysis environment as a config file (similar to a Unity scene file, see \ref{autovis_processing_pipeline}), for example, via cloud storage, to allow remote and independent access.
The others only import the config file into their \tool instance, which reconstructs the 3D analysis environment.

To enable effective interplay between desktop and VR views, \tool provides three visualization concepts that apply to same- and mixed-immersion scenarios:
(1) \tool visualizes other analysts' positions and movements via analyst avatars (see \autoref{fig:case_study_evaluation} a) inspired by \citet{chen2021effect}.
They also display the analysts' POVs via viewing frustums.
In addition, (2) \tool provides visual bookmarks as \textit{ghost} ego-vehicles to indicate the analysis position of collaborators.
By selecting such \textit{ghost} via direct touch or raycast, the analysis replay is set to the selected \textit{ghost}'s position in time and space.
Besides, (3) analysts can communicate via virtual labels and comments.
For example, in asynchronous scenarios, analysts on the desktop could leave comments in the 3D environment for analysts in VR to observe at a different time and vice versa.

\subsubsection{Single-User Scenario}
In single-user scenarios, one analyst uses a combination of VR and desktop.
For example, an analyst could use the desktop view for pre- and post-analysis and the VR view as the main tool in between.
However, when desktop or VR devices are unavailable, analysts can only use one view.

The interaction concept for labels and comments also applies to single-user scenarios.
However, instead of labeling or commenting for others, analysts create self-notes for another session or mark interesting aspects in the desktop view before transitioning to VR.
%Besides, there are no ghost cars as only one user is present in the persistent analysis environment.

%Figure: - showing concrete interaction examples using four images in one figure
%            - two users using the label and annotation feature
%            - two users using the ghost car feature
%            - one user is guiding the other user in VR
%            - one user using the Desktop as pre- and post-analysis and the VR as main analysis tool

%%%%%%%%%%%%%%%%%%%%%%%%%%%%%%%%%%%%%%%%%%%%%%%%%%%%%%%%%%%%%%%%%%%%%%%%%
%%%%%%%%%%%%%%%%%%%%%%%%%%%%%%%%%%%%%%%%%%%%%%%%%%%%%%%%%%%%%%%%%%%%%%%%%
\section{\tool: Prototype Implementation}
\label{autovis_prototype}
The prototype of \tool consists of three sub-prototypes demonstrating the \textit{companion} (see \ref{autovis_processing_pipeline}), the \textit{desktop} view (see \ref{desktop_view}), and the \textit{VR} view (see \ref{vr_view}) including the passthrough VR view (see \ref{autovis_pass-through}).

In our processing pipeline (see \ref{autovis_processing_pipeline}), a (1) \textit{companion} preprocesses the data and creates a JSON config file.
The Python-based \textit{companion} (see \ref{preprocessing}) uses state-of-the-art DL to automatically infer events from datasets and merges the results into the JSON config file.
We employed DeepFace \cite{deepface} for emotion recognition, OpenPose \cite{openpose} for posture/movement recognition, and YOLOv4 \cite{yolo4} for object detection.
Moreover, to infer passenger states, we use Driver-State-Detection \cite{driver-state-detection}, Drowsiness Detection \cite{drowsiness-detection}, and Stress Detector \cite{stress-detector}.
% The model results are merged into the JSON config file.
% Analysts can determine the models to be applied in a setup UI.

A Unity-based (2) \textit{desktop} application generates the 3D analysis environment from spatio-temporal data of the ego vehicle, other road users, and the environment (e.g., weather) given in the JSON config file.
%Besides, it generates 3D visualizations, such as avatars or heatmaps, based on behavioral passenger data (e.g., gazes, hand movements, or touches).
To reconstruct the 3D driving environment, we use Cesium \cite{cesium}, which provides high-resolution real-world photogrammetry in Unity version 2020.3.33f.
Also, the \textit{desktop} application visualizes physiological, event, and metadata returned by the \textit{companion}.
For this, we built a web UI using HTML, JavaScript, D3.js \cite{d3} to visualize detailed graphs, and gridstack.js \cite{gridstack} for panel rearrangement.
We embedded the web UI into Unity using 3D WebView \cite{3d-web-view}.
The \textit{desktop} application can modify the JSON config file and vice versa (e.g., when pausing the replay or selecting objects).

%For remote demonstrations of the \textit{desktop application}, we also built a web-based system that can scale to a client-server architecture.
%To enable a complex 3D environment within a web-based system, we use WebGL 2.0 as the built platform of Unity 2020.3.33f.
%However, most web browsers limit the WebGL application to approx. two gigabytes of RAM.
%Therefore, the remote version of the \textit{desktop application} can only demonstrate parts of a dataset.

Using the same JSON config file, a Unity-based (3) \textit{VR} application generates the analysis environment for the VR and the passthrough VR view.
%The \textit{VR} application displays the same content as the desktop 3D scene panel.
In our prototype, the target platform is the Vive HMD.
However, for the passthrough VR view, we employ the Meta Quest Pro, which provides color passthrough video, gesture, and eye-tracking.
Moreover, the Quest Pro is a mobile VR HMD suitable for use in a real vehicle without access to a desktop system hosting the \textit{VR} application.

To enable multi-user scenarios (same- or mixed-immersion), we employ Mirror \cite{mirror} as a networking library in Unity for live synchronization of object and analyst movements, replay state, 3D labels, comments, and active visualizations.
%One analyst hosts the multiplayer session for the collaborators to join.
%\tool then synchronizes analysts' avatars, ghost cars, and labels/comments between the environments.

%The \tool prototype for desktop and VR, the preprocessing companion, and sample datasets are available as an open-source project\footnote{The \tool repository: \url{https://gitlab.com/Pascal-Jansen/autovis}} on GitLab.
A supplementary video figure illustrates the prototype and the use cases described below.

%%%%%%%%%%%%%%%%%%%%%%%%%%%%%%%%%%%%%%%%%%%%%%%%%%%%%%%%%%%%%%%%%%%%%%%%%
%%%%%%%%%%%%%%%%%%%%%%%%%%%%%%%%%%%%%%%%%%%%%%%%%%%%%%%%%%%%%%%%%%%%%%%%%
\section{Analysis Workflow With Practical Use Cases}
\label{analysis_workflow_and_use_cases}
We evaluate \tool by combining different validation approaches in several AUI analysis use cases.
Our prototype serves as a technical validation that typical AUI study data can be visualized using desktop and current VR devices.
Furthermore, similar to the cognitive walkthrough of \citet{reipschlager2020personal} and \cite{meskens_gummy_2008,houben_watchconnect_2015}, we report on \tool walkthroughs regarding \citet{olsen_evaluating_2007}'s heuristics \textbf{H:} \textbf{Importance}, \textbf{Unsolved Problem}, \textbf{Generality}, \textbf{Reducing Problem Viscosity}, \textbf{Empowering Novices}, \textbf{Power in Combination}, and \textbf{Scalability}.
%With this, we inspect the usability of \tool regarding strengths, weaknesses, and potential value to a broader audience.
%In the following, we will apply Olsen's heuristics \cite{olsen_evaluating_2007} within a user story.
Demonstration and heuristic walkthrough are common toolkit evaluation approaches \cite{ledo_evaluation_2018}.
We have chosen our use cases to show the diverse aspects of \tool in the visual exploration of data, the interplay between desktop and VR, the dataset (pre)processing, and the usage of a real vehicle in the analysis.
%The overall aim was to understand better the interplay between the in-situ (VR) and ex-situ (desktop) analysis and the applicability of \tool.
Similar to \citet{hubenschmid_relive_2022}, we defined three research goals (G1-G3) to guide our evaluation:
\aptLtoX[graphic=no,type=html]{}{\phantomsection}
\label{goal}
\begin{itemize}[noitemsep]
\itema \textbf{Interplay of Immersive and Non-Immersive Analysis:} How do they complement each other in AUI research?
\itemb \textbf{Applicability:} Does \tool meet the requirements of AUI researchers?
\itemc \textbf{Task Allocation:} Which analysis tasks in AUI research benefit from which analysis approach (immersive or non-immersive)?
\end{itemize}

%%%%%%%%%%%%%%%%%%%%%%%%%%%%%%%%%%%%%%%%%%%%%%%%%%%%%%%%%%%%%%%%%%%%%%%%%
\subsection{Use Case: Multimodal Interactions in AVs}
\label{case_study}
We demonstrate the applicability of \tool on the use case of multimodal interactions in AVs (\topic{T5}).
%Multimodal in-vehicle interactions might mitigate the drawbacks of one modality, for example, using gestures to improve touch \cite{ahmad_predictive_2018}, gaze, and speech input \cite{aftab_multimodal_2019} in non-driving related tasks.
%In addition, multimodality can increase reliability when one modality is unavailable \cite{falco2019transfer}.
%Especially during automated driving, passengers have more freedom to interact multimodal in non-driving related tasks.
Evaluating novel input and output modalities (e.g., electrodermal or brain interfaces) and modality combinations in different contexts is challenging.
Thus, analysis of multimodal interactions could significantly benefit from the domain-unique visual exploration of spatio-temporal data provided by \tool.
However, to the best of our knowledge, there is no publicly available dataset suitable for such analysis.
Therefore, we recorded authentic data in an exemplary use case study (see details in Appendix \ref{use_case_study_details}) that allowed multiple input modalities in different contexts during a ride in an AV (SAE 4).

%The use case description is based on \citet{olsen_evaluating_2007} and contains three parts:
%(1) \textit{Users:} Experts and novices in the AUI domain in academia or industry.
%(2) \textit{Situation:} The analysts' goal is to develop multimodal interactions with UIs during a ride in an AV.
%(3) \textit{Tasks:} The analysts' task is to gain insights into the usability of multimodal interactions and investigate the sequence, duration, and selection of modalities.

\begin{figure*}[ht!]
        \centering
        \includegraphics[width=\textwidth]{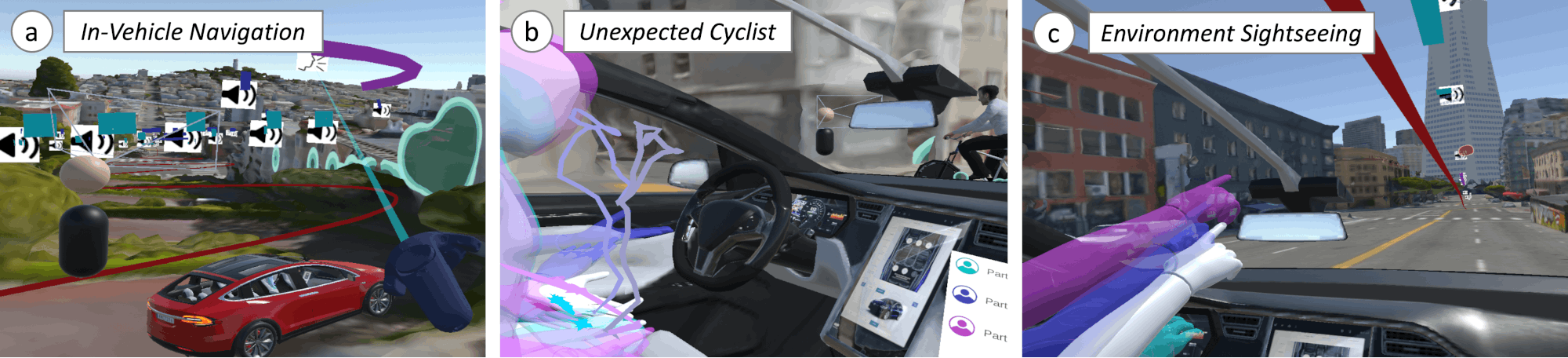}
    \caption{\textbf{(a)} Driving-path events visualize speech interactions' location and duration, while heatmaps visualize gazes. A desktop analyst's avatar (left) is present in this multi-user scenario. \textbf{(b)} Avatars visualize participants' surprising gestures, and the gaze heatmap highlights their focus on the cyclist. \textbf{(c)} Avatars visualize pointing gestures, and a heatmap (red) indicates the target.}
    \label{fig:case_study_evaluation}
    \Description{This figure is split into three sub-figures showcasing events that occurred during the example study. The first shows the in-vehicle navigation task's analysis in a 3D scene at a virtual Lombard Street in San Francisco. The second shows the unexpected cyclist crossing in which a cyclist crosses the street from right to left. Multiple avatars and heatmaps are visible in the area of the driver's seat. The third image shows an environment sightseeing event. On the driver's seat are multiple avatars, each visualizing a pointing gesture toward the sightseeing target, the Transamerica Pyramid in San Francisco. A heatmap is visible on the building}
\end{figure*}

We imagine an AUI researcher, Anna.
%Anna wants to find out how the usability of multimodal interactions changes depending on the environmental context.
She wants to investigate the input modality usage order, the time difference between modality usage, what modalities were used for which tasks, and in which environmental contexts.
Her research has a crucial effect on the passengers of future (automated) vehicles, as the insights could make multimodal interactions more usable and safe (\textbf{H: Importance}).
Anna decided to use \tool as there is currently no other immersive analytics solution enabling her to perform these tasks (\textbf{H: Unsolved Problem}).
%Besides, she anticipates that \tool reduces the analysis effort making her current work practice more efficient.
Before using \tool, she preprocesses the data using the companion (see \ref{preprocessing}).
%Then she loads the resulting dataset into \tool for analysis.

% Environment Sightseeing Task
Anna starts her analysis by looking at the 3D scene view in the desktop application.
This enables her to see a replicated view of the data, providing a much closer match to the original study setting than analyzing the video recording (\textbf{H: Reducing Problem Viscosity}).
She particularly focuses on a detected pointing interaction in the passengers' \textit{environment} task, which she retrieved from selecting the event on the timeline.
The gaze/pointing heatmaps and the avatar visualization help her understand the situation.
However, Anna is not familiar with San Francisco and has difficulties locating the referenced landmark.
Therefore, she selects the pointing event in the spatio-temporal event line to display a \textit{direct context portal}.
Anna immediately locates the landmark through the portal without moving close to it.
In the portal, she notices that some participants gazed directly at the landmark and pointed close to the gaze hit (see \autoref{fig:case_study_evaluation} c).
However, others pointed in the mere direction with bent elbows.
Moreover, Anna sees that all participants first gazed at the target, then pointed, and finally asked about the target via speech.
The \textit{action} event annotations on the timeline show her that the time difference between the interactions was 0.5 seconds on average.
This combination of timeline with 3D scene panel supports Anna in drawing her conclusions (\textbf{H: Power in Combination}).

% Social Setting
%Using a VR HMD, Anna immerses herself in the VR view to gain a deeper understanding of the interplay between the passengers in the \textit{social} setting.
%To investigate the interplay up close, she enters the POV of the passenger who has boarded the vehicle.
%For this, Anna selects a suitable camera perspective by clicking through the predefined positions.
%To find the relevant scenes, she navigates through time using the timeline on the virtual tablet (see \autoref{fig:case_study_evaluation} b).
%She then disables all heatmaps to reduce visual clutter and increase focus on the avatar movements.
%When looking at the corresponding scenes and participants' head movements, Anna noticed that they mostly ignored the other passenger except during a short initial conversation.
%Besides, after scrolling through the 2D panel on the virtual tablet, she notices no major change in the passengers' physiology in the \textit{social} setting.

% Unexpected Situation
Next, Anna wants to explore how participants reacted to the \textit{unexpected} cyclist crossing.
However, she could not find anomalies compared to the average line when she looked at the physiological signals in the 2D panel for this scenario.
%To investigate noticeable changes in distinct participants, Anna cycles through each via the overview panel.
Anna concluded that there was no measurable physiological reaction, likely indicating that participants felt safe and trusted the AV.
The automatic detection of events, such as emotions, helped her select interesting sequences in the physiological streams, as she is unfamiliar with physiology (\textbf{H: Empowering Novices}).
However, Anna assumes that she might have missed something and contacts Jacob, who is an expert in physiology.
Jacob answers her call and loads the shared persistent JSON config file into his \tool desktop application, and takes a closer look at the \textit{unexpected} cyclist sequence.
This transforms Anna's analysis into a distributed, asynchronous multi-user scenario.
Jacob suspects correlations by comparing the line diagrams of the heart rate and the electrodermal activity and creates 3D labels at the relevant spots.
When Anna resumes the analysis a few days later, Jacob's labels point her to the right spots.

Using a VR HMD, Anna immerses herself in the VR view to gain a deeper understanding of the link between in-vehicle and environmental context.
For this, she searches a suitable camera perspective by clicking through the predefined positions and selects the passenger seat POV.
Looking at the avatars, she discovers that some participants raised their hands in surprise while focusing on the cyclist (see \autoref{fig:case_study_evaluation} b).
In addition, they referred to the cyclist in a short conversation with the other passenger.
Anna concludes that AVs must be aware of the intended interaction partner in multimodal interactions (e.g., speech and gaze), especially in social settings, to prevent misunderstandings.

% In-vehicle Navigation Task
The participants could freely choose modalities to query information in the \textit{in-vehicle} navigation task.
Looking at the timeline, Anna noticed that all participants used speech to query their location.
Besides, Anna sees that the in-vehicle touch and gaze heatmaps show a widespread pattern (see \autoref{fig:case_study_evaluation} a).
She concludes that participants likely were curious about the unfamiliar setting and explored the interior by looking around and touching it.
%Similarly, Anna observes in the environment heatmaps that participants' gazes mostly wandered around.
%However, in the \textit{unexpected} crossing, the participants focused on the cyclist.
%Further, she observes that participants often looked at traffic lights and other cars and assumes they checked whether the automated driving worked correctly.
%Overall, Anna successfully retrieved valuable insights on how passengers may behave in multimodal interactions regarding interaction speed, sequence, and context.
%Moreover, she gained useful research insights into how social settings might impact multimodal interactions.

Overall, Anna successfully retrieved valuable insights on passenger behavior in multimodal interactions regarding interaction speed, sequence, and context.
%This scenario described how \tool could be applied to analysis of AUI research on the example of \highlight{\footnotesize \hyperref[T5]{T5}}.
However, \tool also applies to other topics and use cases in which practitioners, such as Anna, are interested (\textbf{H: Generality}).
For example, in the context of \topic{T3}, Anna could also gain a lot of information on the passenger state using the physiological streams overview.
Besides, she is interested in conversational analysis (\topic{T6}) and could use \tool to explore the conversation between participants, vehicle, and other passengers.
Moreover, Anna could evaluate other novel modalities (\topic{T1}, \topic{T2}).

\begin{figure*}[ht!]
        \centering
        \includegraphics[width=\textwidth]{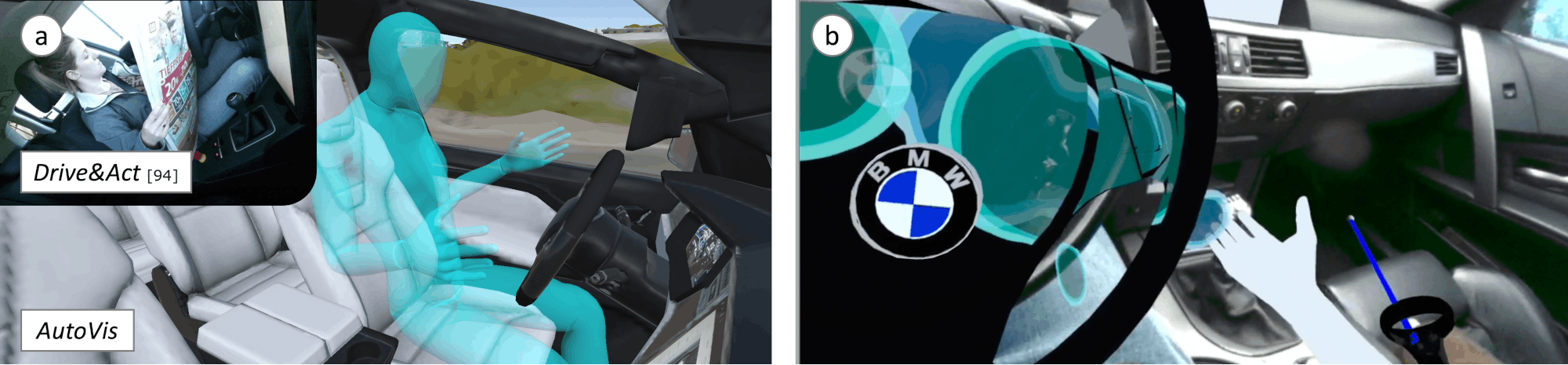}
    \caption{\textbf{(a)} Real-world dataset use case. Screenshots of a source video from Drive\&Act \cite{driveact} and its conversion in \tool. \textbf{(b)} Leveraging a real vehicle (2004 BMW 525i Touring) in the \textit{climate control} use case. The virtual gaze and pointing heatmaps indicate interaction patterns on the real steering wheel and center console.}
    \label{fig:use_case_dataset_real_vehicle}
    \Description{On the left, the virtual recreation of a woman sitting in the driver's seat of a car can be seen. She is reading a newspaper. On the right, touch heatmaps can be seen overlayed on a real vehicle using passthrough VR.}
\end{figure*}

%%%%%%%%%%%%%%%%%%%%%%%%%%%%%%%%%%%%%%%%%%%%%%%%%%%%%%%%%%%%%%%%%%%%%%%%%
\subsection{Use Case: Analyzing a Real-World Dataset}
\label{use_case_dataset}
% co-located synchronous same-immersion multi-user analysis 
We leverage Drive\&Act \cite{driveact} to demonstrate that \tool can convert and visualize real-world datasets. %that were recorded in a real vehicle.
Analysts can also employ similar datasets (e.g., \cite{jegham_mdad_2019}).
However, datasets might be incomplete regarding events, 3D poses, or environment recordings.
For example, we decided against using HARMONY \cite{tavakoli_harmony_2021}, as (in February 2023) only a ten seconds sample is publicly available.

Drive\&Act is a large-scale dataset for driver activity recognition captured in manual and autonomous driving mode (\topic{T3} - \topic{T5}).
It contains data streams of 3D body poses, head poses, interior model, and a camera system with five interior views.
%Besides, the dataset differentiates activities into mid-level (e.g., reading the newspaper), action (e.g., putting down), object (e.g., laptop), and location (e.g., passenger door).
Besides, the dataset differentiates mid-level (e.g., reading the newspaper), action, object, and location activities.
Drive\&Act does not include driving environment data, such as scans or video recordings of the surrounding.
Regardless, \tool can reconstruct a 3D driving environment for Drive\&Act based only on GPS coordinates (\textbf{H: Generality}).
Consequently, this environment lacks 3D heatmaps of gestures, pointing, and other vehicles, as well as context portal visualizations.

Accordingly, AUI researchers Jack and Mia focus their analysis on the vehicle interior.
They want to determine whether passengers' body pose while reading in AVs is similar to non-driving scenarios.
Since Jack and Mia each have a Vive VR HMD (same-immersion), they decide to collaboratively analyze Drive\&Act in the same room (co-located synchronous) for directly discussing their findings (\textbf{H: Scalability}).
Since Drive\&Act contains 3D poses and activity events, Jack and Mia only activate the DeepFace \cite{deepface} option for emotion recognition in the \tool \textit{companion}.

In their analysis, Jack and Mia search for a suitable passenger reading event.
As Mia perceives the event search in the virtual tablet as tedious, she investigates the spatio-temporal event line while driving in the virtual ego vehicle.
When Mia finds a "Newspaper Reading Task" (\textbf{H: Power in Combination}), she tells Jack the time frame.
Jack then jumps there, by selecting the event.
While Mia chooses the passenger seat POV, Jack selects the vehicle's hood POV to have a frontal view on the individual avatars in the driver's seat (see \autoref{fig:use_case_dataset_real_vehicle} a).
After looping the event in VR several times, they notice differing head poses.
While some passengers focused on the reading task and permanently looked down, others looked up more often.
They conclude that these participants frequently checked the driving situation, which might indicate low trust in the AV.
As environment objects were not replicated due to a lack of data, they can make little connection between the driving environment and passenger behavior.
However, the reconstructed 3D environment based on GPS enables them to determine the respective AV driving situations (e.g., highway vs. urban) that could explain the differing head poses (\textbf{H: Power in Combination}).

%%%%%%%%%%%%%%%%%%%%%%%%%%%%%%%%%%%%%%%%%%%%%%%%%%%%%%%%%%%%%%%%%%%%%%%%%
\subsection{Use Case: Leveraging a Real Vehicle}
\label{use_case_real_vehicle}
% distributed synchronous mixed-immersion multi-user analysis
We demonstrate how \tool integrates a real vehicle into an analysis process (\requ{R8}).
For this, we recorded a test dataset in a BMW 525i Touring to analyze a novel UI concept for an air conditioning control specific to the driver's seat area (\topic{T1}, \topic{T2}).
The control UI is located on the driver's door.
As part of a collaborative research project, Simon and Emma aim to develop this UI.
Emma wants to cross-evaluate the results of a user study conducted in a replica of the real vehicle's UI.
However, only Simon has access to the real vehicle.
Therefore, they agree on a distributed synchronous analysis session in \tool so that Emma can remotely guide Simon, who performs in the real vehicle (the BMW) using the passthrough VR view.
The remote access to the real vehicle (\textbf{H: Importance} and \textbf{H: Unsolved Problem}) enables a more efficient cross-evaluation (\textbf{H: Reducing Problem Viscosity}).

Emma creates a mixed-immersion scenario using the \tool desktop application to access the same analysis environment as Simon. 
She wants to investigate the influences of the vehicle replica's button layout on the study results.
Emma guides Simon through the relevant sequences using her analyst avatar and 3D comments (\textbf{H: Power in Combination}).
When Simon is in the right place, she signals him to touch the vehicle's surface to activate the passthrough VR view (see \autoref{fig:use_case_dataset_real_vehicle} b).
Using the heatmaps and the avatars' hand movements, they found that the participants rarely touched and looked at the novel UI on the driver's door.
The participants mainly searched for the UI at the center display.
In the driver avatar POV, Emma also notes a mismatch between the study button layout and the real vehicle, which may have confused participants.
Because Simon can use the real-vehicle's haptics and thus assess spatial relations in VR more effectively (\textbf{H: Power in Combination}), he further notices a greater distance from the driver to the novel UI than the center display.
He concludes that participants had a greater effort and, therefore, they searched the familiar center display first.

%%%%%%%%%%%%%%%%%%%%%%%%%%%%%%%%%%%%%%%%%%%%%%%%%%%%%%%%%%%%%%%%%%%%%%%%%
%%%%%%%%%%%%%%%%%%%%%%%%%%%%%%%%%%%%%%%%%%%%%%%%%%%%%%%%%%%%%%%%%%%%%%%%%
\section{Discussion}
\label{discussion}
We (1) discuss how \tool differentiates from immersive analytics in other domains; (2) elaborate on lessons learned from our demonstration and heuristic evaluation; and (3) describe challenges and insights for future research on immersive analytics for AUIs.

%%%%%%%%%%%%%%%%%%%%%%%%%%%%%%%%%%%%%%%%%%%%%%%%%%%%%%%%%%%%%%%%%%%%%%%%%%%%%%%%%%
\subsection{Differentiation of \tool to Prior Immersive Analytics}
We adapted visualizations from prior research on immersive analytics for mixed-reality usage \cite{hubenschmid_relive_2022}, human motion analysis \cite{reipschlager_avatar_2022}, interactive wall usage \cite{zadow_giant_2017}, and multi-display scenarios \cite{buschel_miria_2021}.
Similar to \tool, they considered multi-user scenarios, various interaction modalities (e.g., touch and gaze), and different analysis devices (VR, AR, and desktop).
In line with \cite{reipschlager_avatar_2022}, we used avatars associated with body part trajectories to replicate user movements.
However, we limited our concept to hand and head trajectories, as other body parts (e.g., feet) are negligible for AUIs.
Besides, trajectories for each body part would increase visual clutter.
Unlike prior work, we introduced an \textit{aggregated} avatar enabling overview movement analysis to mitigate the limitation of visual clutter for many avatars.
Similar to \cite{zadow_giant_2017,reipschlager_avatar_2022,buschel_miria_2021}, we employed heatmaps to visualize interactions with surfaces.
However, we applied the heatmaps as textures to 3D meshes of replicated study environment objects.
This allowed us to increase the level of analysis detail compared to prior immersive analytics and effectively utilize the geometry of the replicated ego vehicle.
A recurring limitation of prior immersive analytics tools is the high level of detail required for datasets (e.g., see \cite{reipschlager_avatar_2022,hubenschmid_relive_2022}).
Therefore, in our processing pipeline, we proposed a companion to preprocess data with insufficient detail (e.g., only video data) into suitable datasets (i.e., including events, objects, and user states).
In contrast to prior immersive analytics \cite{hubenschmid_relive_2022,reipschlager_avatar_2022,buschel_miria_2021,zadow_giant_2017,nebeling_mrat_2020}, \tool includes a real vehicle into immersive analysis processes.
We argue that \tool can serve as a starting point for future immersive analytics considering similar prerequisites, such as the dynamic environment inside and outside an interaction space (e.g., a vehicle).

%%%%%%%%%%%%%%%%%%%%%%%%%%%%%%%%%%%%%%%%%%%%%%%%%%%%%%%%%%%%%%%%%%%%%%%%%%%%%%%%%%
\subsection{Lessons Learned}
Applying Olsen's heuristics (see \ref{case_study}) helped to identify heuristics partially covered by \tool.
According to our research goals \goal{G1}, \goal{G2}, and \goal{G3}, we discuss lessons learned from the implementation and demonstration of \tool.

%%%%%%%%%%%%%%%%%%%%
\paragraph{Task Allocation Between Immersive and Non-Immersive View}
In line with \cite{hubenschmid_relive_2022,kraus_assessing_2020}, our heuristic walkthroughs suggest that the desktop outperforms the VR view for overview tasks, such as understanding the driving environment and in-vehicle events.
In contrast, we assume that analysts will use the VR view for detailed passenger movement analysis.
However, we found that transitions from desktop to VR (e.g., single-user scenario, see \ref{autovis_interplay_vr-desktop}) may be inappropriate due to analysis goals (e.g., analyzing only physiology) and inconvenience (e.g., HMD heat and pressure points).
Therefore, we argue that the desktop view's 3D scene panel suffices for most in-vehicle and driving analysis without transitioning to VR.
Our prototype also showed that such transitions may lead to initial disorientation.
In line with \citet{hubenschmid_relive_2022}, we learned that the 3D labels/comments should be set in the desktop view first to mitigate disorientation.
%Moreover, interactions could be defined that further combine immersive and non-immersive analysis, such as drawing areas of interest in the 3D scene panel, which are then highlighted in VR.
Besides, the \textit{indirect} context portals (see \ref{vr_view}) sparked confusion in the multimodal interaction use case (see \ref{case_study}) as the live-queried contents' locations were unclear.
We learned that these portals should clarify the location textually and link sources.

From our use cases, we also learned that the desktop view is interchangeable with the VR view regarding the amount of accessible data.
%Both views access the same analysis environment.
However, the desktop view is more suitable for retrieving outliers in passengers' physiological data, useful in \topic{T3}, \topic{T4}, and \topic{T7}.
In contrast, the VR view is more appropriate for perceiving spatial distances between driving and interaction events, relevant in \topic{T5}.
However, we found that analysts could quickly get motion sick when scrubbing the timeline in VR moved the ego-vehicle too fast.
Therefore, we argue that the desktop view should be used to prepare driving sections, which are then analyzed in VR.

Overall, we conclude that novices might have problems orienting themselves and, therefore, use inappropriate views for their tasks.
For example, they may spend much time searching for relevant sequences while driving in the VR ego-vehicle in real-time.
Accordingly, \tool should further contribute to \textbf{H: Empowering Novices} via tutorials and hints of unnecessary VR usage.

%%%%%%%%%%%%%%%%%%%%
\paragraph{Collaborative Analysis}
\tool enables collaborative analysis using the desktop and VR view (see \ref{autovis_interplay_vr-desktop}), valuable for the ideation of novel in-vehicle UIs (\topic{T1}).
According to \citet{hubenschmid_relive_2022} and the feedback from our expert interviews (see \ref{autovis_expert_interviews}), analysts would perceive such collaboration as advantageous.
%However, similar to \citet{jansen_share_2020} and in line with the Grand Challenges \cite{ens_grand_2021} of collaborative analytics, challenges arise, such as distribution of information, different perspectives on data, and control of visualizations.
%For example, analysts can have different visual perspectives (immersive and non-immersive) on the data, which may foster diverging conclusions.
%Besides, it is unclear how different analyst roles impact the analysis (e.g., guide and guided).
%In addition, asynchronous collaboration might be less effective than synchronous interaction, as, without verbal communication during analysis, the 3D labels/comments might be insufficient to understand other's work at a different point in time.
%Therefore, future research should investigate the usability of user roles and the effectiveness of collaboration types (e.g., mixed-immersion in desktop and VR vs. same-immersion in VR) for various analysis tasks.
However, we learned that the \textit{ghost} vehicles may significantly occlude the ego vehicle and the 3D visualizations.
For example, when collaborators join a session and want to inspect the current analysis state by navigating to the other analysts' perspectives.
The visual occlusion increases with the number of collaborators, limiting large scale synchronous collaborations.
Additionally, we found that switching between the \textit{ghost} vehicles to inspect the collaborators' analysis states result in context switches, as all comments, labels, and \textit{exploded} driving-path events would reappear in an altered state, requiring analysts to regain situational awareness after switching back.
Therefore, we argue that other analysis objects should remain visible during such switches to preserve the previous analysis context.

Besides, the current \tool prototype provides visualizations to foster collaboration only for the 3D scene panel and the VR view.
Consequently, we learned that collaborative analysis of physiological data and events in the desktop view is challenging, as collaborators do not see each other's mouse positions or selected UI elements.
Therefore, future research should consider incorporating visual aids in the non-immersive panels of the desktop view to facilitate collaborative analysis.

%%%%%%%%%%%%%%%%%%%%
\paragraph{Real-World Study Complexity \& Dataset Size}
Although \tool applies to several use cases (see \ref{case_study}, \ref{use_case_dataset}, and \ref{use_case_real_vehicle}) and is designed to generalize to other AUI domains (\topic{T1} - \topic{T8}), \textbf{H: Scalability} is currently limited.
In our prototype, 3D visualizations (e.g., avatars) work effectively due to the small dataset size.
However, in the AUI domain, (naturalistic) datasets (e.g., required in \topic{T3}) can become large and complex as they include recordings taken over several hours of driving.
In line with \citet{reipschlager_avatar_2022}, we found that avatars overlap, for example, when they are all located on the same seats.
Likewise, heatmaps overlap for environment objects, for example, in traffic jams or pedestrian crowds.
Besides, finding relevant data sequences takes longer as using the timeline becomes inefficient.
For example, analysts using the desktop view would often zoom to perceive events and line diagrams without overlaps, similar to the editing process for larger videos (see \cite{gupta_intelligent_2021}).
Therefore, \tool must provide adaptive visualizations for real-world datasets with arbitrary size, for example, using large timelines (see \cite{aigner_visual_2008}) and intelligent filtering of unnecessarily logged road users and in-vehicle events.
Future research should also consider partitioned dataset analysis and gradually simplify and merge replications of passengers and other road users without losing relevant information to reduce visual clutter.
In addition, topic-specific adaptations might be necessary due to the per-design high generalizability of \tool.
For example, sophisticated in-vehicle conversational analysis (\topic{T6}) would require additional audio lines.

%%%%%%%%%%%%%%%%%%%%
\paragraph{Imperfect Data}
\tool can use datasets that partially meet the data specification (see \ref{data-specification}).
However, if data is missing, the current prototype may not display visualizations (e.g., avatars in case of missing skeletal data). 
In real-world conditions within research topics \topic{T1} - \topic{T8}, datasets can be incomplete.
For example, missing video recordings of the vehicle environment (see Drive\&Act \cite{driveact} and MDAD \cite{jegham_mdad_2019}).
Therefore, \tool needs to be further optimized for datasets of lower fidelity, for example, by interpolating missing motion data.
Besides, automatic inferences (e.g., for events or emotions) in the preprocessing (see \ref{preprocessing}) may misinterpret data due to imperfect DL.
As a result, analysts must review all inference results in the current prototype, which prolongs the analysis.

%%%%%%%%%%%%%%%%%%%%
\paragraph{Leveraging a Real Vehicle}
Evaluations of in-vehicle UIs (\topic{T2}) may benefit from using a real vehicle in the analysis.
However, such a vehicle should be parked or driven in the original study environment to prevent visualizations from interfering with the real environment.
During driving, safe analysis can only be performed in the passenger seat, as analysts cannot simultaneously drive.
This limits the analysis (at least until AVs can be used), as analysts cannot enter the driver's POV while driving.
Besides, current passthrough technology (e.g., less than 720p in the Meta Quest Pro) is not yet advanced enough to provide highly detailed real-world information.
Thus, future work might use augmented reality instead of passthrough VR for leveraging a real vehicle in the analysis.
In this case, the \tool concept of selectively adding the real to the virtual environment would also apply.

%%%%%%%%%%%%%%%%%%%%
\paragraph{Beyond the Vehicle Interior}
Our use cases and most AUI research (\topic{T1} - \topic{T7}) consider the vehicle environment only as additional information for the analysis.
However, \tool's extensive environment replication based on external sensors and automatic object/event recognition can represent interactions outside the vehicle in detail.
This enables examining eHMIs (\topic{T8}) and entering other road users' POVs.%, such as pedestrians or cars.
Thus, \tool is the first tool providing immersive analysis of human-vehicle interactions (in field studies) without interior or exterior restrictions.
However, analyzing UI interactions beyond the interior requires extensive LIDAR or external camera recordings.

\tool also reduces recording efforts for eHMI studies (e.g., \cite{colley_investigating_2021,colley_towards_2020}) by not requiring other road users equipped with sensors. 
However, \tool cannot visualize the physiological data of other road users, as such recordings are impractical in naturalistic driving studies.
Also, 3D movement and intention visualizations are limited by available DL recognition approaches.% used for recognition.

%%%%%%%%%%%%%%%%%%%%%%%%%%%%%%%%%%%%%%%%%%%%%%%%%%%%%%%%%%%%%%%%%%%%%%%%%%%%%%%%%%
\subsection{Limitations \& Future Work}
\label{limitations_future_work}
We demonstrated the potential of \tool, applied heuristics (see \ref{analysis_workflow_and_use_cases}), and presented three use cases in the AUI domain, addressing \topic{T1}, \topic{T2}, \topic{T3}, and \topic{T5}.
We plan to investigate use cases focusing on the research topics of driver distraction (\topic{T4}), conversational UIs (\topic{T6}), takeovers (\topic{T7}), and eHMIs (\topic{T8}).
Besides, \tool could include other forms of mobility, such as urban air mobility (e.g., unmanned air cabs) or micro-mobility (e.g., e-scooters).
Although the technical evaluation of \tool may not require usability studies (see \cite{ledo_evaluation_2018,olsen_evaluating_2007,hudson_concepts_2014}), an automotive domain expert user study might yield additional insights.
Therefore, we will conduct an expert user study to determine how our interaction concepts are used in a real-world analysis, which concepts users prefer, and what flaws or enhancements they identify. 
%In addition, we will conduct (in-the-wild) case studies to investigate how \tool scales to other use cases and derive implications for our design and interaction concept.
Moreover, we want to use \tool for analyses of our own AUI research projects to gain valuable insights while reducing analysis efforts.
As \tool is currently a prototype, we will create new features and 2D/3D visualizations for desktop and VR views in the future.
For this, we plan to improve the replication of 3D environments from driving data, see NVIDIA DRIVE Sim\footnote{\url{https://developer.nvidia.com/drive/drive-sim}; Accessed: 01.02.2023}, and employ procedural building generation (see \cite{salpisti_procedural_2022}).

%%%%%%%%%%%%%%%%%%%%%%%%%%%%%%%%%%%%%%%%%%%%%%%%%%%%%%%%%%%%%%%%%%%%%%%%%
%%%%%%%%%%%%%%%%%%%%%%%%%%%%%%%%%%%%%%%%%%%%%%%%%%%%%%%%%%%%%%%%%%%%%%%%%
\section{Conclusion}
We presented \tool, a mixed-immersion analysis tool combining an immersive VR with a non-immersive desktop view to enable the exploration of AUI interaction studies.
The VR view lets analysts re-experience an interactive recording of the original study.
Complementary, the desktop view provides an overview of study data and facilitates aggregated data analysis.
Both views are synchronized, bridging immersive and non-immersive analysis, and enabling collaborative analysis in multi-user scenarios.
We proposed visualization and tool interaction concepts based on design requirements derived from a literature analysis of AUI research and domain expert interviews.
Our concepts leverage the unique spatiality of AUI interactions with the interplay of in-vehicle and environmental contexts.
We utilize virtual humanoid avatars, 3D trajectories, and heatmap textures embedded in the vehicle interior and environment to visualize the behaviors of passengers and other road users.
In addition, we presented \textit{context portals} and \textit{driving path events} as domain-specific visualizations to link in-vehicle and environmental contexts.
To enable in situ visualizations, we also enable a real vehicle in the analysis via passthrough VR.
\tool could thus speed up the analysis process and preserve valuable contextual and environmental cues.

We demonstrated our concept's applicability to real-world analysis tasks in three use cases: (1) analysis of multimodal interaction in AVs, (2) analyzing a real-world dataset, and (3) leveraging a real vehicle in the analysis by implementing a prototype of \tool.
By applying heuristic evaluations, we could show that, despite currently being a research prototype, \tool can benefit the analysis of AUI interactions.
We plan to extend \tool and further evaluate our system in an expert user study.
Our work contributes to the underexplored field of (immersive) analytics for AUI interactions.
Besides, we tackle Grand Challenges of Immersive Analytics \cite{ens_grand_2021} by accurately placing visualizations in space and supporting transitions between analysis environments.
We are confident that \tool can inspire novel immersive analytics and significantly benefits analysis of human-vehicle interaction.

%%%%%%%%%%%%%%%%%%%%%%%%%% Acknowledgements %%%%%%%%%%%%%%%%%%%%%%%%%%%%%
\begin{acks}
We thank the reviewers for their helpful comments. This work was supported by the project 'SEMULIN' (\textbf{se}lbstunterstützende, \textbf{mul}timodale \textbf{In}teraktion) funded by the Federal Ministry for Economic Affairs and Energy (BMWi).
\end{acks}

\balance
%%
%% The next two lines define the bibliography style to be used, and
%% the bibliography file.
\bibliographystyle{ACM-Reference-Format}
\bibliography{_1_introduction, _2_related_work, _3_research_topics, _4_repos}

%%% -*-BibTeX-*-
%%% Do NOT edit. File created by BibTeX with style
%%% ACM-Reference-Format-Journals [18-Jan-2012].

\begin{thebibliography}{143}

%%% ====================================================================
%%% NOTE TO THE USER: you can override these defaults by providing
%%% customized versions of any of these macros before the \bibliography
%%% command.  Each of them MUST provide its own final punctuation,
%%% except for \shownote{}, \showDOI{}, and \showURL{}.  The latter two
%%% do not use final punctuation, in order to avoid confusing it with
%%% the Web address.
%%%
%%% To suppress output of a particular field, define its macro to expand
%%% to an empty string, or better, \unskip, like this:
%%%
%%% \newcommand{\showDOI}[1]{\unskip}   % LaTeX syntax
%%%
%%% \def \showDOI #1{\unskip}           % plain TeX syntax
%%%
%%% ====================================================================

\ifx \showCODEN    \undefined \def \showCODEN     #1{\unskip}     \fi
\ifx \showDOI      \undefined \def \showDOI       #1{#1}\fi
\ifx \showISBNx    \undefined \def \showISBNx     #1{\unskip}     \fi
\ifx \showISBNxiii \undefined \def \showISBNxiii  #1{\unskip}     \fi
\ifx \showISSN     \undefined \def \showISSN      #1{\unskip}     \fi
\ifx \showLCCN     \undefined \def \showLCCN      #1{\unskip}     \fi
\ifx \shownote     \undefined \def \shownote      #1{#1}          \fi
\ifx \showarticletitle \undefined \def \showarticletitle #1{#1}   \fi
\ifx \showURL      \undefined \def \showURL       {\relax}        \fi
% The following commands are used for tagged output and should be
% invisible to TeX
\providecommand\bibfield[2]{#2}
\providecommand\bibinfo[2]{#2}
\providecommand\natexlab[1]{#1}
\providecommand\showeprint[2][]{arXiv:#2}

\bibitem[Aftab(2019)]%
        {aftab_multimodal_2019}
\bibfield{author}{\bibinfo{person}{Abdul~Rafey Aftab}.}
  \bibinfo{year}{2019}\natexlab{}.
\newblock \showarticletitle{Multimodal {Driver} {Interaction} with {Gesture},
  {Gaze} and {Speech}}. In \bibinfo{booktitle}{\emph{2019 {International}
  {Conference} on {Multimodal} {Interaction}}} \emph{(\bibinfo{series}{{ICMI}
  '19})}. \bibinfo{publisher}{Association for Computing Machinery},
  \bibinfo{address}{New York, NY, USA}, \bibinfo{pages}{487--492}.
\newblock
\showISBNx{978-1-4503-6860-5}
\urldef\tempurl%
\url{https://doi.org/10.1145/3340555.3356093}
\showDOI{\tempurl}


\bibitem[Ahmad et~al\mbox{.}(2018)]%
        {ahmad_predictive_2018}
\bibfield{author}{\bibinfo{person}{Bashar~I. Ahmad}, \bibinfo{person}{Simon~J.
  Godsill}, \bibinfo{person}{Patrick~M. Langdon}, {and} \bibinfo{person}{Lee
  Skrypchuk}.} \bibinfo{year}{2018}\natexlab{}.
\newblock \showarticletitle{Predictive {Touch}: {A} {Novel} {HMI} {Technology}
  for {Intelligent} {Displays} in {Automotive}}. In
  \bibinfo{booktitle}{\emph{You {Have} a {Point} {There} {Object} {Selection}
  {Inside} an}} \emph{(\bibinfo{series}{{AutomotiveUI} '18})}.
  \bibinfo{publisher}{Association for Computing Machinery},
  \bibinfo{address}{New York, NY, USA}, \bibinfo{pages}{259--260}.
\newblock
\showISBNx{978-1-4503-5947-4}
\urldef\tempurl%
\url{https://doi.org/10.1145/3239092.3267103}
\showDOI{\tempurl}


\bibitem[Aigner et~al\mbox{.}(2008)]%
        {aigner_visual_2008}
\bibfield{author}{\bibinfo{person}{Wolfgang Aigner}, \bibinfo{person}{Silvia
  Miksch}, \bibinfo{person}{Wolfgang Müller}, \bibinfo{person}{Heidrun
  Schumann}, {and} \bibinfo{person}{Christian Tominski}.}
  \bibinfo{year}{2008}\natexlab{}.
\newblock \showarticletitle{Visual Methods for Analyzing Time-Oriented Data}.
\newblock \bibinfo{journal}{\emph{IEEE Transactions on Visualization and
  Computer Graphics}} \bibinfo{volume}{14}, \bibinfo{number}{1}
  (\bibinfo{year}{2008}), \bibinfo{pages}{47--60}.
\newblock
\urldef\tempurl%
\url{https://doi.org/10.1109/TVCG.2007.70415}
\showDOI{\tempurl}


\bibitem[{akshaybahadur21}(2022)]%
        {drowsiness-detection}
\bibfield{author}{\bibinfo{person}{{akshaybahadur21}}.}
  \bibinfo{year}{2022}\natexlab{}.
\newblock \bibinfo{booktitle}{\emph{Drowsiness Detection OpenCV}}.
\newblock Drowsiness\_Detection.
\newblock
\urldef\tempurl%
\url{{https://github.com/akshaybahadur21/Drowsiness\_Detection}}
\showURL{%
\tempurl}
\newblock
\shownote{(Accessed on 08/23/2022)}.


\bibitem[Albadawi et~al\mbox{.}(2022)]%
        {albadawi_review_2022}
\bibfield{author}{\bibinfo{person}{Yaman Albadawi}, \bibinfo{person}{Maen
  Takruri}, {and} \bibinfo{person}{Mohammed Awad}.}
  \bibinfo{year}{2022}\natexlab{}.
\newblock \showarticletitle{A Review of Recent Developments in Driver
  Drowsiness Detection Systems}.
\newblock \bibinfo{journal}{\emph{Sensors}} \bibinfo{volume}{22},
  \bibinfo{number}{5} (\bibinfo{year}{2022}), \bibinfo{pages}{41}.
\newblock
\showISSN{1424-8220}
\urldef\tempurl%
\url{https://doi.org/10.3390/s22052069}
\showDOI{\tempurl}


\bibitem[Ayoub et~al\mbox{.}(2019)]%
        {ayoub_10years_2019}
\bibfield{author}{\bibinfo{person}{Jackie Ayoub}, \bibinfo{person}{Feng Zhou},
  \bibinfo{person}{Shan Bao}, {and} \bibinfo{person}{X.~Jessie Yang}.}
  \bibinfo{year}{2019}\natexlab{}.
\newblock \showarticletitle{From Manual Driving to Automated Driving: A Review
  of 10 Years of AutoUI}. In \bibinfo{booktitle}{\emph{Proceedings of the 11th
  International Conference on Automotive User Interfaces and Interactive
  Vehicular Applications}} (Utrecht, Netherlands)
  \emph{(\bibinfo{series}{AutomotiveUI '19})}. \bibinfo{publisher}{Association
  for Computing Machinery}, \bibinfo{address}{New York, NY, USA},
  \bibinfo{pages}{70–90}.
\newblock
\showISBNx{9781450368841}
\urldef\tempurl%
\url{https://doi.org/10.1145/3342197.3344529}
\showDOI{\tempurl}


\bibitem[Bazilinskyy et~al\mbox{.}(2018)]%
        {BAZILINSKYY201882}
\bibfield{author}{\bibinfo{person}{P. Bazilinskyy}, \bibinfo{person}{S.M.
  Petermeijer}, \bibinfo{person}{V. Petrovych}, \bibinfo{person}{D. Dodou},
  {and} \bibinfo{person}{J.C.F. {de Winter}}.} \bibinfo{year}{2018}\natexlab{}.
\newblock \showarticletitle{Take-over requests in highly automated driving: A
  crowdsourcing survey on auditory, vibrotactile, and visual displays}.
\newblock \bibinfo{journal}{\emph{Transportation Research Part F: Traffic
  Psychology and Behaviour}}  \bibinfo{volume}{56} (\bibinfo{year}{2018}),
  \bibinfo{pages}{82--98}.
\newblock
\showISSN{1369-8478}
\urldef\tempurl%
\url{https://doi.org/10.1016/j.trf.2018.04.001}
\showDOI{\tempurl}


\bibitem[Benedetto et~al\mbox{.}(2011)]%
        {benedetto_driver_2011}
\bibfield{author}{\bibinfo{person}{Simone Benedetto}, \bibinfo{person}{Marco
  Pedrotti}, \bibinfo{person}{Luca Minin}, \bibinfo{person}{Thierry Baccino},
  \bibinfo{person}{Alessandra Re}, {and} \bibinfo{person}{Roberto Montanari}.}
  \bibinfo{year}{2011}\natexlab{}.
\newblock \showarticletitle{Driver workload and eye blink duration}.
\newblock \bibinfo{journal}{\emph{Transportation Research Part F: Traffic
  Psychology and Behaviour}} \bibinfo{volume}{14}, \bibinfo{number}{3}
  (\bibinfo{year}{2011}), \bibinfo{pages}{199--208}.
\newblock
\showISSN{1369-8478}
\urldef\tempurl%
\url{https://doi.org/10.1016/j.trf.2010.12.001}
\showDOI{\tempurl}


\bibitem[Bethge et~al\mbox{.}(2021)]%
        {bethge_vemotion_2021}
\bibfield{author}{\bibinfo{person}{David Bethge}, \bibinfo{person}{Thomas
  Kosch}, \bibinfo{person}{Tobias Grosse-Puppendahl}, \bibinfo{person}{Lewis~L.
  Chuang}, \bibinfo{person}{Mohamed Kari}, \bibinfo{person}{Alexander
  Jagaciak}, {and} \bibinfo{person}{Albrecht Schmidt}.}
  \bibinfo{year}{2021}\natexlab{}.
\newblock \showarticletitle{VEmotion: Using Driving Context for Indirect
  Emotion Prediction in Real-Time}. In \bibinfo{booktitle}{\emph{The 34th
  Annual ACM Symposium on User Interface Software and Technology}} (Virtual
  Event, USA) \emph{(\bibinfo{series}{UIST '21})}.
  \bibinfo{publisher}{Association for Computing Machinery},
  \bibinfo{address}{New York, NY, USA}, \bibinfo{pages}{638–651}.
\newblock
\showISBNx{9781450386357}
\urldef\tempurl%
\url{https://doi.org/10.1145/3472749.3474775}
\showDOI{\tempurl}


\bibitem[Billinghurst et~al\mbox{.}(2018)]%
        {billinghurst2018collaborative}
\bibfield{author}{\bibinfo{person}{Mark Billinghurst}, \bibinfo{person}{Maxime
  Cordeil}, \bibinfo{person}{Anastasia Bezerianos}, {and} \bibinfo{person}{Todd
  Margolis}.} \bibinfo{year}{2018}\natexlab{}.
\newblock \showarticletitle{Collaborative immersive analytics}.
\newblock In \bibinfo{booktitle}{\emph{Immersive Analytics}}.
  \bibinfo{publisher}{Springer}, \bibinfo{address}{Cham},
  \bibinfo{pages}{221--257}.
\newblock


\bibitem[{Blickshift}(2022)]%
        {blickshift}
\bibfield{author}{\bibinfo{person}{{Blickshift}}.}
  \bibinfo{year}{2022}\natexlab{}.
\newblock \bibinfo{booktitle}{\emph{Blickshift Analytics}}.
\newblock Blickshift GmbH.
\newblock
\urldef\tempurl%
\url{https://www.blickshift.com/products-services/blickshift-analytics/solutions-sciences-usability/}
\showURL{%
\tempurl}
\newblock
\shownote{(Accessed on 08/23/2022)}.


\bibitem[Bostock et~al\mbox{.}(2011)]%
        {d3}
\bibfield{author}{\bibinfo{person}{Michael Bostock}, \bibinfo{person}{Vadim
  Ogievetsky}, {and} \bibinfo{person}{Jeffrey Heer}.}
  \bibinfo{year}{2011}\natexlab{}.
\newblock \showarticletitle{D³ Data-Driven Documents}.
\newblock \bibinfo{journal}{\emph{IEEE Transactions on Visualization and
  Computer Graphics}} \bibinfo{volume}{17}, \bibinfo{number}{12}
  (\bibinfo{year}{2011}), \bibinfo{pages}{2301--2309}.
\newblock
\urldef\tempurl%
\url{https://doi.org/10.1109/TVCG.2011.185}
\showDOI{\tempurl}


\bibitem[Broy et~al\mbox{.}(2014)]%
        {broy_3d_2014}
\bibfield{author}{\bibinfo{person}{Nora Broy}, \bibinfo{person}{Florian Alt},
  \bibinfo{person}{Stefan Schneegass}, {and} \bibinfo{person}{Bastian
  Pfleging}.} \bibinfo{year}{2014}\natexlab{}.
\newblock \showarticletitle{3D Displays in Cars: Exploring the User Performance
  for a Stereoscopic Instrument Cluster}. In
  \bibinfo{booktitle}{\emph{Proceedings of the 6th International Conference on
  Automotive User Interfaces and Interactive Vehicular Applications}} (Seattle,
  WA, USA) \emph{(\bibinfo{series}{AutomotiveUI '14})}.
  \bibinfo{publisher}{Association for Computing Machinery},
  \bibinfo{address}{New York, NY, USA}, \bibinfo{pages}{1–9}.
\newblock
\showISBNx{9781450332125}
\urldef\tempurl%
\url{https://doi.org/10.1145/2667317.2667319}
\showDOI{\tempurl}


\bibitem[Brudy et~al\mbox{.}(2018)]%
        {brudy_eagle_2018}
\bibfield{author}{\bibinfo{person}{Frederik Brudy}, \bibinfo{person}{Suppachai
  Suwanwatcharachat}, \bibinfo{person}{Wenyu Zhang}, \bibinfo{person}{Steven
  Houben}, {and} \bibinfo{person}{Nicolai Marquardt}.}
  \bibinfo{year}{2018}\natexlab{}.
\newblock \showarticletitle{EagleView: A Video Analysis Tool for Visualising
  and Querying Spatial Interactions of People and Devices}. In
  \bibinfo{booktitle}{\emph{Proceedings of the 2018 ACM International
  Conference on Interactive Surfaces and Spaces}} (Tokyo, Japan)
  \emph{(\bibinfo{series}{ISS '18})}. \bibinfo{publisher}{Association for
  Computing Machinery}, \bibinfo{address}{New York, NY, USA},
  \bibinfo{pages}{61–72}.
\newblock
\showISBNx{9781450356947}
\urldef\tempurl%
\url{https://doi.org/10.1145/3279778.3279795}
\showDOI{\tempurl}


\bibitem[Br\"{u}ning et~al\mbox{.}(2012)]%
        {bruning_pamocat_2012}
\bibfield{author}{\bibinfo{person}{Bernhard Br\"{u}ning},
  \bibinfo{person}{Christian Schnier}, \bibinfo{person}{Karola Pitsch}, {and}
  \bibinfo{person}{Sven Wachsmuth}.} \bibinfo{year}{2012}\natexlab{}.
\newblock \showarticletitle{Integrating PAMOCAT in the Research Cycle: Linking
  Motion Capturing and Conversation Analysis}. In
  \bibinfo{booktitle}{\emph{Proceedings of the 14th ACM International
  Conference on Multimodal Interaction}} (Santa Monica, California, USA)
  \emph{(\bibinfo{series}{ICMI '12})}. \bibinfo{publisher}{Association for
  Computing Machinery}, \bibinfo{address}{New York, NY, USA},
  \bibinfo{pages}{201–208}.
\newblock
\showISBNx{9781450314671}
\urldef\tempurl%
\url{https://doi.org/10.1145/2388676.2388716}
\showDOI{\tempurl}


\bibitem[B\"{u}schel et~al\mbox{.}(2021)]%
        {buschel_miria_2021}
\bibfield{author}{\bibinfo{person}{Wolfgang B\"{u}schel}, \bibinfo{person}{Anke
  Lehmann}, {and} \bibinfo{person}{Raimund Dachselt}.}
  \bibinfo{year}{2021}\natexlab{}.
\newblock \showarticletitle{MIRIA: A Mixed Reality Toolkit for the In-Situ
  Visualization and Analysis of Spatio-Temporal Interaction Data}. In
  \bibinfo{booktitle}{\emph{Proceedings of the 2021 CHI Conference on Human
  Factors in Computing Systems}} (Yokohama, Japan) \emph{(\bibinfo{series}{CHI
  '21})}. \bibinfo{publisher}{Association for Computing Machinery},
  \bibinfo{address}{New York, NY, USA}, Article \bibinfo{articleno}{470},
  \bibinfo{numpages}{15}~pages.
\newblock
\showISBNx{9781450380966}
\urldef\tempurl%
\url{https://doi.org/10.1145/3411764.3445651}
\showDOI{\tempurl}


\bibitem[B\"{u}schel et~al\mbox{.}(2017)]%
        {buschel_investigating_2017}
\bibfield{author}{\bibinfo{person}{Wolfgang B\"{u}schel},
  \bibinfo{person}{Patrick Reipschl\"{a}ger}, \bibinfo{person}{Ricardo
  Langner}, {and} \bibinfo{person}{Raimund Dachselt}.}
  \bibinfo{year}{2017}\natexlab{}.
\newblock \showarticletitle{Investigating the Use of Spatial Interaction for 3D
  Data Visualization on Mobile Devices}. In
  \bibinfo{booktitle}{\emph{Proceedings of the 2017 ACM International
  Conference on Interactive Surfaces and Spaces}} (Brighton, United Kingdom)
  \emph{(\bibinfo{series}{ISS '17})}. \bibinfo{publisher}{Association for
  Computing Machinery}, \bibinfo{address}{New York, NY, USA},
  \bibinfo{pages}{62–71}.
\newblock
\showISBNx{9781450346917}
\urldef\tempurl%
\url{https://doi.org/10.1145/3132272.3134125}
\showDOI{\tempurl}


\bibitem[Butcher et~al\mbox{.}(2021)]%
        {butcher_vria_2021}
\bibfield{author}{\bibinfo{person}{Peter W.~S. Butcher},
  \bibinfo{person}{Nigel~W. John}, {and} \bibinfo{person}{Panagiotis~D.
  Ritsos}.} \bibinfo{year}{2021}\natexlab{}.
\newblock \showarticletitle{VRIA: A Web-Based Framework for Creating Immersive
  Analytics Experiences}.
\newblock \bibinfo{journal}{\emph{IEEE Transactions on Visualization and
  Computer Graphics}} \bibinfo{volume}{27}, \bibinfo{number}{7}
  (\bibinfo{date}{jul} \bibinfo{year}{2021}), \bibinfo{pages}{3213–3225}.
\newblock
\showISSN{1077-2626}
\urldef\tempurl%
\url{https://doi.org/10.1109/TVCG.2020.2965109}
\showDOI{\tempurl}


\bibitem[{Cesium}(2022)]%
        {cesium}
\bibfield{author}{\bibinfo{person}{{Cesium}}.} \bibinfo{year}{2022}\natexlab{}.
\newblock \bibinfo{booktitle}{\emph{Cesium for Unity}}.
\newblock Cesium GS, Inc.
\newblock
\urldef\tempurl%
\url{{https://cesium.com/platform/cesium-for-unity/}}
\showURL{%
\tempurl}
\newblock
\shownote{(Accessed on 08/23/2022)}.


\bibitem[Chang et~al\mbox{.}(2009)]%
        {chang_usability_2009}
\bibfield{author}{\bibinfo{person}{Jackie~C. Chang}, \bibinfo{person}{Annie
  Lien}, \bibinfo{person}{Brian Lathrop}, {and} \bibinfo{person}{Holger Hees}.}
  \bibinfo{year}{2009}\natexlab{}.
\newblock \showarticletitle{Usability Evaluation of a Volkswagen Group
  In-Vehicle Speech System}. In \bibinfo{booktitle}{\emph{Proceedings of the
  1st International Conference on Automotive User Interfaces and Interactive
  Vehicular Applications}} (Essen, Germany)
  \emph{(\bibinfo{series}{AutomotiveUI '09})}. \bibinfo{publisher}{Association
  for Computing Machinery}, \bibinfo{address}{New York, NY, USA},
  \bibinfo{pages}{137–144}.
\newblock
\showISBNx{9781605585710}
\urldef\tempurl%
\url{https://doi.org/10.1145/1620509.1620535}
\showDOI{\tempurl}


\bibitem[Chen et~al\mbox{.}(2021)]%
        {chen2021effect}
\bibfield{author}{\bibinfo{person}{Lei Chen}, \bibinfo{person}{Hai-Ning Liang},
  \bibinfo{person}{Feiyu Lu}, \bibinfo{person}{Jialin Wang},
  \bibinfo{person}{Wenjun Chen}, {and} \bibinfo{person}{Yong Yue}.}
  \bibinfo{year}{2021}\natexlab{}.
\newblock \showarticletitle{Effect of Collaboration Mode and Position
  Arrangement on Immersive Analytics Tasks in Virtual Reality: A Pilot Study}.
\newblock \bibinfo{journal}{\emph{Applied Sciences}} \bibinfo{volume}{11},
  \bibinfo{number}{21} (\bibinfo{year}{2021}), \bibinfo{pages}{10473}.
\newblock


\bibitem[Chittaro et~al\mbox{.}(2006)]%
        {chittaro_vuflow_2006}
\bibfield{author}{\bibinfo{person}{Luca Chittaro}, \bibinfo{person}{Roberto
  Ranon}, {and} \bibinfo{person}{Lucio Ieronutti}.}
  \bibinfo{year}{2006}\natexlab{}.
\newblock \showarticletitle{VU-Flow: A Visualization Tool for Analyzing
  Navigation in Virtual Environments}.
\newblock \bibinfo{journal}{\emph{IEEE Transactions on Visualization and
  Computer Graphics}} \bibinfo{volume}{12}, \bibinfo{number}{6}
  (\bibinfo{date}{nov} \bibinfo{year}{2006}), \bibinfo{pages}{1475–1485}.
\newblock
\showISSN{1077-2626}
\urldef\tempurl%
\url{https://doi.org/10.1109/TVCG.2006.109}
\showDOI{\tempurl}


\bibitem[{CityGen}(2022)]%
        {citygen}
\bibfield{author}{\bibinfo{person}{{CityGen}}.}
  \bibinfo{year}{2022}\natexlab{}.
\newblock \bibinfo{booktitle}{\emph{CityGen3D}}.
\newblock CityGen Technologies Ltd.
\newblock
\urldef\tempurl%
\url{https://www.citygen3d.com/}
\showURL{%
\tempurl}
\newblock
\shownote{(Accessed on 08/23/2022)}.


\bibitem[{CodeChefVIT}(2022)]%
        {stress-detector}
\bibfield{author}{\bibinfo{person}{{CodeChefVIT}}.}
  \bibinfo{year}{2022}\natexlab{}.
\newblock \bibinfo{booktitle}{\emph{An API to detect stress real-time using
  facial recognition}}.
\newblock Stress Detector.
\newblock
\urldef\tempurl%
\url{https://github.com/CodeChefVIT/Stress-Detector}
\showURL{%
\tempurl}
\newblock
\shownote{(Accessed on 08/23/2022)}.


\bibitem[Colley et~al\mbox{.}(2022a)]%
        {colley2022effects}
\bibfield{author}{\bibinfo{person}{Mark Colley}, \bibinfo{person}{Elvedin
  Bajrovic}, {and} \bibinfo{person}{Enrico Rukzio}.}
  \bibinfo{year}{2022}\natexlab{a}.
\newblock \showarticletitle{Effects of Pedestrian Behavior, Time Pressure, and
  Repeated Exposure on Crossing Decisions in Front of Automated Vehicles
  Equipped with External Communication}. In
  \bibinfo{booktitle}{\emph{Proceedings of the 2022 CHI Conference on Human
  Factors in Computing Systems}} (New Orleans, LA, USA)
  \emph{(\bibinfo{series}{CHI '22})}. \bibinfo{publisher}{Association for
  Computing Machinery}, \bibinfo{address}{New York, NY, USA}, Article
  \bibinfo{articleno}{367}, \bibinfo{numpages}{11}~pages.
\newblock
\showISBNx{9781450391573}
\urldef\tempurl%
\url{https://doi.org/10.1145/3491102.3517571}
\showDOI{\tempurl}


\bibitem[Colley et~al\mbox{.}(2021a)]%
        {colley_investigating_2021}
\bibfield{author}{\bibinfo{person}{Mark Colley}, \bibinfo{person}{Jan~Henry
  Belz}, {and} \bibinfo{person}{Enrico Rukzio}.}
  \bibinfo{year}{2021}\natexlab{a}.
\newblock \showarticletitle{Investigating the {Effects} of {Feedback}
  {Communication} of {Autonomous} {Vehicles}}. In
  \bibinfo{booktitle}{\emph{13th {International} {Conference} on {Automotive}
  {User} {Interfaces} and {Interactive} {Vehicular} {Applications}}}.
  \bibinfo{publisher}{ACM}, \bibinfo{address}{Leeds United Kingdom},
  \bibinfo{pages}{263--273}.
\newblock
\showISBNx{978-1-4503-8063-8}
\urldef\tempurl%
\url{https://doi.org/10.1145/3409118.3475133}
\showDOI{\tempurl}


\bibitem[Colley et~al\mbox{.}(2020a)]%
        {10.1145/3409120.3410648}
\bibfield{author}{\bibinfo{person}{Mark Colley}, \bibinfo{person}{Christian
  Br\"{a}uner}, \bibinfo{person}{Mirjam Lanzer}, \bibinfo{person}{Marcel
  Walch}, \bibinfo{person}{Martin Baumann}, {and} \bibinfo{person}{Enrico
  Rukzio}.} \bibinfo{year}{2020}\natexlab{a}.
\newblock \showarticletitle{Effect of Visualization of Pedestrian Intention
  Recognition on Trust and Cognitive Load}. In \bibinfo{booktitle}{\emph{12th
  International Conference on Automotive User Interfaces and Interactive
  Vehicular Applications}} (Virtual Event, DC, USA)
  \emph{(\bibinfo{series}{AutomotiveUI '20})}. \bibinfo{publisher}{Association
  for Computing Machinery}, \bibinfo{address}{New York, NY, USA},
  \bibinfo{pages}{181–191}.
\newblock
\showISBNx{9781450380652}
\urldef\tempurl%
\url{https://doi.org/10.1145/3409120.3410648}
\showDOI{\tempurl}


\bibitem[Colley et~al\mbox{.}(2021b)]%
        {10.1145/3411764.3445351}
\bibfield{author}{\bibinfo{person}{Mark Colley}, \bibinfo{person}{Benjamin
  Eder}, \bibinfo{person}{Jan~Ole Rixen}, {and} \bibinfo{person}{Enrico
  Rukzio}.} \bibinfo{year}{2021}\natexlab{b}.
\newblock \showarticletitle{Effects of Semantic Segmentation Visualization on
  Trust, Situation Awareness, and Cognitive Load in Highly Automated Vehicles}.
  In \bibinfo{booktitle}{\emph{Proceedings of the 2021 CHI Conference on Human
  Factors in Computing Systems}} (Yokohama, Japan) \emph{(\bibinfo{series}{CHI
  '21})}. \bibinfo{publisher}{Association for Computing Machinery},
  \bibinfo{address}{New York, NY, USA}, Article \bibinfo{articleno}{155},
  \bibinfo{numpages}{11}~pages.
\newblock
\showISBNx{9781450380966}
\urldef\tempurl%
\url{https://doi.org/10.1145/3411764.3445351}
\showDOI{\tempurl}


\bibitem[Colley et~al\mbox{.}(2022b)]%
        {colley2022investigating}
\bibfield{author}{\bibinfo{person}{Mark Colley}, \bibinfo{person}{Tim Fabian},
  {and} \bibinfo{person}{Enrico Rukzio}.} \bibinfo{year}{2022}\natexlab{b}.
\newblock \showarticletitle{Investigating the Effects of External Communication
  and Automation Behavior on Manual Drivers at Intersections}.
\newblock \bibinfo{journal}{\emph{Proc. ACM Hum.-Comput. Interact.}}
  \bibinfo{volume}{6}, \bibinfo{number}{MHCI}, Article \bibinfo{articleno}{176}
  (\bibinfo{date}{sep} \bibinfo{year}{2022}), \bibinfo{numpages}{16}~pages.
\newblock
\urldef\tempurl%
\url{https://doi.org/10.1145/3546711}
\showDOI{\tempurl}


\bibitem[Colley et~al\mbox{.}(2022c)]%
        {colley_swivrcarseat_2022}
\bibfield{author}{\bibinfo{person}{Mark Colley}, \bibinfo{person}{Pascal
  Jansen}, \bibinfo{person}{Enrico Rukzio}, {and} \bibinfo{person}{Jan
  Gugenheimer}.} \bibinfo{year}{2022}\natexlab{c}.
\newblock \showarticletitle{SwiVR-Car-Seat: Exploring Vehicle Motion Effects on
  Interaction Quality in Virtual Reality Automated Driving Using a Motorized
  Swivel Seat}.
\newblock \bibinfo{journal}{\emph{Proc. ACM Interact. Mob. Wearable Ubiquitous
  Technol.}} \bibinfo{volume}{5}, \bibinfo{number}{4}, Article
  \bibinfo{articleno}{150} (\bibinfo{date}{dec} \bibinfo{year}{2022}),
  \bibinfo{numpages}{26}~pages.
\newblock
\urldef\tempurl%
\url{https://doi.org/10.1145/3494968}
\showDOI{\tempurl}


\bibitem[Colley et~al\mbox{.}(2021c)]%
        {colley_increasing_2021}
\bibfield{author}{\bibinfo{person}{Mark Colley}, \bibinfo{person}{Surong Li},
  {and} \bibinfo{person}{Enrico Rukzio}.} \bibinfo{year}{2021}\natexlab{c}.
\newblock \showarticletitle{Increasing {Pedestrian} {Safety} {Using} {External}
  {Communication} of {Autonomous} {Vehicles} for {Signalling} {Hazards}}. In
  \bibinfo{booktitle}{\emph{Proceedings of the 23rd {International}
  {Conference} on {Mobile} {Human}-{Computer} {Interaction}}}.
  \bibinfo{publisher}{ACM}, \bibinfo{address}{Toulouse \& Virtual France},
  \bibinfo{pages}{1--10}.
\newblock
\showISBNx{978-1-4503-8328-8}
\urldef\tempurl%
\url{https://doi.org/10.1145/3447526.3472024}
\showDOI{\tempurl}


\bibitem[Colley et~al\mbox{.}(2022d)]%
        {10.1145/3534609}
\bibfield{author}{\bibinfo{person}{Mark Colley}, \bibinfo{person}{Max
  R\"{a}dler}, \bibinfo{person}{Jonas Glimmann}, {and} \bibinfo{person}{Enrico
  Rukzio}.} \bibinfo{year}{2022}\natexlab{d}.
\newblock \showarticletitle{Effects of Scene Detection, Scene Prediction, and
  Maneuver Planning Visualizations on Trust, Situation Awareness, and Cognitive
  Load in Highly Automated Vehicles}.
\newblock \bibinfo{journal}{\emph{Proc. ACM Interact. Mob. Wearable Ubiquitous
  Technol.}} \bibinfo{volume}{6}, \bibinfo{number}{2}, Article
  \bibinfo{articleno}{49} (\bibinfo{date}{jul} \bibinfo{year}{2022}),
  \bibinfo{numpages}{21}~pages.
\newblock
\urldef\tempurl%
\url{https://doi.org/10.1145/3534609}
\showDOI{\tempurl}


\bibitem[Colley et~al\mbox{.}(2020b)]%
        {colley_towards_2020}
\bibfield{author}{\bibinfo{person}{Mark Colley}, \bibinfo{person}{Marcel
  Walch}, \bibinfo{person}{Jan Gugenheimer}, \bibinfo{person}{Ali Askari},
  {and} \bibinfo{person}{Enrico Rukzio}.} \bibinfo{year}{2020}\natexlab{b}.
\newblock \showarticletitle{Towards {Inclusive} {External} {Communication} of
  {Autonomous} {Vehicles} for {Pedestrians} with {Vision} {Impairments}}. In
  \bibinfo{booktitle}{\emph{Proceedings of the 2020 {CHI} {Conference} on
  {Human} {Factors} in {Computing} {Systems}}}. \bibinfo{publisher}{ACM},
  \bibinfo{address}{Honolulu HI USA}, \bibinfo{pages}{1--14}.
\newblock
\showISBNx{978-1-4503-6708-0}
\urldef\tempurl%
\url{https://doi.org/10.1145/3313831.3376472}
\showDOI{\tempurl}


\bibitem[Colley et~al\mbox{.}(2022e)]%
        {colley_user_2022}
\bibfield{author}{\bibinfo{person}{Mark Colley}, \bibinfo{person}{Bastian
  Wankmüller}, \bibinfo{person}{Tim Mend}, \bibinfo{person}{Thomas Väth},
  \bibinfo{person}{Enrico Rukzio}, {and} \bibinfo{person}{Jan Gugenheimer}.}
  \bibinfo{year}{2022}\natexlab{e}.
\newblock \showarticletitle{User gesticulation inside an automated vehicle with
  external communication can cause confusion in pedestrians and a lower
  willingness to cross}.
\newblock \bibinfo{journal}{\emph{Transportation Research Part F: Traffic
  Psychology and Behaviour}}  \bibinfo{volume}{87} (\bibinfo{date}{May}
  \bibinfo{year}{2022}), \bibinfo{pages}{120--137}.
\newblock
\showISSN{13698478}
\urldef\tempurl%
\url{https://doi.org/10.1016/j.trf.2022.03.011}
\showDOI{\tempurl}


\bibitem[De~Clercq et~al\mbox{.}(2019)]%
        {de_external_2019}
\bibfield{author}{\bibinfo{person}{Koen De~Clercq}, \bibinfo{person}{Andre
  Dietrich}, \bibinfo{person}{Juan~Pablo N{\'u}{\~n}ez~Velasco},
  \bibinfo{person}{Joost De~Winter}, {and} \bibinfo{person}{Riender Happee}.}
  \bibinfo{year}{2019}\natexlab{}.
\newblock \showarticletitle{External human-machine interfaces on automated
  vehicles: effects on pedestrian crossing decisions}.
\newblock \bibinfo{journal}{\emph{Human factors}} \bibinfo{volume}{61},
  \bibinfo{number}{8} (\bibinfo{year}{2019}), \bibinfo{pages}{1353--1370}.
\newblock


\bibitem[DeCamp et~al\mbox{.}(2010)]%
        {decamp_immersive_2010}
\bibfield{author}{\bibinfo{person}{Philip DeCamp}, \bibinfo{person}{George
  Shaw}, \bibinfo{person}{Rony Kubat}, {and} \bibinfo{person}{Deb Roy}.}
  \bibinfo{year}{2010}\natexlab{}.
\newblock \showarticletitle{An Immersive System for Browsing and Visualizing
  Surveillance Video}. In \bibinfo{booktitle}{\emph{Proceedings of the 18th ACM
  International Conference on Multimedia}} (Firenze, Italy)
  \emph{(\bibinfo{series}{MM '10})}. \bibinfo{publisher}{Association for
  Computing Machinery}, \bibinfo{address}{New York, NY, USA},
  \bibinfo{pages}{371–380}.
\newblock
\showISBNx{9781605589336}
\urldef\tempurl%
\url{https://doi.org/10.1145/1873951.1874002}
\showDOI{\tempurl}


\bibitem[Detjen et~al\mbox{.}(2021a)]%
        {detjen_how_2021}
\bibfield{author}{\bibinfo{person}{Henrik Detjen}, \bibinfo{person}{Sarah
  Faltaous}, \bibinfo{person}{Bastian Pfleging}, \bibinfo{person}{Stefan
  Geisler}, {and} \bibinfo{person}{Stefan Schneegass}.}
  \bibinfo{year}{2021}\natexlab{a}.
\newblock \showarticletitle{How to {Increase} {Automated} {Vehicles}’
  {Acceptance} through {In}-{Vehicle} {Interaction} {Design}: {A} {Review}}.
\newblock \bibinfo{journal}{\emph{International Journal of Human–Computer
  Interaction}} \bibinfo{volume}{37}, \bibinfo{number}{4} (\bibinfo{date}{Feb.}
  \bibinfo{year}{2021}), \bibinfo{pages}{308--330}.
\newblock
\showISSN{1044-7318}
\urldef\tempurl%
\url{https://doi.org/10.1080/10447318.2020.1860517}
\showDOI{\tempurl}
\newblock
\shownote{Publisher: Taylor \& Francis \_eprint:
  https://doi.org/10.1080/10447318.2020.1860517}.


\bibitem[Detjen et~al\mbox{.}(2021b)]%
        {detjen_driving_2021}
\bibfield{author}{\bibinfo{person}{Henrik Detjen}, \bibinfo{person}{Stefan
  Geisler}, {and} \bibinfo{person}{Stefan Schneegass}.}
  \bibinfo{year}{2021}\natexlab{b}.
\newblock \showarticletitle{Driving as {Side} {Task}: {Exploring} {Intuitive}
  {Input} {Modalities} for {Multitasking} in {Automated} {Vehicles}}.
\newblock In \bibinfo{booktitle}{\emph{A {Simulation} {Study} {Examining}
  {Smartphone} {Destinatio}}}. \bibinfo{publisher}{Association for Computing
  Machinery}, \bibinfo{address}{New York, NY, USA}.
\newblock
\showISBNx{978-1-4503-8095-9}
\urldef\tempurl%
\url{https://doi.org/10.1145/3411763.3451803}
\showURL{%
\tempurl}


\bibitem[Detjen et~al\mbox{.}(2020)]%
        {detjen_wizard_2020}
\bibfield{author}{\bibinfo{person}{Henrik Detjen}, \bibinfo{person}{Bastian
  Pfleging}, {and} \bibinfo{person}{Stefan Schneegass}.}
  \bibinfo{year}{2020}\natexlab{}.
\newblock \showarticletitle{A Wizard of Oz Field Study to Understand
  Non-Driving-Related Activities, Trust, and Acceptance of Automated Vehicles}.
  In \bibinfo{booktitle}{\emph{12th International Conference on Automotive User
  Interfaces and Interactive Vehicular Applications}} (Virtual Event, DC, USA)
  \emph{(\bibinfo{series}{AutomotiveUI '20})}. \bibinfo{publisher}{Association
  for Computing Machinery}, \bibinfo{address}{New York, NY, USA},
  \bibinfo{pages}{19–29}.
\newblock
\showISBNx{9781450380652}
\urldef\tempurl%
\url{https://doi.org/10.1145/3409120.3410662}
\showDOI{\tempurl}


\bibitem[Dettmann et~al\mbox{.}(2021)]%
        {dettman_comfort_2021}
\bibfield{author}{\bibinfo{person}{Andre Dettmann}, \bibinfo{person}{Franziska
  Hartwich}, \bibinfo{person}{Patrick Roßner}, \bibinfo{person}{Matthias
  Beggiato}, \bibinfo{person}{Konstantin Felbel}, \bibinfo{person}{Josef
  Krems}, {and} \bibinfo{person}{Angelika~C. Bullinger}.}
  \bibinfo{year}{2021}\natexlab{}.
\newblock \showarticletitle{Comfort or Not? Automated Driving Style and User
  Characteristics Causing Human Discomfort in Automated Driving}.
\newblock \bibinfo{journal}{\emph{International Journal of Human–Computer
  Interaction}} \bibinfo{volume}{37}, \bibinfo{number}{4}
  (\bibinfo{year}{2021}), \bibinfo{pages}{331--339}.
\newblock
\urldef\tempurl%
\url{https://doi.org/10.1080/10447318.2020.1860518}
\showDOI{\tempurl}
\showeprint{https://doi.org/10.1080/10447318.2020.1860518}


\bibitem[Dey et~al\mbox{.}(2018)]%
        {dey_interface_2018}
\bibfield{author}{\bibinfo{person}{Debargha Dey}, \bibinfo{person}{Marieke
  Martens}, \bibinfo{person}{Chao Wang}, \bibinfo{person}{Felix Ros}, {and}
  \bibinfo{person}{Jacques Terken}.} \bibinfo{year}{2018}\natexlab{}.
\newblock \showarticletitle{Interface Concepts for Intent Communication from
  Autonomous Vehicles to Vulnerable Road Users}. In
  \bibinfo{booktitle}{\emph{Adjunct Proceedings of the 10th International
  Conference on Automotive User Interfaces and Interactive Vehicular
  Applications}} (Toronto, ON, Canada) \emph{(\bibinfo{series}{AutomotiveUI
  '18})}. \bibinfo{publisher}{Association for Computing Machinery},
  \bibinfo{address}{New York, NY, USA}, \bibinfo{pages}{82–86}.
\newblock
\showISBNx{9781450359474}
\urldef\tempurl%
\url{https://doi.org/10.1145/3239092.3265946}
\showDOI{\tempurl}


\bibitem[Dey and Terken(2017)]%
        {dey_pedestrian_2017}
\bibfield{author}{\bibinfo{person}{Debargha Dey} {and} \bibinfo{person}{Jacques
  Terken}.} \bibinfo{year}{2017}\natexlab{}.
\newblock \showarticletitle{Pedestrian Interaction with Vehicles: Roles of
  Explicit and Implicit Communication}. In
  \bibinfo{booktitle}{\emph{Proceedings of the 9th International Conference on
  Automotive User Interfaces and Interactive Vehicular Applications}}
  (Oldenburg, Germany) \emph{(\bibinfo{series}{AutomotiveUI '17})}.
  \bibinfo{publisher}{Association for Computing Machinery},
  \bibinfo{address}{New York, NY, USA}, \bibinfo{pages}{109–113}.
\newblock
\showISBNx{9781450351508}
\urldef\tempurl%
\url{https://doi.org/10.1145/3122986.3123009}
\showDOI{\tempurl}


\bibitem[Dillen et~al\mbox{.}(2020)]%
        {dillen_keep_2020}
\bibfield{author}{\bibinfo{person}{Nicole Dillen}, \bibinfo{person}{Marko
  Ilievski}, \bibinfo{person}{Edith Law}, \bibinfo{person}{Lennart~E. Nacke},
  \bibinfo{person}{Krzysztof Czarnecki}, {and} \bibinfo{person}{Oliver
  Schneider}.} \bibinfo{year}{2020}\natexlab{}.
\newblock \showarticletitle{Keep Calm and Ride Along: Passenger Comfort and
  Anxiety as Physiological Responses to Autonomous Driving Styles}. In
  \bibinfo{booktitle}{\emph{Proceedings of the 2020 CHI Conference on Human
  Factors in Computing Systems}} (Honolulu, HI, USA)
  \emph{(\bibinfo{series}{CHI '20})}. \bibinfo{publisher}{Association for
  Computing Machinery}, \bibinfo{address}{New York, NY, USA},
  \bibinfo{pages}{1–13}.
\newblock
\showISBNx{9781450367080}
\urldef\tempurl%
\url{https://doi.org/10.1145/3313831.3376247}
\showDOI{\tempurl}


\bibitem[Dmitrenko et~al\mbox{.}(2020)]%
        {dimitrenko_caroma_2020}
\bibfield{author}{\bibinfo{person}{Dmitrijs Dmitrenko},
  \bibinfo{person}{Emanuela Maggioni}, \bibinfo{person}{Giada Brianza},
  \bibinfo{person}{Brittany~E. Holthausen}, \bibinfo{person}{Bruce~N. Walker},
  {and} \bibinfo{person}{Marianna Obrist}.} \bibinfo{year}{2020}\natexlab{}.
\newblock \showarticletitle{CARoma Therapy: Pleasant Scents Promote Safer
  Driving, Better Mood, and Improved Well-Being in Angry Drivers}. In
  \bibinfo{booktitle}{\emph{Proceedings of the 2020 CHI Conference on Human
  Factors in Computing Systems}} (Honolulu, HI, USA)
  \emph{(\bibinfo{series}{CHI '20})}. \bibinfo{publisher}{Association for
  Computing Machinery}, \bibinfo{address}{New York, NY, USA},
  \bibinfo{pages}{1–13}.
\newblock
\showISBNx{9781450367080}
\urldef\tempurl%
\url{https://doi.org/10.1145/3313831.3376176}
\showDOI{\tempurl}


\bibitem[{Dovetail}(2022)]%
        {dovetail}
\bibfield{author}{\bibinfo{person}{{Dovetail}}.}
  \bibinfo{year}{2022}\natexlab{}.
\newblock \bibinfo{booktitle}{\emph{Markup}}.
\newblock Dovetail Research Pty. Ltd.
\newblock
\urldef\tempurl%
\url{https://dovetailapp.com/products/markup}
\showURL{%
\tempurl}
\newblock
\shownote{(Accessed on 08/23/2022)}.


\bibitem[Dumesny et~al\mbox{.}(2022)]%
        {gridstack}
\bibfield{author}{\bibinfo{person}{Alain Dumesny}, \bibinfo{person}{Dylan
  Weiss}, {and} \bibinfo{person}{Pavel Reznikov}.}
  \bibinfo{year}{2022}\natexlab{}.
\newblock \bibinfo{booktitle}{\emph{gridstack.js - Build interactive dashboards
  in minutes}}.
\newblock gridstack.js.
\newblock
\urldef\tempurl%
\url{https://gridstackjs.com/}
\showURL{%
\tempurl}
\newblock
\shownote{(Accessed on 08/23/2022)}.


\bibitem[{e-candeloro}(2022)]%
        {driver-state-detection}
\bibfield{author}{\bibinfo{person}{{e-candeloro}}.}
  \bibinfo{year}{2022}\natexlab{}.
\newblock \bibinfo{booktitle}{\emph{Real Time Driver State Detection}}.
\newblock Driver-State-Detection.
\newblock
\urldef\tempurl%
\url{https://github.com/e-candeloro/Driver-State-Detection}
\showURL{%
\tempurl}
\newblock
\shownote{(Accessed on 08/23/2022)}.


\bibitem[Ebel et~al\mbox{.}(2021)]%
        {ebel_visualizing_2021}
\bibfield{author}{\bibinfo{person}{Patrick Ebel}, \bibinfo{person}{Christoph
  Lingenfelder}, {and} \bibinfo{person}{Andreas Vogelsang}.}
  \bibinfo{year}{2021}\natexlab{}.
\newblock \showarticletitle{Visualizing Event Sequence Data for User Behavior
  Evaluation of In-Vehicle Information Systems}. In
  \bibinfo{booktitle}{\emph{13th International Conference on Automotive User
  Interfaces and Interactive Vehicular Applications}} (Leeds, United Kingdom)
  \emph{(\bibinfo{series}{AutomotiveUI '21})}. \bibinfo{publisher}{Association
  for Computing Machinery}, \bibinfo{address}{New York, NY, USA},
  \bibinfo{pages}{219–229}.
\newblock
\showISBNx{9781450380638}
\urldef\tempurl%
\url{https://doi.org/10.1145/3409118.3475140}
\showDOI{\tempurl}


\bibitem[Eisma et~al\mbox{.}(2020)]%
        {eisma_external_2020}
\bibfield{author}{\bibinfo{person}{Y.~B. Eisma}, \bibinfo{person}{S. van
  Bergen}, \bibinfo{person}{S.~M. ter Brake}, \bibinfo{person}{M.~T.~T.
  Hensen}, \bibinfo{person}{W.~J. Tempelaar}, {and} \bibinfo{person}{J.~C.~F.
  de Winter}.} \bibinfo{year}{2020}\natexlab{}.
\newblock \showarticletitle{External Human–Machine Interfaces: The Effect of
  Display Location on Crossing Intentions and Eye Movements}.
\newblock \bibinfo{journal}{\emph{Information}} \bibinfo{volume}{11},
  \bibinfo{number}{1} (\bibinfo{year}{2020}), \bibinfo{pages}{18}.
\newblock
\showISSN{2078-2489}
\urldef\tempurl%
\url{https://doi.org/10.3390/info11010013}
\showDOI{\tempurl}


\bibitem[{Empatica}(2022)]%
        {empatica}
\bibfield{author}{\bibinfo{person}{{Empatica}}.}
  \bibinfo{year}{2022}\natexlab{}.
\newblock \bibinfo{booktitle}{\emph{E4 Wristband}}.
\newblock Empatica Inc.
\newblock
\urldef\tempurl%
\url{https://www.empatica.com/research/e4/}
\showURL{%
\tempurl}
\newblock
\shownote{(Accessed on 08/23/2022)}.


\bibitem[Ens et~al\mbox{.}(2021)]%
        {ens_grand_2021}
\bibfield{author}{\bibinfo{person}{Barrett Ens}, \bibinfo{person}{Benjamin
  Bach}, \bibinfo{person}{Maxime Cordeil}, \bibinfo{person}{Ulrich Engelke},
  \bibinfo{person}{Marcos Serrano}, \bibinfo{person}{Wesley Willett},
  \bibinfo{person}{Arnaud Prouzeau}, \bibinfo{person}{Christoph Anthes},
  \bibinfo{person}{Wolfgang B\"{u}schel}, \bibinfo{person}{Cody Dunne},
  \bibinfo{person}{Tim Dwyer}, \bibinfo{person}{Jens Grubert},
  \bibinfo{person}{Jason~H. Haga}, \bibinfo{person}{Nurit Kirshenbaum},
  \bibinfo{person}{Dylan Kobayashi}, \bibinfo{person}{Tica Lin},
  \bibinfo{person}{Monsurat Olaosebikan}, \bibinfo{person}{Fabian Pointecker},
  \bibinfo{person}{David Saffo}, \bibinfo{person}{Nazmus Saquib},
  \bibinfo{person}{Dieter Schmalstieg}, \bibinfo{person}{Danielle~Albers
  Szafir}, \bibinfo{person}{Matt Whitlock}, {and} \bibinfo{person}{Yalong
  Yang}.} \bibinfo{year}{2021}\natexlab{}.
\newblock \showarticletitle{Grand Challenges in Immersive Analytics}. In
  \bibinfo{booktitle}{\emph{Proceedings of the 2021 CHI Conference on Human
  Factors in Computing Systems}} (Yokohama, Japan) \emph{(\bibinfo{series}{CHI
  '21})}. \bibinfo{publisher}{Association for Computing Machinery},
  \bibinfo{address}{New York, NY, USA}, Article \bibinfo{articleno}{459},
  \bibinfo{numpages}{17}~pages.
\newblock
\showISBNx{9781450380966}
\urldef\tempurl%
\url{https://doi.org/10.1145/3411764.3446866}
\showDOI{\tempurl}


\bibitem[Filho et~al\mbox{.}(2020)]%
        {filho_evaluating_2020}
\bibfield{author}{\bibinfo{person}{Jorge A.~Wagner Filho},
  \bibinfo{person}{Wolfgang Stuerzlinger}, {and} \bibinfo{person}{Luciana
  Nedel}.} \bibinfo{year}{2020}\natexlab{}.
\newblock \showarticletitle{Evaluating an Immersive Space-Time Cube
  Geovisualization for Intuitive Trajectory Data Exploration}.
\newblock \bibinfo{journal}{\emph{IEEE Transactions on Visualization and
  Computer Graphics}} \bibinfo{volume}{26}, \bibinfo{number}{1}
  (\bibinfo{year}{2020}), \bibinfo{pages}{514--524}.
\newblock
\urldef\tempurl%
\url{https://doi.org/10.1109/TVCG.2019.2934415}
\showDOI{\tempurl}


\bibitem[Fouch\'{e} et~al\mbox{.}(2022)]%
        {fouche2022timeline}
\bibfield{author}{\bibinfo{person}{Gwendal Fouch\'{e}}, \bibinfo{person}{Ferran
  Argelaguet~Sanz}, \bibinfo{person}{Emmanuel Faure}, {and}
  \bibinfo{person}{Charles Kervrann}.} \bibinfo{year}{2022}\natexlab{}.
\newblock \showarticletitle{Timeline Design Space for Immersive Exploration of
  Time-Varying Spatial 3D Data}. In \bibinfo{booktitle}{\emph{Proceedings of
  the 28th ACM Symposium on Virtual Reality Software and Technology}} (Tsukuba,
  Japan) \emph{(\bibinfo{series}{VRST '22})}. \bibinfo{publisher}{Association
  for Computing Machinery}, \bibinfo{address}{New York, NY, USA}, Article
  \bibinfo{articleno}{21}, \bibinfo{numpages}{11}~pages.
\newblock
\showISBNx{9781450398893}
\urldef\tempurl%
\url{https://doi.org/10.1145/3562939.3565612}
\showDOI{\tempurl}


\bibitem[Foundation(2022)]%
        {r}
\bibfield{author}{\bibinfo{person}{R Foundation}.}
  \bibinfo{year}{2022}\natexlab{}.
\newblock \bibinfo{booktitle}{\emph{R: {The} {R} {Project} for {Statistical}
  {Computing}}}.
\newblock The R Foundation.
\newblock
\urldef\tempurl%
\url{https://www.tibco.com/products/tibco-spotfire}
\showURL{%
\tempurl}
\newblock
\shownote{(Accessed on 08/23/2022)}.


\bibitem[Funk et~al\mbox{.}(2020)]%
        {funk_nonverbal_2020}
\bibfield{author}{\bibinfo{person}{Markus Funk}, \bibinfo{person}{Vanessa
  Tobisch}, {and} \bibinfo{person}{Adam Emfield}.}
  \bibinfo{year}{2020}\natexlab{}.
\newblock \showarticletitle{Non-Verbal Auditory Input for Controlling Binary,
  Discrete, and Continuous Input in Automotive User Interfaces}. In
  \bibinfo{booktitle}{\emph{Proceedings of the 2020 CHI Conference on Human
  Factors in Computing Systems}} (Honolulu, HI, USA)
  \emph{(\bibinfo{series}{CHI '20})}. \bibinfo{publisher}{Association for
  Computing Machinery}, \bibinfo{address}{New York, NY, USA},
  \bibinfo{pages}{1–13}.
\newblock
\showISBNx{9781450367080}
\urldef\tempurl%
\url{https://doi.org/10.1145/3313831.3376816}
\showDOI{\tempurl}


\bibitem[Goedicke et~al\mbox{.}(2018)]%
        {goedicke_vroom_2018}
\bibfield{author}{\bibinfo{person}{David Goedicke}, \bibinfo{person}{Jamy Li},
  \bibinfo{person}{Vanessa Evers}, {and} \bibinfo{person}{Wendy Ju}.}
  \bibinfo{year}{2018}\natexlab{}.
\newblock \showarticletitle{VR-OOM: Virtual Reality On-ROad Driving
  SiMulation}. In \bibinfo{booktitle}{\emph{Proceedings of the 2018 CHI
  Conference on Human Factors in Computing Systems}} (Montreal QC, Canada)
  \emph{(\bibinfo{series}{CHI '18})}. \bibinfo{publisher}{Association for
  Computing Machinery}, \bibinfo{address}{New York, NY, USA},
  \bibinfo{pages}{1–11}.
\newblock
\showISBNx{9781450356206}
\urldef\tempurl%
\url{https://doi.org/10.1145/3173574.3173739}
\showDOI{\tempurl}


\bibitem[Gomaa et~al\mbox{.}(2020)]%
        {gomaa_studying_2020}
\bibfield{author}{\bibinfo{person}{Amr Gomaa}, \bibinfo{person}{Guillermo
  Reyes}, \bibinfo{person}{Alexandra Alles}, \bibinfo{person}{Lydia Rupp},
  {and} \bibinfo{person}{Michael Feld}.} \bibinfo{year}{2020}\natexlab{}.
\newblock \showarticletitle{Studying Person-Specific Pointing and Gaze Behavior
  for Multimodal Referencing of Outside Objects from a Moving Vehicle}. In
  \bibinfo{booktitle}{\emph{Proceedings of the 2020 International Conference on
  Multimodal Interaction}} (Virtual Event, Netherlands)
  \emph{(\bibinfo{series}{ICMI '20})}. \bibinfo{publisher}{Association for
  Computing Machinery}, \bibinfo{address}{New York, NY, USA},
  \bibinfo{pages}{501–509}.
\newblock
\showISBNx{9781450375818}
\urldef\tempurl%
\url{https://doi.org/10.1145/3382507.3418817}
\showDOI{\tempurl}


\bibitem[Guo et~al\mbox{.}(2021)]%
        {guo_driving_2021}
\bibfield{author}{\bibinfo{person}{Yuxi Guo}, \bibinfo{person}{Qinyu Sun},
  \bibinfo{person}{Yanqi Su}, \bibinfo{person}{Yingshi Guo}, {and}
  \bibinfo{person}{Chang Wang}.} \bibinfo{year}{2021}\natexlab{}.
\newblock \showarticletitle{Can driving condition prompt systems improve
  passenger comfort of intelligent vehicles? A driving simulator study}.
\newblock \bibinfo{journal}{\emph{Transportation Research Part F: Traffic
  Psychology and Behaviour}}  \bibinfo{volume}{81} (\bibinfo{year}{2021}),
  \bibinfo{pages}{240--250}.
\newblock
\showISSN{1369-8478}
\urldef\tempurl%
\url{https://doi.org/10.1016/j.trf.2021.06.007}
\showDOI{\tempurl}


\bibitem[Gupta et~al\mbox{.}(2021)]%
        {gupta_intelligent_2021}
\bibfield{author}{\bibinfo{person}{Anchit Gupta},
  \bibinfo{person}{Faizan~Farooq Khan}, \bibinfo{person}{Rudrabha
  Mukhopadhyay}, \bibinfo{person}{Vinay~P. Namboodiri}, {and}
  \bibinfo{person}{C.~V. Jawahar}.} \bibinfo{year}{2021}\natexlab{}.
\newblock \showarticletitle{Intelligent Video Editing: Incorporating Modern
  Talking Face Generation Algorithms in a Video Editor}. In
  \bibinfo{booktitle}{\emph{Proceedings of the Twelfth Indian Conference on
  Computer Vision, Graphics and Image Processing}} (Jodhpur, India)
  \emph{(\bibinfo{series}{ICVGIP '21})}. \bibinfo{publisher}{Association for
  Computing Machinery}, \bibinfo{address}{New York, NY, USA}, Article
  \bibinfo{articleno}{25}, \bibinfo{numpages}{9}~pages.
\newblock
\showISBNx{9781450375962}
\urldef\tempurl%
\url{https://doi.org/10.1145/3490035.3490284}
\showDOI{\tempurl}


\bibitem[Haimerl et~al\mbox{.}(2022)]%
        {haimerl2022evaluation}
\bibfield{author}{\bibinfo{person}{Mathias Haimerl}, \bibinfo{person}{Mark
  Colley}, {and} \bibinfo{person}{Andreas Riener}.}
  \bibinfo{year}{2022}\natexlab{}.
\newblock \showarticletitle{Evaluation of Common External Communication
  Concepts of Automated Vehicles for People With Intellectual Disabilities}. In
  \bibinfo{booktitle}{\emph{Proc. ACM Hum.-Comput. Interact.}}
  \emph{(\bibinfo{series}{MobileHCI ’22})}. \bibinfo{publisher}{Association
  for Computing Machinery}, \bibinfo{address}{New York, NY, USA},
  \bibinfo{numpages}{13}~pages.
\newblock
\showISBNx{9781450367080}
\urldef\tempurl%
\url{https://doi.org/10.1145/3546717}
\showDOI{\tempurl}


\bibitem[Hecht et~al\mbox{.}(2020)]%
        {hecht_what_2020}
\bibfield{author}{\bibinfo{person}{Tobias Hecht}, \bibinfo{person}{Anna
  Feldh{\"u}tter}, \bibinfo{person}{Kathrin Draeger}, {and}
  \bibinfo{person}{Klaus Bengler}.} \bibinfo{year}{2020}\natexlab{}.
\newblock \showarticletitle{What Do You Do? An Analysis of Non-driving Related
  Activities During a 60 Minutes Conditionally Automated Highway Drive}. In
  \bibinfo{booktitle}{\emph{Human Interaction and Emerging Technologies}},
  \bibfield{editor}{\bibinfo{person}{Tareq Ahram}, \bibinfo{person}{Redha
  Taiar}, \bibinfo{person}{Serge Colson}, {and} \bibinfo{person}{Arnaud
  Choplin}} (Eds.). \bibinfo{publisher}{Springer International Publishing},
  \bibinfo{address}{Cham}, \bibinfo{pages}{28--34}.
\newblock
\showISBNx{978-3-030-25629-6}


\bibitem[Hernandez et~al\mbox{.}(2014)]%
        {hernandez_autoemotive_2014}
\bibfield{author}{\bibinfo{person}{Javier Hernandez}, \bibinfo{person}{Daniel
  McDuff}, \bibinfo{person}{Xavier Benavides}, \bibinfo{person}{Judith Amores},
  \bibinfo{person}{Pattie Maes}, {and} \bibinfo{person}{Rosalind Picard}.}
  \bibinfo{year}{2014}\natexlab{}.
\newblock \showarticletitle{AutoEmotive: Bringing Empathy to the Driving
  Experience to Manage Stress}. In \bibinfo{booktitle}{\emph{Proceedings of the
  2014 Companion Publication on Designing Interactive Systems}} (Vancouver, BC,
  Canada) \emph{(\bibinfo{series}{DIS Companion '14})}.
  \bibinfo{publisher}{Association for Computing Machinery},
  \bibinfo{address}{New York, NY, USA}, \bibinfo{pages}{53–56}.
\newblock
\showISBNx{9781450329033}
\urldef\tempurl%
\url{https://doi.org/10.1145/2598784.2602780}
\showDOI{\tempurl}


\bibitem[Hidalgo et~al\mbox{.}(2022)]%
        {openpose}
\bibfield{author}{\bibinfo{person}{Ginés Hidalgo}, \bibinfo{person}{Zhe Cao},
  \bibinfo{person}{Tomas Simon}, \bibinfo{person}{Shih-En Wei},
  \bibinfo{person}{Yaadhav Raaj}, \bibinfo{person}{Hanbyul Joo}, {and}
  \bibinfo{person}{Yaser Sheikh}.} \bibinfo{year}{2022}\natexlab{}.
\newblock \bibinfo{booktitle}{\emph{OpenPose}}.
\newblock CMU-Perceptual-Computing-Lab.
\newblock
\urldef\tempurl%
\url{https://github.com/CMU-Perceptual-Computing-Lab/openpose}
\showURL{%
\tempurl}
\newblock
\shownote{(Accessed on 08/23/2022)}.


\bibitem[Hock et~al\mbox{.}(2022)]%
        {hock2022vampire}
\bibfield{author}{\bibinfo{person}{Philipp Hock}, \bibinfo{person}{Mark
  Colley}, \bibinfo{person}{Ali Askari}, \bibinfo{person}{Tobias Wagner},
  \bibinfo{person}{Martin Baumann}, {and} \bibinfo{person}{Enrico Rukzio}.}
  \bibinfo{year}{2022}\natexlab{}.
\newblock \showarticletitle{Introducing VAMPIRE – Using Kinaesthetic Feedback
  in Virtual Reality for Automated Driving Experiments}. In
  \bibinfo{booktitle}{\emph{Proceedings of the 14th International Conference on
  Automotive User Interfaces and Interactive Vehicular Applications}} (Seoul,
  Republic of Korea) \emph{(\bibinfo{series}{AutomotiveUI '22})}.
  \bibinfo{publisher}{Association for Computing Machinery},
  \bibinfo{address}{New York, NY, USA}, \bibinfo{pages}{204–214}.
\newblock
\showISBNx{9781450394154}
\urldef\tempurl%
\url{https://doi.org/10.1145/3543174.3545252}
\showDOI{\tempurl}


\bibitem[Holl\"{a}nder et~al\mbox{.}(2021)]%
        {10.1145/3411764.3445480}
\bibfield{author}{\bibinfo{person}{Kai Holl\"{a}nder}, \bibinfo{person}{Mark
  Colley}, \bibinfo{person}{Enrico Rukzio}, {and} \bibinfo{person}{Andreas
  Butz}.} \bibinfo{year}{2021}\natexlab{}.
\newblock \showarticletitle{A Taxonomy of Vulnerable Road Users for HCI Based
  On A Systematic Literature Review}. In \bibinfo{booktitle}{\emph{Proceedings
  of the 2021 CHI Conference on Human Factors in Computing Systems}} (Yokohama,
  Japan) \emph{(\bibinfo{series}{CHI '21})}. \bibinfo{publisher}{Association
  for Computing Machinery}, \bibinfo{address}{New York, NY, USA}, Article
  \bibinfo{articleno}{158}, \bibinfo{numpages}{13}~pages.
\newblock
\showISBNx{9781450380966}
\urldef\tempurl%
\url{https://doi.org/10.1145/3411764.3445480}
\showDOI{\tempurl}


\bibitem[Houben and Marquardt(2015)]%
        {houben_watchconnect_2015}
\bibfield{author}{\bibinfo{person}{Steven Houben} {and}
  \bibinfo{person}{Nicolai Marquardt}.} \bibinfo{year}{2015}\natexlab{}.
\newblock \showarticletitle{WatchConnect: A Toolkit for Prototyping
  Smartwatch-Centric Cross-Device Applications}. In
  \bibinfo{booktitle}{\emph{Proceedings of the 33rd Annual ACM Conference on
  Human Factors in Computing Systems}} (Seoul, Republic of Korea)
  \emph{(\bibinfo{series}{CHI '15})}. \bibinfo{publisher}{Association for
  Computing Machinery}, \bibinfo{address}{New York, NY, USA},
  \bibinfo{pages}{1247–1256}.
\newblock
\showISBNx{9781450331456}
\urldef\tempurl%
\url{https://doi.org/10.1145/2702123.2702215}
\showDOI{\tempurl}


\bibitem[Hubenschmid et~al\mbox{.}(2022)]%
        {hubenschmid_relive_2022}
\bibfield{author}{\bibinfo{person}{Sebastian Hubenschmid},
  \bibinfo{person}{Jonathan Wieland}, \bibinfo{person}{Daniel~Immanuel Fink},
  \bibinfo{person}{Andrea Batch}, \bibinfo{person}{Johannes Zagermann},
  \bibinfo{person}{Niklas Elmqvist}, {and} \bibinfo{person}{Harald Reiterer}.}
  \bibinfo{year}{2022}\natexlab{}.
\newblock \showarticletitle{ReLive: Bridging In-Situ and Ex-Situ Visual
  Analytics for Analyzing Mixed Reality User Studies}. In
  \bibinfo{booktitle}{\emph{Proceedings of the 2022 CHI Conference on Human
  Factors in Computing Systems}} (New Orleans, LA, USA)
  \emph{(\bibinfo{series}{CHI '22})}. \bibinfo{publisher}{Association for
  Computing Machinery}, \bibinfo{address}{New York, NY, USA}, Article
  \bibinfo{articleno}{24}, \bibinfo{numpages}{20}~pages.
\newblock
\showISBNx{9781450391573}
\urldef\tempurl%
\url{https://doi.org/10.1145/3491102.3517550}
\showDOI{\tempurl}


\bibitem[Hubenschmid et~al\mbox{.}(2021)]%
        {hubenschmid2021stream}
\bibfield{author}{\bibinfo{person}{Sebastian Hubenschmid},
  \bibinfo{person}{Johannes Zagermann}, \bibinfo{person}{Simon Butscher}, {and}
  \bibinfo{person}{Harald Reiterer}.} \bibinfo{year}{2021}\natexlab{}.
\newblock \showarticletitle{STREAM: Exploring the Combination of
  Spatially-Aware Tablets with Augmented Reality Head-Mounted Displays for
  Immersive Analytics}. In \bibinfo{booktitle}{\emph{Proceedings of the 2021
  CHI Conference on Human Factors in Computing Systems}} (Yokohama, Japan)
  \emph{(\bibinfo{series}{CHI '21})}. \bibinfo{publisher}{Association for
  Computing Machinery}, \bibinfo{address}{New York, NY, USA}, Article
  \bibinfo{articleno}{469}, \bibinfo{numpages}{14}~pages.
\newblock
\showISBNx{9781450380966}
\urldef\tempurl%
\url{https://doi.org/10.1145/3411764.3445298}
\showDOI{\tempurl}


\bibitem[Hudson and Mankoff(2014)]%
        {hudson_concepts_2014}
\bibfield{author}{\bibinfo{person}{Scott~E. Hudson} {and}
  \bibinfo{person}{Jennifer Mankoff}.} \bibinfo{year}{2014}\natexlab{}.
\newblock \bibinfo{booktitle}{\emph{Concepts, Values, and Methods for Technical
  Human--Computer Interaction Research}}.
\newblock \bibinfo{publisher}{Springer New York}, \bibinfo{address}{New York,
  NY}, \bibinfo{pages}{69--93}.
\newblock
\showISBNx{978-1-4939-0378-8}
\urldef\tempurl%
\url{https://doi.org/10.1007/978-1-4939-0378-8_4}
\showDOI{\tempurl}


\bibitem[Isenberg et~al\mbox{.}(2011)]%
        {isenberg2011collaborative}
\bibfield{author}{\bibinfo{person}{Petra Isenberg}, \bibinfo{person}{Niklas
  Elmqvist}, \bibinfo{person}{Jean Scholtz}, \bibinfo{person}{Daniel Cernea},
  \bibinfo{person}{Kwan-Liu Ma}, {and} \bibinfo{person}{Hans Hagen}.}
  \bibinfo{year}{2011}\natexlab{}.
\newblock \showarticletitle{Collaborative visualization: Definition,
  challenges, and research agenda}.
\newblock \bibinfo{journal}{\emph{Information Visualization}}
  \bibinfo{volume}{10}, \bibinfo{number}{4} (\bibinfo{year}{2011}),
  \bibinfo{pages}{310--326}.
\newblock


\bibitem[Jakobsen and Hornb\AE{}k(2014)]%
        {jakobsen_upclose_2014}
\bibfield{author}{\bibinfo{person}{Mikkel~R. Jakobsen} {and}
  \bibinfo{person}{Kasper Hornb\AE{}k}.} \bibinfo{year}{2014}\natexlab{}.
\newblock \showarticletitle{Up Close and Personal: Collaborative Work on a
  High-Resolution Multitouch Wall Display}.
\newblock \bibinfo{journal}{\emph{ACM Trans. Comput.-Hum. Interact.}}
  \bibinfo{volume}{21}, \bibinfo{number}{2}, Article \bibinfo{articleno}{11}
  (\bibinfo{date}{feb} \bibinfo{year}{2014}), \bibinfo{numpages}{34}~pages.
\newblock
\showISSN{1073-0516}
\urldef\tempurl%
\url{https://doi.org/10.1145/2576099}
\showDOI{\tempurl}


\bibitem[Jansen et~al\mbox{.}(2022)]%
        {jansen_design_2022}
\bibfield{author}{\bibinfo{person}{Pascal Jansen}, \bibinfo{person}{Mark
  Colley}, {and} \bibinfo{person}{Enrico Rukzio}.}
  \bibinfo{year}{2022}\natexlab{}.
\newblock \showarticletitle{A Design Space for Human Sensor and Actuator
  Focused In-Vehicle Interaction Based on a Systematic Literature Review}.
\newblock \bibinfo{journal}{\emph{Proc. ACM Interact. Mob. Wearable Ubiquitous
  Technol.}} \bibinfo{volume}{6}, \bibinfo{number}{2}, Article
  \bibinfo{articleno}{56} (\bibinfo{date}{jul} \bibinfo{year}{2022}),
  \bibinfo{numpages}{51}~pages.
\newblock
\urldef\tempurl%
\url{https://doi.org/10.1145/3534617}
\showDOI{\tempurl}


\bibitem[Jegham et~al\mbox{.}(2019)]%
        {jegham_mdad_2019}
\bibfield{author}{\bibinfo{person}{Imen Jegham}, \bibinfo{person}{Anouar
  Ben~Khalifa}, \bibinfo{person}{Ihsen Alouani}, {and}
  \bibinfo{person}{Mohamed~Ali Mahjoub}.} \bibinfo{year}{2019}\natexlab{}.
\newblock \showarticletitle{MDAD: A Multimodal and Multiview in-Vehicle Driver
  Action Dataset}. In \bibinfo{booktitle}{\emph{Computer Analysis of Images and
  Patterns}}, \bibfield{editor}{\bibinfo{person}{Mario Vento} {and}
  \bibinfo{person}{Gennaro Percannella}} (Eds.). \bibinfo{publisher}{Springer
  International Publishing}, \bibinfo{address}{Cham},
  \bibinfo{pages}{518--529}.
\newblock


\bibitem[Kennedy et~al\mbox{.}(1993)]%
        {kennedy_simulator_1993}
\bibfield{author}{\bibinfo{person}{Robert~S. Kennedy},
  \bibinfo{person}{Norman~E. Lane}, \bibinfo{person}{Kevin~S. Berbaum}, {and}
  \bibinfo{person}{Michael~G. Lilienthal}.} \bibinfo{year}{1993}\natexlab{}.
\newblock \showarticletitle{Simulator Sickness Questionnaire: An Enhanced
  Method for Quantifying Simulator Sickness}.
\newblock \bibinfo{journal}{\emph{The International Journal of Aviation
  Psychology}} \bibinfo{volume}{3}, \bibinfo{number}{3} (\bibinfo{year}{1993}),
  \bibinfo{pages}{203--220}.
\newblock
\urldef\tempurl%
\url{https://doi.org/10.1207/s15327108ijap0303\_3}
\showDOI{\tempurl}


\bibitem[Kim et~al\mbox{.}(2019)]%
        {kim_predicting_2019}
\bibfield{author}{\bibinfo{person}{Auk Kim}, \bibinfo{person}{Woohyeok Choi},
  \bibinfo{person}{Jungmi Park}, \bibinfo{person}{Kyeyoon Kim}, {and}
  \bibinfo{person}{Uichin Lee}.} \bibinfo{year}{2019}\natexlab{}.
\newblock \showarticletitle{Predicting Opportune Moments for In-Vehicle
  Proactive Speech Services}. In \bibinfo{booktitle}{\emph{Adjunct Proceedings
  of the 2019 ACM International Joint Conference on Pervasive and Ubiquitous
  Computing and Proceedings of the 2019 ACM International Symposium on Wearable
  Computers}} (London, United Kingdom) \emph{(\bibinfo{series}{UbiComp/ISWC '19
  Adjunct})}. \bibinfo{publisher}{Association for Computing Machinery},
  \bibinfo{address}{New York, NY, USA}, \bibinfo{pages}{101–104}.
\newblock
\showISBNx{9781450368698}
\urldef\tempurl%
\url{https://doi.org/10.1145/3341162.3343841}
\showDOI{\tempurl}


\bibitem[Kim et~al\mbox{.}(2020a)]%
        {kim_seated_2020}
\bibfield{author}{\bibinfo{person}{Eunyeong Kim}, \bibinfo{person}{Mohammad
  Fard}, {and} \bibinfo{person}{Kazuhito Kato}.}
  \bibinfo{year}{2020}\natexlab{a}.
\newblock \showarticletitle{A seated human model for predicting the coupled
  human-seat transmissibility exposed to fore-aft whole-body vibration}.
\newblock \bibinfo{journal}{\emph{Applied Ergonomics}}  \bibinfo{volume}{84}
  (\bibinfo{date}{April} \bibinfo{year}{2020}), \bibinfo{pages}{102929}.
\newblock
\showISSN{0003-6870}
\urldef\tempurl%
\url{https://doi.org/10.1016/j.apergo.2019.102929}
\showDOI{\tempurl}


\bibitem[Kim et~al\mbox{.}(2018)]%
        {kim_design_2018}
\bibfield{author}{\bibinfo{person}{Kyongho Kim}, \bibinfo{person}{HyunKyun
  Choi}, {and} \bibinfo{person}{Byungtae Jang}.}
  \bibinfo{year}{2018}\natexlab{}.
\newblock \showarticletitle{Design of the Driver-Adaptive Vehicle Interaction
  System}. In \bibinfo{booktitle}{\emph{2018 International Conference on
  Information and Communication Technology Convergence (ICTC)}}.
  \bibinfo{publisher}{IEEE}, \bibinfo{address}{3 Park Ave, New York City},
  \bibinfo{pages}{297--299}.
\newblock
\urldef\tempurl%
\url{https://doi.org/10.1109/ICTC.2018.8539526}
\showDOI{\tempurl}


\bibitem[Kim et~al\mbox{.}(2020b)]%
        {kim_cascaded_2020}
\bibfield{author}{\bibinfo{person}{Myeongseop Kim}, \bibinfo{person}{Eunjin
  Seong}, \bibinfo{person}{Younkyung Jwa}, \bibinfo{person}{Jieun Lee}, {and}
  \bibinfo{person}{Seungjun Kim}.} \bibinfo{year}{2020}\natexlab{b}.
\newblock \showarticletitle{A {Cascaded} {Multimodal} {Natural} {User}
  {Interface} to {Reduce} {Driver} {Distraction}}.
\newblock \bibinfo{journal}{\emph{IEEE Access}}  \bibinfo{volume}{8}
  (\bibinfo{year}{2020}), \bibinfo{pages}{112969--112984}.
\newblock
\showISSN{2169-3536}
\urldef\tempurl%
\url{https://doi.org/10.1109/ACCESS.2020.3002775}
\showDOI{\tempurl}


\bibitem[Kloiber et~al\mbox{.}(2020)]%
        {kloiber_immersive_2020}
\bibfield{author}{\bibinfo{person}{Simon Kloiber}, \bibinfo{person}{Volker
  Settgast}, \bibinfo{person}{Christoph Schinko}, \bibinfo{person}{Martin
  Weinzerl}, \bibinfo{person}{Johannes Fritz}, \bibinfo{person}{Tobias
  Schreck}, {and} \bibinfo{person}{Reinhold Preiner}.}
  \bibinfo{year}{2020}\natexlab{}.
\newblock \showarticletitle{Immersive analysis of user motion in VR
  applications}.
\newblock \bibinfo{journal}{\emph{The Visual Computer}} \bibinfo{volume}{36},
  \bibinfo{number}{10} (\bibinfo{year}{2020}), \bibinfo{pages}{1937--1949}.
\newblock


\bibitem[Kraus et~al\mbox{.}(2020a)]%
        {kraus_assessing_2020}
\bibfield{author}{\bibinfo{person}{Matthias Kraus}, \bibinfo{person}{Katrin
  Angerbauer}, \bibinfo{person}{Juri Buchm\"{u}ller}, \bibinfo{person}{Daniel
  Schweitzer}, \bibinfo{person}{Daniel~A. Keim}, \bibinfo{person}{Michael
  Sedlmair}, {and} \bibinfo{person}{Johannes Fuchs}.}
  \bibinfo{year}{2020}\natexlab{a}.
\newblock \showarticletitle{Assessing 2D and 3D Heatmaps for Comparative
  Analysis: An Empirical Study}. In \bibinfo{booktitle}{\emph{Proceedings of
  the 2020 CHI Conference on Human Factors in Computing Systems}} (Honolulu,
  HI, USA) \emph{(\bibinfo{series}{CHI '20})}. \bibinfo{publisher}{Association
  for Computing Machinery}, \bibinfo{address}{New York, NY, USA},
  \bibinfo{pages}{1–14}.
\newblock
\showISBNx{9781450367080}
\urldef\tempurl%
\url{https://doi.org/10.1145/3313831.3376675}
\showDOI{\tempurl}


\bibitem[Kraus et~al\mbox{.}(2020b)]%
        {kraus_impact_2020}
\bibfield{author}{\bibinfo{person}{M. Kraus}, \bibinfo{person}{N. Weiler},
  \bibinfo{person}{D. Oelke}, \bibinfo{person}{J. Kehrer},
  \bibinfo{person}{D.~A. Keim}, {and} \bibinfo{person}{J. Fuchs}.}
  \bibinfo{year}{2020}\natexlab{b}.
\newblock \showarticletitle{The Impact of Immersion on Cluster Identification
  Tasks}.
\newblock \bibinfo{journal}{\emph{IEEE Transactions on Visualization and
  Computer Graphics}} \bibinfo{volume}{26}, \bibinfo{number}{1}
  (\bibinfo{year}{2020}), \bibinfo{pages}{525--535}.
\newblock
\urldef\tempurl%
\url{https://doi.org/10.1109/TVCG.2019.2934395}
\showDOI{\tempurl}


\bibitem[Kun et~al\mbox{.}(2009)]%
        {kun_glancing_2009}
\bibfield{author}{\bibinfo{person}{Andrew~L. Kun}, \bibinfo{person}{Tim Paek},
  \bibinfo{person}{\v{Z}eljko Medenica}, \bibinfo{person}{Nemanja
  Memarovi\'{c}}, {and} \bibinfo{person}{Oskar Palinko}.}
  \bibinfo{year}{2009}\natexlab{}.
\newblock \showarticletitle{Glancing at Personal Navigation Devices Can Affect
  Driving: Experimental Results and Design Implications}. In
  \bibinfo{booktitle}{\emph{Proceedings of the 1st International Conference on
  Automotive User Interfaces and Interactive Vehicular Applications}} (Essen,
  Germany) \emph{(\bibinfo{series}{AutomotiveUI '09})}.
  \bibinfo{publisher}{Association for Computing Machinery},
  \bibinfo{address}{New York, NY, USA}, \bibinfo{pages}{129–136}.
\newblock
\showISBNx{9781605585710}
\urldef\tempurl%
\url{https://doi.org/10.1145/1620509.1620534}
\showDOI{\tempurl}


\bibitem[Kundinger et~al\mbox{.}(2021)]%
        {kundinger_performance_2021}
\bibfield{author}{\bibinfo{person}{Thomas Kundinger}, \bibinfo{person}{Andreas
  Riener}, {and} \bibinfo{person}{Ramyashree Bhat}.}
  \bibinfo{year}{2021}\natexlab{}.
\newblock \showarticletitle{Performance and Acceptance Evaluation of a Driver
  Drowsiness Detection System Based on Smart Wearables}. In
  \bibinfo{booktitle}{\emph{13th International Conference on Automotive User
  Interfaces and Interactive Vehicular Applications}} (Leeds, United Kingdom)
  \emph{(\bibinfo{series}{AutomotiveUI '21})}. \bibinfo{publisher}{Association
  for Computing Machinery}, \bibinfo{address}{New York, NY, USA},
  \bibinfo{pages}{49–58}.
\newblock
\showISBNx{9781450380638}
\urldef\tempurl%
\url{https://doi.org/10.1145/3409118.3475141}
\showDOI{\tempurl}


\bibitem[Langner et~al\mbox{.}(2021)]%
        {langner_marvis_2021}
\bibfield{author}{\bibinfo{person}{Ricardo Langner}, \bibinfo{person}{Marc
  Satkowski}, \bibinfo{person}{Wolfgang B\"{u}schel}, {and}
  \bibinfo{person}{Raimund Dachselt}.} \bibinfo{year}{2021}\natexlab{}.
\newblock \showarticletitle{MARVIS: Combining Mobile Devices and Augmented
  Reality for Visual Data Analysis}. In \bibinfo{booktitle}{\emph{Proceedings
  of the 2021 CHI Conference on Human Factors in Computing Systems}} (Yokohama,
  Japan) \emph{(\bibinfo{series}{CHI '21})}. \bibinfo{publisher}{Association
  for Computing Machinery}, \bibinfo{address}{New York, NY, USA}, Article
  \bibinfo{articleno}{468}, \bibinfo{numpages}{17}~pages.
\newblock
\showISBNx{9781450380966}
\urldef\tempurl%
\url{https://doi.org/10.1145/3411764.3445593}
\showDOI{\tempurl}


\bibitem[Ledo et~al\mbox{.}(2018)]%
        {ledo_evaluation_2018}
\bibfield{author}{\bibinfo{person}{David Ledo}, \bibinfo{person}{Steven
  Houben}, \bibinfo{person}{Jo Vermeulen}, \bibinfo{person}{Nicolai Marquardt},
  \bibinfo{person}{Lora Oehlberg}, {and} \bibinfo{person}{Saul Greenberg}.}
  \bibinfo{year}{2018}\natexlab{}.
\newblock \showarticletitle{Evaluation Strategies for HCI Toolkit Research}. In
  \bibinfo{booktitle}{\emph{Proceedings of the 2018 CHI Conference on Human
  Factors in Computing Systems}} (Montreal QC, Canada)
  \emph{(\bibinfo{series}{CHI '18})}. \bibinfo{publisher}{Association for
  Computing Machinery}, \bibinfo{address}{New York, NY, USA},
  \bibinfo{pages}{1–17}.
\newblock
\showISBNx{9781450356206}
\urldef\tempurl%
\url{https://doi.org/10.1145/3173574.3173610}
\showDOI{\tempurl}


\bibitem[Lee et~al\mbox{.}(2021)]%
        {lee_data_2021}
\bibfield{author}{\bibinfo{person}{Benjamin Lee}, \bibinfo{person}{Dave Brown},
  \bibinfo{person}{Bongshin Lee}, \bibinfo{person}{Christophe Hurter},
  \bibinfo{person}{Steven Drucker}, {and} \bibinfo{person}{Tim Dwyer}.}
  \bibinfo{year}{2021}\natexlab{}.
\newblock \showarticletitle{Data Visceralization: Enabling Deeper Understanding
  of Data Using Virtual Reality}.
\newblock \bibinfo{journal}{\emph{IEEE Transactions on Visualization and
  Computer Graphics}} \bibinfo{volume}{27}, \bibinfo{number}{2}
  (\bibinfo{year}{2021}), \bibinfo{pages}{1095--1105}.
\newblock
\urldef\tempurl%
\url{https://doi.org/10.1109/TVCG.2020.3030435}
\showDOI{\tempurl}


\bibitem[Lee et~al\mbox{.}(2019)]%
        {lee_understanding_2019}
\bibfield{author}{\bibinfo{person}{Yee~Mun Lee}, \bibinfo{person}{Ruth
  Madigan}, \bibinfo{person}{Jorge Garcia}, \bibinfo{person}{Andrew Tomlinson},
  \bibinfo{person}{Albert Solernou}, \bibinfo{person}{Richard Romano},
  \bibinfo{person}{Gustav Markkula}, \bibinfo{person}{Natasha Merat}, {and}
  \bibinfo{person}{Jim Uttley}.} \bibinfo{year}{2019}\natexlab{}.
\newblock \showarticletitle{Understanding the Messages Conveyed by Automated
  Vehicles}. In \bibinfo{booktitle}{\emph{Proceedings of the 11th International
  Conference on Automotive User Interfaces and Interactive Vehicular
  Applications}} (Utrecht, Netherlands) \emph{(\bibinfo{series}{AutomotiveUI
  '19})}. \bibinfo{publisher}{Association for Computing Machinery},
  \bibinfo{address}{New York, NY, USA}, \bibinfo{pages}{134–143}.
\newblock
\showISBNx{9781450368841}
\urldef\tempurl%
\url{https://doi.org/10.1145/3342197.3344546}
\showDOI{\tempurl}


\bibitem[Lilija et~al\mbox{.}(2020)]%
        {lilija_who_2020}
\bibfield{author}{\bibinfo{person}{Klemen Lilija}, \bibinfo{person}{Henning
  Pohl}, {and} \bibinfo{person}{Kasper Hornb\ae{}k}.}
  \bibinfo{year}{2020}\natexlab{}.
\newblock \showarticletitle{Who Put That There? Temporal Navigation of Spatial
  Recordings by Direct Manipulation}. In \bibinfo{booktitle}{\emph{Proceedings
  of the 2020 CHI Conference on Human Factors in Computing Systems}} (Honolulu,
  HI, USA) \emph{(\bibinfo{series}{CHI '20})}. \bibinfo{publisher}{Association
  for Computing Machinery}, \bibinfo{address}{New York, NY, USA},
  \bibinfo{pages}{1–11}.
\newblock
\showISBNx{9781450367080}
\urldef\tempurl%
\url{https://doi.org/10.1145/3313831.3376604}
\showDOI{\tempurl}


\bibitem[Lin et~al\mbox{.}(2018)]%
        {lin_adasa_2018}
\bibfield{author}{\bibinfo{person}{Shih-Chieh Lin}, \bibinfo{person}{Chang-Hong
  Hsu}, \bibinfo{person}{Walter Talamonti}, \bibinfo{person}{Yunqi Zhang},
  \bibinfo{person}{Steve Oney}, \bibinfo{person}{Jason Mars}, {and}
  \bibinfo{person}{Lingjia Tang}.} \bibinfo{year}{2018}\natexlab{}.
\newblock \showarticletitle{Adasa: {A} {Conversational} {In}-{Vehicle}
  {Digital} {Assistant} for {Advanced} {Driver} {Assistance} {Features}}. In
  \bibinfo{booktitle}{\emph{Proceedings of the 31st {Annual} {ACM} {Symposium}
  on {User} {Interface} {Software} and {Technology}}}
  \emph{(\bibinfo{series}{{UIST} '18})}. \bibinfo{publisher}{Association for
  Computing Machinery}, \bibinfo{address}{New York, NY, USA},
  \bibinfo{pages}{531--542}.
\newblock
\showISBNx{978-1-4503-5948-1}
\urldef\tempurl%
\url{https://doi.org/10.1145/3242587.3242593}
\showDOI{\tempurl}


\bibitem[Liu et~al\mbox{.}(2021)]%
        {liu_empathic_2021}
\bibfield{author}{\bibinfo{person}{Shu Liu}, \bibinfo{person}{Kevin Koch},
  \bibinfo{person}{Zimu Zhou}, \bibinfo{person}{Simon F\"{o}ll},
  \bibinfo{person}{Xiaoxi He}, \bibinfo{person}{Tina Menke},
  \bibinfo{person}{Elgar Fleisch}, {and} \bibinfo{person}{Felix Wortmann}.}
  \bibinfo{year}{2021}\natexlab{}.
\newblock \showarticletitle{The Empathetic Car: Exploring Emotion Inference via
  Driver Behaviour and Traffic Context}.
\newblock \bibinfo{journal}{\emph{Proc. ACM Interact. Mob. Wearable Ubiquitous
  Technol.}} \bibinfo{volume}{5}, \bibinfo{number}{3}, Article
  \bibinfo{articleno}{117} (\bibinfo{date}{sep} \bibinfo{year}{2021}),
  \bibinfo{numpages}{34}~pages.
\newblock
\urldef\tempurl%
\url{https://doi.org/10.1145/3478078}
\showDOI{\tempurl}


\bibitem[Lohani et~al\mbox{.}(2019)]%
        {lohani2019review}
\bibfield{author}{\bibinfo{person}{Monika Lohani}, \bibinfo{person}{Brennan~R
  Payne}, {and} \bibinfo{person}{David~L Strayer}.}
  \bibinfo{year}{2019}\natexlab{}.
\newblock \showarticletitle{A review of psychophysiological measures to assess
  cognitive states in real-world driving}.
\newblock \bibinfo{journal}{\emph{Frontiers in human neuroscience}}
  \bibinfo{volume}{13} (\bibinfo{year}{2019}), \bibinfo{pages}{57}.
\newblock


\bibitem[Lv et~al\mbox{.}(2018)]%
        {chen_takeover_2018}
\bibfield{author}{\bibinfo{person}{Chen Lv}, \bibinfo{person}{Huaji Wang},
  \bibinfo{person}{Dongpu Cao}, \bibinfo{person}{Yifan Zhao},
  \bibinfo{person}{Mark Sullman}, \bibinfo{person}{Daniel~J. Auger},
  \bibinfo{person}{James Brighton}, \bibinfo{person}{Rebecca Matthias},
  \bibinfo{person}{Lee Skrypchuk}, {and} \bibinfo{person}{Alexandros
  Mouzakitis}.} \bibinfo{year}{2018}\natexlab{}.
\newblock \showarticletitle{A Novel Control Framework of Haptic Take-Over
  System for Automated Vehicles}. In \bibinfo{booktitle}{\emph{2018 IEEE
  Intelligent Vehicles Symposium (IV)}}. \bibinfo{publisher}{IEEE},
  \bibinfo{address}{New York, NY, USA}, \bibinfo{pages}{1596--1601}.
\newblock
\urldef\tempurl%
\url{https://doi.org/10.1109/IVS.2018.8500480}
\showDOI{\tempurl}


\bibitem[Madigan et~al\mbox{.}(2019)]%
        {madigan_understanding_2019}
\bibfield{author}{\bibinfo{person}{Ruth Madigan}, \bibinfo{person}{Sina
  Nordhoff}, \bibinfo{person}{Charles Fox}, \bibinfo{person}{Roja {Ezzati
  Amini}}, \bibinfo{person}{Tyron Louw}, \bibinfo{person}{Marc Wilbrink},
  \bibinfo{person}{Anna Schieben}, {and} \bibinfo{person}{Natasha Merat}.}
  \bibinfo{year}{2019}\natexlab{}.
\newblock \showarticletitle{Understanding interactions between Automated Road
  Transport Systems and other road users: A video analysis}.
\newblock \bibinfo{journal}{\emph{Transportation Research Part F: Traffic
  Psychology and Behaviour}}  \bibinfo{volume}{66} (\bibinfo{year}{2019}),
  \bibinfo{pages}{196--213}.
\newblock
\showISSN{1369-8478}
\urldef\tempurl%
\url{https://doi.org/10.1016/j.trf.2019.09.006}
\showDOI{\tempurl}


\bibitem[Marquardt et~al\mbox{.}(2015)]%
        {marquardt_excite_2015}
\bibfield{author}{\bibinfo{person}{Nicolai Marquardt},
  \bibinfo{person}{Frederico Schardong}, {and} \bibinfo{person}{Anthony Tang}.}
  \bibinfo{year}{2015}\natexlab{}.
\newblock \showarticletitle{EXCITE: Exploring collaborative interaction in
  tracked environments}. In \bibinfo{booktitle}{\emph{IFIP Conference on
  Human-Computer Interaction}}. Springer, \bibinfo{publisher}{Springer},
  \bibinfo{address}{Cham}, \bibinfo{pages}{89--97}.
\newblock


\bibitem[Martin et~al\mbox{.}(2019)]%
        {driveact}
\bibfield{author}{\bibinfo{person}{Manuel Martin}, \bibinfo{person}{Alina
  Roitberg}, \bibinfo{person}{Monica Haurilet}, \bibinfo{person}{Matthias
  Horne}, \bibinfo{person}{Simon Reiß}, \bibinfo{person}{Michael Voit}, {and}
  \bibinfo{person}{Rainer Stiefelhagen}.} \bibinfo{year}{2019}\natexlab{}.
\newblock \showarticletitle{Drive\&Act: A Multi-Modal Dataset for Fine-Grained
  Driver Behavior Recognition in Autonomous Vehicles}. In
  \bibinfo{booktitle}{\emph{2019 IEEE/CVF International Conference on Computer
  Vision (ICCV)}}. \bibinfo{publisher}{IEEE}, \bibinfo{address}{New York, NY,
  USA}, \bibinfo{pages}{2801--2810}.
\newblock
\urldef\tempurl%
\url{https://doi.org/10.1109/ICCV.2019.00289}
\showDOI{\tempurl}


\bibitem[Meskens et~al\mbox{.}(2008)]%
        {meskens_gummy_2008}
\bibfield{author}{\bibinfo{person}{Jan Meskens}, \bibinfo{person}{Jo
  Vermeulen}, \bibinfo{person}{Kris Luyten}, {and} \bibinfo{person}{Karin
  Coninx}.} \bibinfo{year}{2008}\natexlab{}.
\newblock \showarticletitle{Gummy for Multi-Platform User Interface Designs:
  Shape Me, Multiply Me, Fix Me, Use Me}. In
  \bibinfo{booktitle}{\emph{Proceedings of the Working Conference on Advanced
  Visual Interfaces}} (Napoli, Italy) \emph{(\bibinfo{series}{AVI '08})}.
  \bibinfo{publisher}{Association for Computing Machinery},
  \bibinfo{address}{New York, NY, USA}, \bibinfo{pages}{233–240}.
\newblock
\showISBNx{9781605581415}
\urldef\tempurl%
\url{https://doi.org/10.1145/1385569.1385607}
\showDOI{\tempurl}


\bibitem[Morales-Alvarez et~al\mbox{.}(2020)]%
        {morales_takeover_2020}
\bibfield{author}{\bibinfo{person}{Walter Morales-Alvarez},
  \bibinfo{person}{Oscar Sipele}, \bibinfo{person}{Régis Léberon},
  \bibinfo{person}{Hadj~Hamma Tadjine}, {and} \bibinfo{person}{Cristina
  Olaverri-Monreal}.} \bibinfo{year}{2020}\natexlab{}.
\newblock \showarticletitle{Automated Driving: A Literature Review of the Take
  over Request in Conditional Automation}.
\newblock \bibinfo{journal}{\emph{Electronics}} \bibinfo{volume}{9},
  \bibinfo{number}{12} (\bibinfo{year}{2020}), \bibinfo{pages}{34}.
\newblock
\showISSN{2079-9292}
\urldef\tempurl%
\url{https://doi.org/10.3390/electronics9122087}
\showDOI{\tempurl}


\bibitem[Nebeling et~al\mbox{.}(2020)]%
        {nebeling_mrat_2020}
\bibfield{author}{\bibinfo{person}{Michael Nebeling},
  \bibinfo{person}{Maximilian Speicher}, \bibinfo{person}{Xizi Wang},
  \bibinfo{person}{Shwetha Rajaram}, \bibinfo{person}{Brian~D. Hall},
  \bibinfo{person}{Zijian Xie}, \bibinfo{person}{Alexander R.~E. Raistrick},
  \bibinfo{person}{Michelle Aebersold}, \bibinfo{person}{Edward~G. Happ},
  \bibinfo{person}{Jiayin Wang}, \bibinfo{person}{Yanan Sun},
  \bibinfo{person}{Lotus Zhang}, \bibinfo{person}{Leah~E. Ramsier}, {and}
  \bibinfo{person}{Rhea Kulkarni}.} \bibinfo{year}{2020}\natexlab{}.
\newblock \showarticletitle{MRAT: The Mixed Reality Analytics Toolkit}. In
  \bibinfo{booktitle}{\emph{Proceedings of the 2020 CHI Conference on Human
  Factors in Computing Systems}} (Honolulu, HI, USA)
  \emph{(\bibinfo{series}{CHI '20})}. \bibinfo{publisher}{Association for
  Computing Machinery}, \bibinfo{address}{New York, NY, USA},
  \bibinfo{pages}{1–12}.
\newblock
\showISBNx{9781450367080}
\urldef\tempurl%
\url{https://doi.org/10.1145/3313831.3376330}
\showDOI{\tempurl}


\bibitem[Normark(2015)]%
        {normark_design_2015}
\bibfield{author}{\bibinfo{person}{Carl~Jörgen Normark}.}
  \bibinfo{year}{2015}\natexlab{}.
\newblock \showarticletitle{Design and Evaluation of a Touch-Based
  Personalizable In-Vehicle User Interface}.
\newblock \bibinfo{journal}{\emph{International Journal of Human–Computer
  Interaction}} \bibinfo{volume}{31}, \bibinfo{number}{11}
  (\bibinfo{year}{2015}), \bibinfo{pages}{731--745}.
\newblock
\urldef\tempurl%
\url{https://doi.org/10.1080/10447318.2015.1045240}
\showDOI{\tempurl}
\showeprint{https://doi.org/10.1080/10447318.2015.1045240}


\bibitem[Olsen(2007)]%
        {olsen_evaluating_2007}
\bibfield{author}{\bibinfo{person}{Dan~R. Olsen}.}
  \bibinfo{year}{2007}\natexlab{}.
\newblock \showarticletitle{Evaluating User Interface Systems Research}. In
  \bibinfo{booktitle}{\emph{Proceedings of the 20th Annual ACM Symposium on
  User Interface Software and Technology}} (Newport, Rhode Island, USA)
  \emph{(\bibinfo{series}{UIST '07})}. \bibinfo{publisher}{Association for
  Computing Machinery}, \bibinfo{address}{New York, NY, USA},
  \bibinfo{pages}{251–258}.
\newblock
\showISBNx{9781595936790}
\urldef\tempurl%
\url{https://doi.org/10.1145/1294211.1294256}
\showDOI{\tempurl}


\bibitem[Osswald et~al\mbox{.}(2012)]%
        {osswald_predicting_2012}
\bibfield{author}{\bibinfo{person}{Sebastian Osswald}, \bibinfo{person}{Daniela
  Wurhofer}, \bibinfo{person}{Sandra Tr\"{o}sterer}, \bibinfo{person}{Elke
  Beck}, {and} \bibinfo{person}{Manfred Tscheligi}.}
  \bibinfo{year}{2012}\natexlab{}.
\newblock \showarticletitle{Predicting Information Technology Usage in the Car:
  Towards a Car Technology Acceptance Model}. In
  \bibinfo{booktitle}{\emph{Proceedings of the 4th International Conference on
  Automotive User Interfaces and Interactive Vehicular Applications}}
  (Portsmouth, New Hampshire) \emph{(\bibinfo{series}{AutomotiveUI '12})}.
  \bibinfo{publisher}{Association for Computing Machinery},
  \bibinfo{address}{New York, NY, USA}, \bibinfo{pages}{51–58}.
\newblock
\showISBNx{9781450317511}
\urldef\tempurl%
\url{https://doi.org/10.1145/2390256.2390264}
\showDOI{\tempurl}


\bibitem[Pakdamanian et~al\mbox{.}(2021)]%
        {pakdamanian_deeptake_2021}
\bibfield{author}{\bibinfo{person}{Erfan Pakdamanian}, \bibinfo{person}{Shili
  Sheng}, \bibinfo{person}{Sonia Baee}, \bibinfo{person}{Seongkook Heo},
  \bibinfo{person}{Sarit Kraus}, {and} \bibinfo{person}{Lu Feng}.}
  \bibinfo{year}{2021}\natexlab{}.
\newblock \showarticletitle{DeepTake: Prediction of Driver Takeover Behavior
  Using Multimodal Data}. In \bibinfo{booktitle}{\emph{Proceedings of the 2021
  CHI Conference on Human Factors in Computing Systems}} (Yokohama, Japan)
  \emph{(\bibinfo{series}{CHI '21})}. \bibinfo{publisher}{Association for
  Computing Machinery}, \bibinfo{address}{New York, NY, USA}, Article
  \bibinfo{articleno}{103}, \bibinfo{numpages}{14}~pages.
\newblock
\showISBNx{9781450380966}
\urldef\tempurl%
\url{https://doi.org/10.1145/3411764.3445563}
\showDOI{\tempurl}


\bibitem[Politis et~al\mbox{.}(2015)]%
        {politis_language_2015}
\bibfield{author}{\bibinfo{person}{Ioannis Politis}, \bibinfo{person}{Stephen
  Brewster}, {and} \bibinfo{person}{Frank Pollick}.}
  \bibinfo{year}{2015}\natexlab{}.
\newblock \showarticletitle{Language-Based Multimodal Displays for the Handover
  of Control in Autonomous Cars}. In \bibinfo{booktitle}{\emph{Proceedings of
  the 7th International Conference on Automotive User Interfaces and
  Interactive Vehicular Applications}} (Nottingham, United Kingdom)
  \emph{(\bibinfo{series}{AutomotiveUI '15})}. \bibinfo{publisher}{Association
  for Computing Machinery}, \bibinfo{address}{New York, NY, USA},
  \bibinfo{pages}{3–10}.
\newblock
\showISBNx{9781450337366}
\urldef\tempurl%
\url{https://doi.org/10.1145/2799250.2799262}
\showDOI{\tempurl}


\bibitem[Prilla and R{\"u}hmann(2018)]%
        {prilla_analysis_2018}
\bibfield{author}{\bibinfo{person}{Michael Prilla} {and}
  \bibinfo{person}{Lisa~M R{\"u}hmann}.} \bibinfo{year}{2018}\natexlab{}.
\newblock \showarticletitle{An Analysis Tool for Cooperative Mixed Reality
  Scenarios}. In \bibinfo{booktitle}{\emph{2018 IEEE International Symposium on
  Mixed and Augmented Reality Adjunct (ISMAR-Adjunct)}}. IEEE,
  \bibinfo{publisher}{IEEE}, \bibinfo{address}{New YOrk, NY, USA},
  \bibinfo{pages}{31--35}.
\newblock


\bibitem[Radlmayr et~al\mbox{.}(2014)]%
        {radlmayr_how_2014}
\bibfield{author}{\bibinfo{person}{Jonas Radlmayr}, \bibinfo{person}{Christian
  Gold}, \bibinfo{person}{Lutz Lorenz}, \bibinfo{person}{Mehdi Farid}, {and}
  \bibinfo{person}{Klaus Bengler}.} \bibinfo{year}{2014}\natexlab{}.
\newblock \showarticletitle{How Traffic Situations and Non-Driving Related
  Tasks Affect the Take-Over Quality in Highly Automated Driving}.
\newblock \bibinfo{journal}{\emph{Proceedings of the Human Factors and
  Ergonomics Society Annual Meeting}} \bibinfo{volume}{58}, \bibinfo{number}{1}
  (\bibinfo{year}{2014}), \bibinfo{pages}{2063--2067}.
\newblock
\urldef\tempurl%
\url{https://doi.org/10.1177/1541931214581434}
\showDOI{\tempurl}
\showeprint{https://doi.org/10.1177/1541931214581434}


\bibitem[Reichherzer et~al\mbox{.}(2021)]%
        {reichherzer_bringing_2021}
\bibfield{author}{\bibinfo{person}{Carolin Reichherzer},
  \bibinfo{person}{Andrew Cunningham}, \bibinfo{person}{Tracey Coleman},
  \bibinfo{person}{Ruochen Cao}, \bibinfo{person}{Kurt McManus},
  \bibinfo{person}{Dion Sheppard}, \bibinfo{person}{Mark Kohler},
  \bibinfo{person}{Mark Billinghurst}, {and} \bibinfo{person}{Bruce~H Thomas}.}
  \bibinfo{year}{2021}\natexlab{}.
\newblock \showarticletitle{Bringing the Jury to the Scene of the Crime: Memory
  and Decision-Making in a Simulated Crime Scene}. In
  \bibinfo{booktitle}{\emph{Proceedings of the 2021 CHI Conference on Human
  Factors in Computing Systems}} (Yokohama, Japan) \emph{(\bibinfo{series}{CHI
  '21})}. \bibinfo{publisher}{Association for Computing Machinery},
  \bibinfo{address}{New York, NY, USA}, Article \bibinfo{articleno}{709},
  \bibinfo{numpages}{12}~pages.
\newblock
\showISBNx{9781450380966}
\urldef\tempurl%
\url{https://doi.org/10.1145/3411764.3445464}
\showDOI{\tempurl}


\bibitem[Reipschl\"{a}ger et~al\mbox{.}(2022)]%
        {reipschlager_avatar_2022}
\bibfield{author}{\bibinfo{person}{Patrick Reipschl\"{a}ger},
  \bibinfo{person}{Frederik Brudy}, \bibinfo{person}{Raimund Dachselt},
  \bibinfo{person}{Justin Matejka}, \bibinfo{person}{George Fitzmaurice}, {and}
  \bibinfo{person}{Fraser Anderson}.} \bibinfo{year}{2022}\natexlab{}.
\newblock \showarticletitle{AvatAR: An Immersive Analysis Environment for Human
  Motion Data Combining Interactive 3D Avatars and Trajectories}. In
  \bibinfo{booktitle}{\emph{Proceedings of the 2022 CHI Conference on Human
  Factors in Computing Systems}} (New Orleans, LA, USA)
  \emph{(\bibinfo{series}{CHI '22})}. \bibinfo{publisher}{Association for
  Computing Machinery}, \bibinfo{address}{New York, NY, USA}, Article
  \bibinfo{articleno}{23}, \bibinfo{numpages}{15}~pages.
\newblock
\showISBNx{9781450391573}
\urldef\tempurl%
\url{https://doi.org/10.1145/3491102.3517676}
\showDOI{\tempurl}


\bibitem[Reipschlager et~al\mbox{.}(2020)]%
        {reipschlager2020personal}
\bibfield{author}{\bibinfo{person}{Patrick Reipschlager},
  \bibinfo{person}{Tamara Flemisch}, {and} \bibinfo{person}{Raimund Dachselt}.}
  \bibinfo{year}{2020}\natexlab{}.
\newblock \showarticletitle{Personal augmented reality for information
  visualization on large interactive displays}.
\newblock \bibinfo{journal}{\emph{IEEE Transactions on Visualization and
  Computer Graphics}} \bibinfo{volume}{27}, \bibinfo{number}{2}
  (\bibinfo{year}{2020}), \bibinfo{pages}{1182--1192}.
\newblock


\bibitem[Rittger et~al\mbox{.}(2022)]%
        {rittger_adaptive_2022}
\bibfield{author}{\bibinfo{person}{Lena Rittger}, \bibinfo{person}{Doreen
  Engelhardt}, {and} \bibinfo{person}{Robert Schwartz}.}
  \bibinfo{year}{2022}\natexlab{}.
\newblock \showarticletitle{Adaptive User Experience in the Car—Levels of
  Adaptivity and Adaptive HMI Design}.
\newblock \bibinfo{journal}{\emph{IEEE Transactions on Intelligent
  Transportation Systems}} \bibinfo{volume}{23}, \bibinfo{number}{5}
  (\bibinfo{year}{2022}), \bibinfo{pages}{4866--4876}.
\newblock
\urldef\tempurl%
\url{https://doi.org/10.1109/TITS.2021.3124990}
\showDOI{\tempurl}


\bibitem[Roider et~al\mbox{.}(2018)]%
        {roider_just_2018}
\bibfield{author}{\bibinfo{person}{Florian Roider}, \bibinfo{person}{Lars
  Reisig}, {and} \bibinfo{person}{Tom Gross}.} \bibinfo{year}{2018}\natexlab{}.
\newblock \showarticletitle{Just {Look}: {The} {Benefits} of {Gaze}-{Activated}
  {Voice} {Input} in the {Car}}. In \bibinfo{booktitle}{\emph{Adjunct
  {Proceedings} of the 10th {International} {Conference} on {Automotive} {User}
  {Interfaces} and {Interactive} {Vehicular} {Applications}}}
  \emph{(\bibinfo{series}{{AutomotiveUI} '18})}.
  \bibinfo{publisher}{Association for Computing Machinery},
  \bibinfo{address}{New York, NY, USA}, \bibinfo{pages}{210--214}.
\newblock
\showISBNx{978-1-4503-5947-4}
\urldef\tempurl%
\url{https://doi.org/10.1145/3239092.3265968}
\showDOI{\tempurl}


\bibitem[Rosenbaum et~al\mbox{.}(2011)]%
        {rosenbaum2011involve}
\bibfield{author}{\bibinfo{person}{Ren{\'e} Rosenbaum}, \bibinfo{person}{Jeremy
  Bottleson}, \bibinfo{person}{Zhuiguang Liu}, {and} \bibinfo{person}{Bernd
  Hamann}.} \bibinfo{year}{2011}\natexlab{}.
\newblock \showarticletitle{Involve Me and I Will Understand!--Abstract Data
  Visualization in Immersive Environments}. In
  \bibinfo{booktitle}{\emph{International Symposium on Visual Computing}}.
  Springer, \bibinfo{publisher}{Springer}, \bibinfo{address}{Cham},
  \bibinfo{pages}{530--540}.
\newblock


\bibitem[SAE-International(2021)]%
        {sae_levels}
\bibfield{author}{\bibinfo{person}{SAE-International}.}
  \bibinfo{year}{2021}\natexlab{}.
\newblock \bibinfo{title}{Taxonomy and Definitions for Terms Related to Driving
  Automation Systems for On-Road Motor Vehicles}.
\newblock
  \bibinfo{howpublished}{\url{https://www.sae.org/standards/content/j3016_202104/}}.
\newblock
\newblock
\shownote{[Online; accessed: 24-August-2021]}.


\bibitem[Saenz et~al\mbox{.}(2017)]%
        {saenz2017reexamining}
\bibfield{author}{\bibinfo{person}{Michael Saenz}, \bibinfo{person}{Ali
  Baigelenov}, \bibinfo{person}{Ya-Hsin Hung}, {and} \bibinfo{person}{Paul
  Parsons}.} \bibinfo{year}{2017}\natexlab{}.
\newblock \showarticletitle{Reexamining the cognitive utility of 3D
  visualizations using augmented reality holograms}. In
  \bibinfo{booktitle}{\emph{IEEE VIS Workshop on Immersive Analytics: Exploring
  Future Interaction and Visualization Technologies for Data Analytics}}.
  \bibinfo{publisher}{ACM}, \bibinfo{address}{New York, NY, USA},
  \bibinfo{pages}{5}.
\newblock


\bibitem[{Salesforce}(2022)]%
        {tableau}
\bibfield{author}{\bibinfo{person}{{Salesforce}}.}
  \bibinfo{year}{2022}\natexlab{}.
\newblock \bibinfo{booktitle}{\emph{Tableau: {Business} {Intelligence} and
  {Analytics} {Software}}}.
\newblock Salesforce Inc.
\newblock
\urldef\tempurl%
\url{https://www.tableau.com/}
\showURL{%
\tempurl}
\newblock
\shownote{(Accessed on 08/23/2022)}.


\bibitem[Salpisti et~al\mbox{.}(2022)]%
        {salpisti_procedural_2022}
\bibfield{author}{\bibinfo{person}{Despoina Salpisti},
  \bibinfo{person}{Matthias de Clerk}, \bibinfo{person}{Sebastian Hinz},
  \bibinfo{person}{Frank Henkies}, {and} \bibinfo{person}{Gudrun Klinker}.}
  \bibinfo{year}{2022}\natexlab{}.
\newblock \showarticletitle{A Procedural Building Generator Based on Real-World
  Data Enabling Designers to Create Context for XR Automotive Design
  Experiences}. In \bibinfo{booktitle}{\emph{Virtual Reality and Mixed
  Reality}}, \bibfield{editor}{\bibinfo{person}{Gabriel Zachmann},
  \bibinfo{person}{Mariano Alca{\~{n}}iz~Raya}, \bibinfo{person}{Partrick
  Bourdot}, \bibinfo{person}{Maud Marchal}, \bibinfo{person}{Jeanine
  Stefanucci}, {and} \bibinfo{person}{Xubo Yang}} (Eds.).
  \bibinfo{publisher}{Springer International Publishing},
  \bibinfo{address}{Cham}, \bibinfo{pages}{149--170}.
\newblock
\showISBNx{978-3-031-16234-3}


\bibitem[Schmidt et~al\mbox{.}(2010)]%
        {schmidt_driving_2010}
\bibfield{author}{\bibinfo{person}{Albrecht Schmidt}, \bibinfo{person}{Wolfgang
  Spiessl}, {and} \bibinfo{person}{Dagmar Kern}.}
  \bibinfo{year}{2010}\natexlab{}.
\newblock \showarticletitle{Driving Automotive User Interface Research}.
\newblock \bibinfo{journal}{\emph{IEEE Pervasive Computing}}
  \bibinfo{volume}{9}, \bibinfo{number}{1} (\bibinfo{year}{2010}),
  \bibinfo{pages}{85--88}.
\newblock
\urldef\tempurl%
\url{https://doi.org/10.1109/MPRV.2010.3}
\showDOI{\tempurl}


\bibitem[Sch\"{o}lkopf et~al\mbox{.}(2021)]%
        {scholkopf_conception_2021}
\bibfield{author}{\bibinfo{person}{Lasse Sch\"{o}lkopf},
  \bibinfo{person}{Maria-Magdalena Wolf}, \bibinfo{person}{Veronika Hutmann},
  {and} \bibinfo{person}{Frank Diermeyer}.} \bibinfo{year}{2021}\natexlab{}.
\newblock \showarticletitle{Conception, Development and First Evaluation of a
  Context-Adaptive User Interface for Commercial Vehicles}. In
  \bibinfo{booktitle}{\emph{13th International Conference on Automotive User
  Interfaces and Interactive Vehicular Applications}} (Leeds, United Kingdom)
  \emph{(\bibinfo{series}{AutomotiveUI '21 Adjunct})}.
  \bibinfo{publisher}{Association for Computing Machinery},
  \bibinfo{address}{New York, NY, USA}, \bibinfo{pages}{21–25}.
\newblock
\showISBNx{9781450386418}
\urldef\tempurl%
\url{https://doi.org/10.1145/3473682.3480256}
\showDOI{\tempurl}


\bibitem[Sedlmair et~al\mbox{.}(2012)]%
        {sedlmair_design_2012}
\bibfield{author}{\bibinfo{person}{Michael Sedlmair}, \bibinfo{person}{Miriah
  Meyer}, {and} \bibinfo{person}{Tamara Munzner}.}
  \bibinfo{year}{2012}\natexlab{}.
\newblock \showarticletitle{Design Study Methodology: Reflections from the
  Trenches and the Stacks}.
\newblock \bibinfo{journal}{\emph{IEEE Transactions on Visualization and
  Computer Graphics}} \bibinfo{volume}{18}, \bibinfo{number}{12}
  (\bibinfo{year}{2012}), \bibinfo{pages}{2431--2440}.
\newblock
\urldef\tempurl%
\url{https://doi.org/10.1109/TVCG.2012.213}
\showDOI{\tempurl}


\bibitem[{Sefik Ilkin Serengil}(2022)]%
        {deepface}
\bibfield{author}{\bibinfo{person}{{Sefik Ilkin Serengil}}.}
  \bibinfo{year}{2022}\natexlab{}.
\newblock \bibinfo{booktitle}{\emph{DeepFace: A Facial Recognition Library for
  Python}}.
\newblock DeepFace.
\newblock
\urldef\tempurl%
\url{https://github.com/serengil/deepface}
\showURL{%
\tempurl}
\newblock
\shownote{(Accessed on 08/23/2022)}.


\bibitem[Sereno et~al\mbox{.}(2022)]%
        {sereno2022hybrid}
\bibfield{author}{\bibinfo{person}{Mickael Sereno},
  \bibinfo{person}{St{\'e}phane Gosset}, \bibinfo{person}{Lonni
  Besan{\c{c}}on}, {and} \bibinfo{person}{Tobias Isenberg}.}
  \bibinfo{year}{2022}\natexlab{}.
\newblock \showarticletitle{Hybrid touch/tangible spatial selection in
  augmented reality}. In \bibinfo{booktitle}{\emph{Computer Graphics Forum}}.
  Wiley Online Library, \bibinfo{publisher}{Wiley}, \bibinfo{address}{Hoboken,
  New Jersey, USA}, \bibinfo{pages}{403--415}.
\newblock


\bibitem[Shen et~al\mbox{.}(2022)]%
        {shen_gesture_2022}
\bibfield{author}{\bibinfo{person}{Junxiao Shen}, \bibinfo{person}{John
  Dudley}, \bibinfo{person}{George Mo}, {and} \bibinfo{person}{Per~Ola
  Kristensson}.} \bibinfo{year}{2022}\natexlab{}.
\newblock \showarticletitle{Gesture Spotter: A Rapid Prototyping Tool for Key
  Gesture Spotting in Virtual and Augmented Reality Applications}.
\newblock \bibinfo{journal}{\emph{IEEE Transactions on Visualization and
  Computer Graphics}} \bibinfo{volume}{28}, \bibinfo{number}{11}
  (\bibinfo{year}{2022}), \bibinfo{pages}{1--11}.
\newblock
\urldef\tempurl%
\url{https://doi.org/10.1109/TVCG.2022.3203004}
\showDOI{\tempurl}


\bibitem[Tang et~al\mbox{.}(2010)]%
        {tang_vistaco_2010}
\bibfield{author}{\bibinfo{person}{Anthony Tang}, \bibinfo{person}{Michel
  Pahud}, \bibinfo{person}{Sheelagh Carpendale}, {and} \bibinfo{person}{Bill
  Buxton}.} \bibinfo{year}{2010}\natexlab{}.
\newblock \showarticletitle{VisTACO: Visualizing Tabletop Collaboration}. In
  \bibinfo{booktitle}{\emph{ACM International Conference on Interactive
  Tabletops and Surfaces}} (Saarbr\"{u}cken, Germany)
  \emph{(\bibinfo{series}{ITS '10})}. \bibinfo{publisher}{Association for
  Computing Machinery}, \bibinfo{address}{New York, NY, USA},
  \bibinfo{pages}{29–38}.
\newblock
\showISBNx{9781450303996}
\urldef\tempurl%
\url{https://doi.org/10.1145/1936652.1936659}
\showDOI{\tempurl}


\bibitem[Tavakoli et~al\mbox{.}(2021)]%
        {tavakoli_harmony_2021}
\bibfield{author}{\bibinfo{person}{Arash Tavakoli}, \bibinfo{person}{Shashwat
  Kumar}, \bibinfo{person}{Xiang Guo}, \bibinfo{person}{Vahid Balali},
  \bibinfo{person}{Mehdi Boukhechba}, {and} \bibinfo{person}{Arsalan
  Heydarian}.} \bibinfo{year}{2021}\natexlab{}.
\newblock \showarticletitle{HARMONY: A Human-Centered Multimodal Driving Study
  in the Wild}.
\newblock \bibinfo{journal}{\emph{IEEE Access}}  \bibinfo{volume}{9}
  (\bibinfo{year}{2021}), \bibinfo{pages}{23956--23978}.
\newblock
\urldef\tempurl%
\url{https://doi.org/10.1109/ACCESS.2021.3056007}
\showDOI{\tempurl}


\bibitem[{Tianxiaomo}(2022)]%
        {yolo4}
\bibfield{author}{\bibinfo{person}{{Tianxiaomo}}.}
  \bibinfo{year}{2022}\natexlab{}.
\newblock \bibinfo{booktitle}{\emph{Pytorch-YOLOv4}}.
\newblock YOLOv4.
\newblock
\urldef\tempurl%
\url{https://github.com/Tianxiaomo/pytorch-YOLOv4}
\showURL{%
\tempurl}
\newblock
\shownote{(Accessed on 08/23/2022)}.


\bibitem[{TIBCO}(2022)]%
        {spotfire}
\bibfield{author}{\bibinfo{person}{{TIBCO}}.} \bibinfo{year}{2022}\natexlab{}.
\newblock \bibinfo{booktitle}{\emph{{TIBCO} {Spotfire}®}}.
\newblock TIBCO Software Inc.
\newblock
\urldef\tempurl%
\url{https://www.tibco.com/products/tibco-spotfire}
\showURL{%
\tempurl}
\newblock
\shownote{(Accessed on 08/23/2022)}.


\bibitem[Trenta et~al\mbox{.}(2019)]%
        {trenta_advanced_2019}
\bibfield{author}{\bibinfo{person}{Francesca Trenta}, \bibinfo{person}{Sabrina
  Conoci}, \bibinfo{person}{Francesco Rundo}, {and} \bibinfo{person}{Sebastiano
  Battiato}.} \bibinfo{year}{2019}\natexlab{}.
\newblock \showarticletitle{Advanced Motion-Tracking System with Multi-Layers
  Deep Learning Framework for Innovative Car-Driver Drowsiness Monitoring}. In
  \bibinfo{booktitle}{\emph{2019 14th IEEE International Conference on
  Automatic Face \& Gesture Recognition (FG 2019)}}. \bibinfo{publisher}{IEEE},
  \bibinfo{address}{Cham}, \bibinfo{pages}{1--5}.
\newblock
\urldef\tempurl%
\url{https://doi.org/10.1109/FG.2019.8756566}
\showDOI{\tempurl}


\bibitem[Triess et~al\mbox{.}(2021)]%
        {lidar}
\bibfield{author}{\bibinfo{person}{Larissa~T. Triess},
  \bibinfo{person}{Mariella Dreissig}, \bibinfo{person}{Christoph~B. Rist},
  {and} \bibinfo{person}{J. Marius~Zöllner}.} \bibinfo{year}{2021}\natexlab{}.
\newblock \showarticletitle{A Survey on Deep Domain Adaptation for LiDAR
  Perception}. In \bibinfo{booktitle}{\emph{2021 IEEE Intelligent Vehicles
  Symposium Workshops (IV Workshops)}}. \bibinfo{publisher}{IEEE},
  \bibinfo{address}{New York, NY, USA}, \bibinfo{pages}{350--357}.
\newblock
\urldef\tempurl%
\url{https://doi.org/10.1109/IVWorkshops54471.2021.9669228}
\showDOI{\tempurl}


\bibitem[Upton and Cook(1996)]%
        {upton1996understanding}
\bibfield{author}{\bibinfo{person}{Graham Upton} {and} \bibinfo{person}{Ian
  Cook}.} \bibinfo{year}{1996}\natexlab{}.
\newblock \bibinfo{booktitle}{\emph{Understanding statistics}}.
\newblock \bibinfo{publisher}{Oxford University Press},
  \bibinfo{address}{Oxford, United Kingdom}.
\newblock


\bibitem[{vis2k}(2022)]%
        {mirror}
\bibfield{author}{\bibinfo{person}{{vis2k}}.} \bibinfo{year}{2022}\natexlab{}.
\newblock \bibinfo{booktitle}{\emph{Mirror}}.
\newblock vis2k.
\newblock
\urldef\tempurl%
\url{{https://assetstore.unity.com/packages/tools/network/mirror-129321}}
\showURL{%
\tempurl}
\newblock
\shownote{(Accessed on 08/23/2022)}.


\bibitem[von Zadow and Dachselt(2017)]%
        {zadow_giant_2017}
\bibfield{author}{\bibinfo{person}{Ulrich von Zadow} {and}
  \bibinfo{person}{Raimund Dachselt}.} \bibinfo{year}{2017}\natexlab{}.
\newblock \showarticletitle{GIAnT: Visualizing Group Interaction at Large Wall
  Displays}. In \bibinfo{booktitle}{\emph{Proceedings of the 2017 CHI
  Conference on Human Factors in Computing Systems}} (Denver, Colorado, USA)
  \emph{(\bibinfo{series}{CHI '17})}. \bibinfo{publisher}{Association for
  Computing Machinery}, \bibinfo{address}{New York, NY, USA},
  \bibinfo{pages}{2639–2647}.
\newblock
\showISBNx{9781450346559}
\urldef\tempurl%
\url{https://doi.org/10.1145/3025453.3026006}
\showDOI{\tempurl}


\bibitem[{Vuplex}(2022)]%
        {3d-web-view}
\bibfield{author}{\bibinfo{person}{{Vuplex}}.} \bibinfo{year}{2022}\natexlab{}.
\newblock \bibinfo{booktitle}{\emph{3D WebView for Windows and macOS (Web
  Browser)}}.
\newblock Vuplex.
\newblock
\urldef\tempurl%
\url{{https://assetstore.unity.com/packages/tools/gui/3d-webview-for-windows-and-macos-web-browser-154144}}
\showURL{%
\tempurl}
\newblock
\shownote{(Accessed on 08/23/2022)}.


\bibitem[Wagner~Filho et~al\mbox{.}(2018)]%
        {wagner2018immersive}
\bibfield{author}{\bibinfo{person}{Jorge~A Wagner~Filho},
  \bibinfo{person}{Marina~F Rey}, \bibinfo{person}{Carla~MDS Freitas}, {and}
  \bibinfo{person}{Luciana Nedel}.} \bibinfo{year}{2018}\natexlab{}.
\newblock \showarticletitle{Immersive visualization of abstract information: An
  evaluation on dimensionally-reduced data scatterplots}. In
  \bibinfo{booktitle}{\emph{2018 IEEE Conference on Virtual Reality and 3D User
  Interfaces (VR)}}. IEEE, \bibinfo{publisher}{IEEE}, \bibinfo{address}{New
  YOrk, NY, USA}, \bibinfo{pages}{483--490}.
\newblock


\bibitem[Wang et~al\mbox{.}(2022)]%
        {wang_conversational_2022}
\bibfield{author}{\bibinfo{person}{Manhua Wang}, \bibinfo{person}{Seul~Chan
  Lee}, \bibinfo{person}{Genevieve Montavon}, \bibinfo{person}{Jiakang Qin},
  {and} \bibinfo{person}{Myounghoon Jeon}.} \bibinfo{year}{2022}\natexlab{}.
\newblock \showarticletitle{Conversational Voice Agents Are Preferred and Lead
  to Better Driving Performance in Conditionally Automated Vehicles}. In
  \bibinfo{booktitle}{\emph{Proceedings of the 14th International Conference on
  Automotive User Interfaces and Interactive Vehicular Applications}} (Seoul,
  Republic of Korea) \emph{(\bibinfo{series}{AutomotiveUI '22})}.
  \bibinfo{publisher}{Association for Computing Machinery},
  \bibinfo{address}{New York, NY, USA}, \bibinfo{pages}{86–95}.
\newblock
\showISBNx{9781450394154}
\urldef\tempurl%
\url{https://doi.org/10.1145/3543174.3546830}
\showDOI{\tempurl}


\bibitem[Wang et~al\mbox{.}(2020)]%
        {wang2020towards}
\bibfield{author}{\bibinfo{person}{Xiyao Wang}, \bibinfo{person}{Lonni
  Besan{\c{c}}on}, \bibinfo{person}{David Rousseau}, \bibinfo{person}{Mickael
  Sereno}, \bibinfo{person}{Mehdi Ammi}, {and} \bibinfo{person}{Tobias
  Isenberg}.} \bibinfo{year}{2020}\natexlab{}.
\newblock \showarticletitle{Towards an understanding of augmented reality
  extensions for existing 3D data analysis tools}. In
  \bibinfo{booktitle}{\emph{Proceedings of the 2020 CHI Conference on Human
  Factors in Computing Systems}}. \bibinfo{publisher}{ACM},
  \bibinfo{address}{New York, NY, USA}, \bibinfo{pages}{1--13}.
\newblock


\bibitem[Weng et~al\mbox{.}(2016)]%
        {weng_conversational_2016}
\bibfield{author}{\bibinfo{person}{Fuliang Weng}, \bibinfo{person}{Pongtep
  Angkititrakul}, \bibinfo{person}{Elizabeth~E. Shriberg},
  \bibinfo{person}{Larry Heck}, \bibinfo{person}{Stanley Peters}, {and}
  \bibinfo{person}{John~H.L. Hansen}.} \bibinfo{year}{2016}\natexlab{}.
\newblock \showarticletitle{Conversational In-Vehicle Dialog Systems: The past,
  present, and future}.
\newblock \bibinfo{journal}{\emph{IEEE Signal Processing Magazine}}
  \bibinfo{volume}{33}, \bibinfo{number}{6} (\bibinfo{year}{2016}),
  \bibinfo{pages}{49--60}.
\newblock
\urldef\tempurl%
\url{https://doi.org/10.1109/MSP.2016.2599201}
\showDOI{\tempurl}


\bibitem[Wintersberger et~al\mbox{.}(2018a)]%
        {wintersberger_man_2018}
\bibfield{author}{\bibinfo{person}{Philipp Wintersberger},
  \bibinfo{person}{Anna-Katharina Frison}, {and} \bibinfo{person}{Andreas
  Riener}.} \bibinfo{year}{2018}\natexlab{a}.
\newblock \showarticletitle{Man vs. Machine: Comparing a Fully Automated Bus
  Shuttle with a Manually Driven Group Taxi in a Field Study}. In
  \bibinfo{booktitle}{\emph{Adjunct Proceedings of the 10th International
  Conference on Automotive User Interfaces and Interactive Vehicular
  Applications}} (Toronto, ON, Canada) \emph{(\bibinfo{series}{AutomotiveUI
  '18})}. \bibinfo{publisher}{Association for Computing Machinery},
  \bibinfo{address}{New York, NY, USA}, \bibinfo{pages}{215–220}.
\newblock
\showISBNx{9781450359474}
\urldef\tempurl%
\url{https://doi.org/10.1145/3239092.3265969}
\showDOI{\tempurl}


\bibitem[Wintersberger et~al\mbox{.}(2018b)]%
        {wintersberger_fostering_2018}
\bibfield{author}{\bibinfo{person}{Philipp Wintersberger},
  \bibinfo{person}{Anna-Katharina Frison}, \bibinfo{person}{Andreas Riener},
  {and} \bibinfo{person}{Tamara~von Sawitzky}.}
  \bibinfo{year}{2018}\natexlab{b}.
\newblock \showarticletitle{Fostering {User} {Acceptance} and {Trust} in
  {Fully} {Automated} {Vehicles}: {Evaluating} the {Potential} of {Augmented}
  {Reality}}.
\newblock \bibinfo{journal}{\emph{Presence: Teleoperators and Virtual
  Environments}} \bibinfo{volume}{27}, \bibinfo{number}{1}
  (\bibinfo{date}{Feb.} \bibinfo{year}{2018}), \bibinfo{pages}{46--62}.
\newblock
\urldef\tempurl%
\url{https://doi.org/10.1162/pres_a_00320}
\showDOI{\tempurl}
\newblock
\shownote{\_eprint:
  https://direct.mit.edu/pvar/article-pdf/27/1/46/2003540/pres\_a\_00320.pdf}.


\bibitem[Winzer et~al\mbox{.}(2018)]%
        {winzer_intersection_2018}
\bibfield{author}{\bibinfo{person}{Oliver~M. Winzer},
  \bibinfo{person}{Antonia~S. Conti-Kufner}, {and} \bibinfo{person}{Klaus
  Bengler}.} \bibinfo{year}{2018}\natexlab{}.
\newblock \showarticletitle{Intersection Traffic Light Assistant – An
  Evaluation of the Suitability of Two Human Machine Interfaces}. In
  \bibinfo{booktitle}{\emph{2018 21st International Conference on Intelligent
  Transportation Systems (ITSC)}} (Maui, HI, USA). \bibinfo{publisher}{IEEE
  Press}, \bibinfo{address}{New York, NY, USA}, \bibinfo{pages}{261–265}.
\newblock
\showISBNx{978-1-7281-0321-1}
\urldef\tempurl%
\url{https://doi.org/10.1109/ITSC.2018.8569708}
\showDOI{\tempurl}


\bibitem[Xu et~al\mbox{.}(2021)]%
        {xu_when_2021}
\bibfield{author}{\bibinfo{person}{Zhigang Xu}, \bibinfo{person}{Zijun Jiang},
  \bibinfo{person}{Guanqun Wang}, \bibinfo{person}{Runmin Wang},
  \bibinfo{person}{Tingting Li}, \bibinfo{person}{Jinting Liu},
  \bibinfo{person}{Yijing Zhang}, {and} \bibinfo{person}{Peng Liu}.}
  \bibinfo{year}{2021}\natexlab{}.
\newblock \showarticletitle{When the automated driving system fails: Dynamics
  of public responses to automated vehicles}.
\newblock \bibinfo{journal}{\emph{Transportation Research Part C: Emerging
  Technologies}}  \bibinfo{volume}{129} (\bibinfo{year}{2021}),
  \bibinfo{pages}{103271}.
\newblock
\showISSN{0968-090X}
\urldef\tempurl%
\url{https://doi.org/10.1016/j.trc.2021.103271}
\showDOI{\tempurl}


\bibitem[Yan et~al\mbox{.}(2019)]%
        {yan_effects_2019}
\bibfield{author}{\bibinfo{person}{Yan Yan}, \bibinfo{person}{Ke Chen},
  \bibinfo{person}{Yu Xie}, \bibinfo{person}{Yiming Song}, {and}
  \bibinfo{person}{Yonghong Liu}.} \bibinfo{year}{2019}\natexlab{}.
\newblock \showarticletitle{The Effects of Weight on Comfort of Virtual Reality
  Devices}. In \bibinfo{booktitle}{\emph{Advances in Ergonomics in Design}},
  \bibfield{editor}{\bibinfo{person}{Francisco Rebelo} {and}
  \bibinfo{person}{Marcelo~M. Soares}} (Eds.). \bibinfo{publisher}{Springer
  International Publishing}, \bibinfo{address}{Cham},
  \bibinfo{pages}{239--248}.
\newblock
\showISBNx{978-3-319-94706-8}


\bibitem[Yi et~al\mbox{.}(2007)]%
        {yi_towards_2007}
\bibfield{author}{\bibinfo{person}{Ji~Soo Yi}, \bibinfo{person}{Youn~ah Kang},
  \bibinfo{person}{John Stasko}, {and} \bibinfo{person}{J.A. Jacko}.}
  \bibinfo{year}{2007}\natexlab{}.
\newblock \showarticletitle{Toward a Deeper Understanding of the Role of
  Interaction in Information Visualization}.
\newblock \bibinfo{journal}{\emph{IEEE Transactions on Visualization and
  Computer Graphics}} \bibinfo{volume}{13}, \bibinfo{number}{6}
  (\bibinfo{year}{2007}), \bibinfo{pages}{1224--1231}.
\newblock
\urldef\tempurl%
\url{https://doi.org/10.1109/TVCG.2007.70515}
\showDOI{\tempurl}


\bibitem[Zhang et~al\mbox{.}(2015)]%
        {zhang_using_2015}
\bibfield{author}{\bibinfo{person}{Meng-Jia Zhang}, \bibinfo{person}{Jie Li},
  {and} \bibinfo{person}{Kang Zhang}.} \bibinfo{year}{2015}\natexlab{}.
\newblock \showarticletitle{Using Virtual Reality Technique to Enhance
  Experience of Exploring 3D Trajectory Visualizations}. In
  \bibinfo{booktitle}{\emph{Proceedings of the 8th International Symposium on
  Visual Information Communication and Interaction}} (Tokyo, AA, Japan)
  \emph{(\bibinfo{series}{VINCI '15})}. \bibinfo{publisher}{Association for
  Computing Machinery}, \bibinfo{address}{New York, NY, USA},
  \bibinfo{pages}{168–169}.
\newblock
\showISBNx{9781450334822}
\urldef\tempurl%
\url{https://doi.org/10.1145/2801040.2801072}
\showDOI{\tempurl}


\bibitem[Zhong et~al\mbox{.}(2022)]%
        {qi_dynamic_2022}
\bibfield{author}{\bibinfo{person}{Qi Zhong}, \bibinfo{person}{Jinyi Zhi},
  {and} \bibinfo{person}{Gang Guo}.} \bibinfo{year}{2022}\natexlab{}.
\newblock \showarticletitle{Dynamic is optimal: Effect of three alternative
  auto-complete on the usability of in-vehicle dialing displays and driver
  distraction}.
\newblock \bibinfo{journal}{\emph{Traffic Injury Prevention}}
  \bibinfo{volume}{23}, \bibinfo{number}{1} (\bibinfo{year}{2022}),
  \bibinfo{pages}{51--56}.
\newblock
\urldef\tempurl%
\url{https://doi.org/10.1080/15389588.2021.2010052}
\showDOI{\tempurl}
\showeprint{https://doi.org/10.1080/15389588.2021.2010052}
\newblock
\shownote{PMID: 34937441}.


\end{thebibliography}

%%
%% If your work has an appendix, this is the place to put it.
\appendix
\section{Coded Themes from the AUI Domain Expert Interviews}
\label{interview_appendix}
In the following, we present the results from the automotive domain expert interviews coded in common themes.
Five experts (E1-E5) participated in the semi-structured interviews (see \ref{autovis_expert_interviews}).

\subsection{Visualization of Object Positions and Movements}
\begin{itemize}
\item[E1] "There are many objects inside and outside the vehicle, which can be recorded in sometimes hours of video footage. These are always difficult to keep track of. Manual coding of videos is often time-consuming, and interesting aspects could get overlooked."
\item[E2] "The positions and movements of passengers within the vehicle should be displayed so that it can be understood when, how, and where interaction took place."
\item[E3] "Many interactions with AUIs are triggered by external factors of the vehicle environment, for example, the driver might gesture towards or look at other vehicles or pedestrians at a crosswalk. Therefore, it is essential that the positions and movements of other objects in the environment, such as cars or pedestrians, are visible."
\item[E4] "For eHMI research, a replication of the exterior of a vehicle (or several vehicles) in the same simulation environment is of key importance. In this context, the vehicle's interior would be less relevant."
\end{itemize}

\subsection{Collaborative Analysis}
\begin{itemize}
\item[E2] "Most of the time, many researchers collaborate in an AUI research project. They often collaborate across locations and different time zones. One difficulty is working together on an analysis at the same time. That's why people use online notebooks that persistently store notes and analysis without needing someone to share them live, so they can work independently on the same file."
\item[E5] "The data from automotive studies are usually analyzed in a team, with one person doing the main work, such as preparing the data for analysis, and then later the other authors iterating together over the initially collected results and annotating anomalies."
\item[E3] "For collaboration to be effective, the data should be stored persistently (like an online notebook) so that each collaborator can access the data independently of the others (in terms of time) and mark relevant passages for later discussions."
\item[E4] "One's own knowledge sometimes reaches its limits when evaluating extensive AUI studies, for example, if when analyzing physiological data. In this case, we often consult other researchers in this case, who are more experienced in these areas. It would be great if these additional experts could be also part of the analysis environment to foster interactions."
\end{itemize}

\subsection{Visualizing Data Interdependencies}
\begin{itemize}
\item[E2] "When we record data that represent different passenger states, these are stored in logs that are difficult to read manually. A common problem is that the correlations between the position logs of the passengers and the event logs are not readable and thus have to be reconstructed in a cumbersome way."
\item[E4] "If video recordings and other measurements, such as the heart rate, are added to the event logs, then correlations can no longer be analyzed without additional effort."
\end{itemize}

\subsection{Dataset Annotations}
\begin{itemize}
\item[E4] "Independent of the specific research topic, one should be able to annotate data. This allows one to describe the data for collaborators. In addition, annotated datasets can be effectively used in further processing, such as DL."
\end{itemize}

\subsection{Data Filtering and Modularity}
\begin{itemize}
\item[E1] "Self-recorded but also public datasets are often very extensive. Therefore, it would be handy if these could be filtered so that, for example, only interactions with a specific dashboard element are displayed, although the dataset contains logs about interactions with the entire interior."
\item[E5] "Analogous to development environments like Unity or analysis environments like RStudio, it would be helpful to combine different visualizations and view modules. For example, I would not always be interested in a video of the driver, or I would like the view of the events to be larger to increase visibility."
\end{itemize}

\subsection{Leveraging a Real Vehicle in the Analysis}
\label{expert_comment_real_vehicle}
\begin{itemize}
\item[E1] "An analysis in a real vehicle only makes sense if the same vehicle has been used across all participants in a study. In naturalistic driving studies, each study participant drives his or her own vehicle. Nevertheless, it would be conceivable to use the surfaces of a real vehicle in the analysis, e.g., to visualize the interaction pattern directly on the surfaces."
\item[E3] "I consider the analysis in a real vehicle, while it is driving, to be impractical and not purposeful. The driving environment (apart from the landscape and the city) is constantly changing and therefore not usable in the analysis. Thus, I would rather find the movements and the changing background while driving distracting for a focused analysis."
\item[E4] "In studies that take place in a real car or even in a driving simulator, the study supervisor often can't or isn't allowed to be there and ride in the car and observe the participants up close. It would be great if exactly that were possible in VR, e.g., there could be a detailed simulation of the ride that analysts could relive."
\end{itemize}

\subsection{Enabling Mixed-Immersion Analysis}
\label{expert_results_enabling-mixed-immersion}
\begin{itemize}
\item[E5] "Not everyone has a VR HMD beside their table, so it would be handy to have a desktop environment where you can also analyze and, in the best case, collaborate with the others and not be excluded."
\item[E2] "I think that the immersive analysis using a VR HMD is overhead, for example, if one would like to have only an overview of the events, retrieving outliers, or other measuring errors in the physiological data recordings. Likewise, I don't always find an immersive 3D environment helpful, because I often have to jump to different places and, consequently, have a hard time keeping track of a participant's entire ride."
\item[E4] "In a VR view, there is often a lack of overview of the situation because the viewpoint is usually the first-person perspective of a passenger or bystanders (other cars, pedestrians, etc.). In addition, the feeling of missing something is increased by the fact that I have to turn my head to capture all the information in a 3D space that can appear in 360 degrees."
\end{itemize}

\subsection{Immersive Analysis via VR}
\begin{itemize}
\item[E2] "When participants make certain interactions, I always find it hard to assess why they just happened the way they did. Taking the perspective of the participants would help to see the study from their POV, and then perhaps have a new perspective on the data."
\item[E3] "When analyzing videos and interaction logs, I always have to imagine in my head what the (3D) driving environment just looked like, so that I have a better context when analyzing data. If I designed the study myself and implemented it (e.g., in Unity), I can usually remember the driving environment or just look it up. With other people's studies, I can't do that as easily anymore and have trouble establishing a spatial relationship between the data."
\end{itemize}

\subsection{Interplay Between Immersive and Non-Immersive Analysis}
\begin{itemize}
\item[E4] "I assume that the transition between a VR and desktop application is too cumbersome. If I were analyzing data, I might only see the desktop environment for data preparation or post-processing. If it makes sense, then, I would only use the VR environment as the main analysis tool and not constantly switch between VR and desktop."
\item[E2] "If desktop and VR can be used at the same time, then one person could guide the other via desktop, analogous to the VR preview in desktop in SteamVR."
\end{itemize}

\subsection{Gradual Control Over the Visualization}
\begin{itemize}
\item[E1] "It is very difficult to understand how many things are happening simultaneously in automotive scenarios (e.g., many passengers inside the vehicle, multiple operations outside, complicated relationships between inside and outside, etc.). Therefore, step-by-step time-lapse control would allow a more thorough examination of a scenario at slower or faster playback speeds."
\end{itemize}

\subsection{Linking In-Vehicle and Environment Contexts}
\begin{itemize}
\item[E3] "When a landmark is referenced in a gesture interaction, one must search for the reference in the environment. Such "search" can be time-consuming and difficult if the scene is unknown. In addition, the referenced object could be too far away to see in detail in VR. Similarly, voice interactions are difficult to analyze, if the referenced location/object is not even close to the current street."
\end{itemize}

\subsection{Automatic Conversion of Datasets and Data Preprocessing}
\begin{itemize}
\item[E5] "An analysis tool in the automotive area should merge all available data of a dataset so that one can easily access video, audio, and other logs without having to open three different programs and switch back and forth between windows."
\item[E4] "As there are always many different sensor recordings, logs, video recordings, etc., it would be an enormous relief if a tool could automatically convert and prepare all this data to use it in a plug-and-play environment."
\end{itemize}

\section{Non-Immersive Desktop View Details}
\label{desktop_view_details}
In this section, we describe the detailed concepts for the non-immersive desktop view (see \ref{desktop_view}).
The desktop view is divided into five panels (see \autoref{fig:desktop_view}): \textit{2D panel} (A), \textit{3D scene panel} (B), \textit{video} (C), \textit{inspector} (D), \textit{overview} (E), and \textit{timeline} (F).

The \textbf{timeline panel} consists of: a timeline, event line, and control elements (see \autoref{fig:desktop_view} F).
Similar to ReLive \cite{hubenschmid_relive_2022}, analysts can examine events \requ{R2} and control the tool-wide study playback, for example, to directly jump to interesting points within the data.
Analysts can also annotate sections on the timeline with labels that are automatically added to the dataset and visualized as tags on the event line \requ{R6}.
The timeline enables audio playback and shows when participants spoke as audio events \requ{R7}.
The event line hosts four types of events: \textit{(inter)action}, \textit{emotion}, \textit{driving}, and \textit{activity}, which can be automatically derived from a dataset (see \ref{preprocessing}).
The participants' events are color-coded to reduce visual clutter (see \autoref{fig:desktop_view} F).
\textit{Interaction} events describe passenger interactions, such as touches or gestures.
In contrast, \textit{activity} events are any passenger activity unrelated to UI interactions.
Lastly, \textit{driving} events describe the driving scenario (e.g., red traffic lights or accelerations).

While the timeline was adopted from \citet{hubenschmid_relive_2022}, we added line diagrams in the \textbf{2D panel} for non-spatial temporal data, such as physiological data, to overcome the AUI-specific challenge of ubiquitous physiological measures (see \ref{autovis_expert_interviews}).
The 2D panel provides an overview of 2D data and events (see \autoref{fig:desktop_view} A) enabling quick identification of relevant data sequences \requ{R3} for further examinations on the timeline, in the 3D scene, or VR.
At the bottom, an event line facilitates the \textit{vertical} recognition of correlations between data streams and events.
In addition, a red horizontal line helps detect deviations from the mean and, combined with an outlier detection (in yellow, based on the \textit{1.5xIQR rule} \cite{upton1996understanding}), hints at data anomalies.
However, like other outlier detectors, this should be used cautiously, due to false positives.
Besides, analysts can mark sections in the line diagrams to zoom \requ{R5}.

\begin{figure*}
        \centering
        \includegraphics[width=\textwidth]{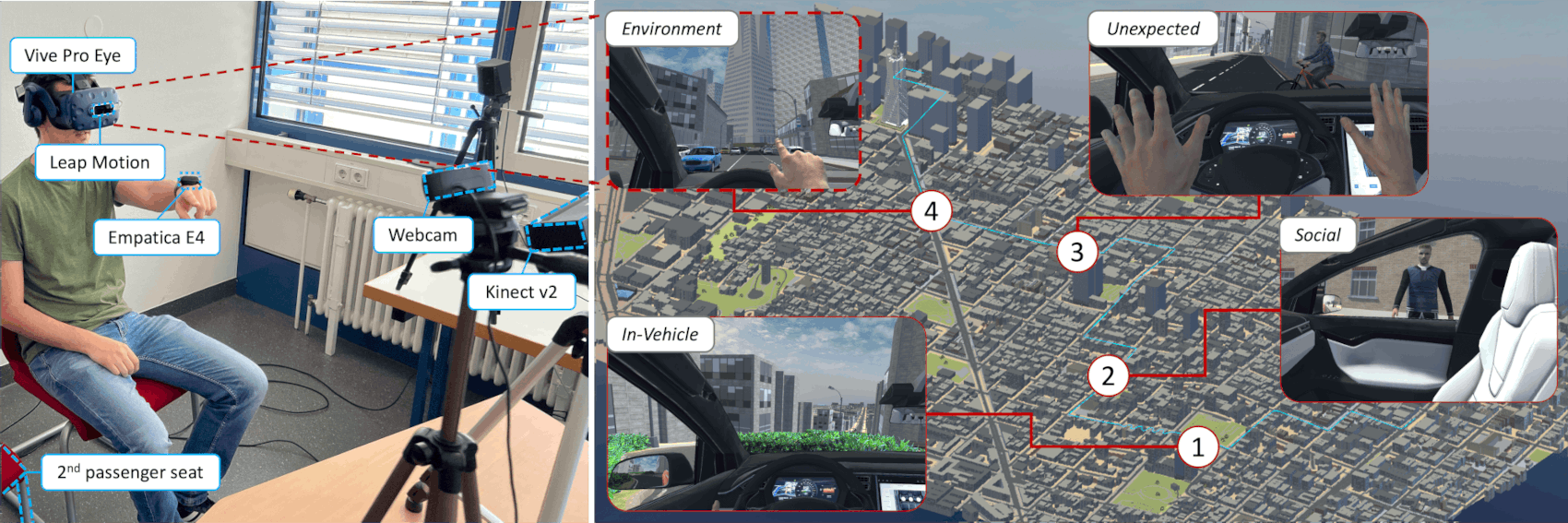}
    \caption{\textbf{Left}, use case study setup. \textbf{Right}, test track in San Francisco with: (1) \textit{in-vehicle} navigation task at Lombard Street, (2) passenger in the \textit{social} event, (3) \textit{unexpected} cyclist crossing, and (4) \textit{environment} sightseeing at the Transamerica Pyramid.}
    \label{fig:case_study}
    \Description{The figure is split into two images. On the left, a real picture of a person sitting on a chair can be seen. The person is wearing a VR headset with an additional hand-tracking sensor glued to it, as well as an Empatica E4 wearable sensor on their left hand. The person is pointing at something. Additional sensors (Kinect v2) and a webcam point at the person. The picture on the right shows an aerial view of a virtual city. A cyan-colored line represents a path a car took through the city. Along the path, four markings show special events with close-up images of said events. All event-related images are shot from the POV of the person in the driver's seat. The first event is the car driving through a winding road with hedges. In the second event, another passenger enters the car. In the third picture, a cyclist unexpectedly appears in front of the vehicle. The passenger has his arms raised in shock. In the fourth picture, the passenger points at a building the car is driving towards.}
\end{figure*}

To address \requ{R1} and \requ{R4}, the \textbf{3D scene panel} replicates the original study environment and visualizes the study vehicles' movements using a virtual ego-vehicle (see \autoref{fig:desktop_view} B).
%The goal is to provide a realistic replication of passengers' interactions with the vehicle and the environment \requ{R4}.
The 3D panel also replicates objects (in the range of vehicle sensors), such as buildings, pedestrians, cars, and cyclists \requ{R1}.
In contrast to ReLive \cite{hubenschmid_relive_2022}, which replicates smaller environments (e.g., a room), the 3D scene panel replicates large environments (using GPS logs) to overcome the AUI-specific problem of large distances between objects of interest (see \ref{autovis_expert_interviews}).
Similar to immersive analytics tools, such as MIRIA \cite{buschel_miria_2021} and ReLive \cite{hubenschmid_relive_2022}, analysts can re-experience the scene from any POV using free movement and predefined virtual (isometric) camera positions.
However, to account for the AUI-specific challenge of volatile in-vehicle and environmental contexts (see \ref{autovis_expert_interviews}), analysts can slice the virtual ego-vehicle at any axis to see inside the interior (see \autoref{fig:desktop_view} B) and select objects to track them through the scene \requ{R5}.
The 3D scene panel shows the same content as the VR view, as they are generated from the same JSON config file.

The \textbf{inspector panel} provides meta information on 3D scene objects upon selection in the 3D scene panel (e.g., the ego vehicle's speed) and study-specific metrics, such as demographic data (see \autoref{fig:desktop_view} D).
%Selecting a 3D scene object shows more detailed information in the inspector, such as the ego vehicle's position and speed.
The \textbf{overview panel} lists all 3D scene objects, available participants samples, and visualization settings addressing \requ{R5} (see \autoref{fig:desktop_view} E).
Participants have a unique adjustable color throughout \tool.
In addition, analysts can select which subsets of participants are visualized tool-wide.
They can also toggle the visibility of avatars, trajectories, heatmaps, and events.

\section{Use Case Study Details: Multimodal Interactions in AVs}
\label{use_case_study_details}
We provide the details on the apparatus and procedure of the use case study on multimodal in-vehicle interactions (see \ref{case_study}).

\paragraph{\textbf{Apparatus}}
We leverage a static VR simulator using the Vive Pro Eye, as safety regulations forbid a real-world AV study at our institution.
We placed two chairs next to each other to resemble the front seats (see \autoref{fig:case_study}).
%AUI studies commonly use touch, gesture, gaze, and speech for multimodal interactions (see \cite{colley_swivrcarseat_2022}).
%Therefore, we set up a study considering these modalities and multiple sensors, see \autoref{fig:case_study}.
We employed the Empatica E4 \cite{empatica} wristband to record physiological signals (blood volume pulse, inter-beat interval, skin temperature, electrodermal activity, and acceleration).
For body tracking, we used Microsoft Kinect for Windows v2, and for hand tracking, we used the Leap Motion attached to the Vive HMD.
In a Wizard of Oz approach, the study supervisor manually labeled hand gestures, such as pointing, by watching the participants and pressing a button.
In contrast to camera-based methods, this approach can recognize gestures only understandable to humans.
However, the timing could be slightly off.
The Vive microphone was used to record speech input, and the built-in eye-tracking recorded gaze and pupil size.
We also captured the participants' behavior using a webcam. %to provide an overview of the in-vehicle situation.
%We instrumented a Tesla Model 3 with a virtual 15.4-inch center display as a virtual vehicle.
As a virtual ego-vehicle, we instrumented a Tesla Model X capable of automated driving.
We detected touch inputs via Unity GameObject collisions with the virtual hands.

The virtual test track has a total length of about 2400 meters and resembles the downtown of San Francisco, see \autoref{fig:case_study} right.
We selected this environment as Waymo currently employs fully automated taxis in that area\footnote{\url{https://waymo.com/sf/}; Accessed: 01.02.2023}, making it a realistic testbed.
The environment was created in Unity 2020.3.37f1, and the city layout was generated with CityGen3D \cite{citygen} using OpenStreetMap data of a 3.9 $km^{2}$ area.
%We manually inserted landmarks, such as the Transamerica Pyramid and Lombard Street.
%Blood volume pulse,@64Hz
%Inter beat interval: time, IBI(time) pair
%Electrodermal activity @4 Hz
%XYZ raw acceleration @32Hz
%Skin temperature @4Hz*

\paragraph{\textbf{Procedure}}
For a demonstration of \tool, we recorded data from only three participants.
One recording session lasted 12 minutes.
In the real world, the participants wore a VR HMD and sat on the left chair, resembling the driver's seat.
The automated vehicle in the VR simulation traveled at a speed of approx. 35 $km/h$.
%Initially, we informed participants about the AV capabilities and the use case scenario.
Overall, the participants performed four tasks, see \autoref{fig:case_study} (1-4), \textit{in-vehicle}, \textit{social}, \textit{unexpected}, and \textit{environment}, covering important aspects for the analysis of multimodal interactions.
First, they queried their current location (Lombard Street) in an \textit{in-vehicle} navigation task.
Then, another passenger entered the AV in the VR simulation, transforming the vehicle into a \textit{social} place.
However, in the real world, this passenger just sat down on a chair next to the participant.
After that, the AV performed an \textit{unexpected} emergency break because a cyclist crossed the street.
Finally, the participants referred to a landmark in the \textit{environment} (Transamerica Pyramid).
In each task, participants could freely interact with one or multiple modalities (i.e., touch, gesture, gaze, and speech) simultaneously.

\end{document}